\documentclass[fleqn,usenatbib]{mnras}

\usepackage{newtxtext,newtxmath}
\usepackage[T1]{fontenc}

\DeclareRobustCommand{\VAN}[3]{#2}
\let\VANthebibliography\thebibliography
\def\thebibliography{\DeclareRobustCommand{\VAN}[3]{##3}\VANthebibliography}

\usepackage{graphicx}	% Including figure files
\usepackage{amsmath}	% Advanced maths commands

\usepackage[normalem]{ulem}

\title[The hot nuclear regions of NGC~7582]{Shock-driven heating in the circumnuclear star-forming regions of NGC~7582: Insights from JWST NIRSpec and MIRI/MRS spectroscopy}

\author[O. Veenema et al.]{
Oscar Veenema,$^{1}$\thanks{E-mail: oscar.veenema@physics.ox.ac.uk}
Niranjan Thatte,$^{1}$
Dimitra Rigopoulou,$^{1, 2}$
Ismael Garc\'{i}a-Bernete,$^{3}$
\newauthor
Almudena Alonso-Herrero,$^{3}$
Anelise Audibert,$^{4, 5}$
Enrica Bellocchi,$^{6}$
Andrew J. Bunker,$^{1}$
\newauthor
Steph Campbell,$^{7}$
Francoise Combes,$^{8}$
Ric I. Davies,$^{9}$
Daniel Delaney,$^{10, 11}$
Fergus Donnan,$^{1}$
\newauthor
Federico Esposito,$^{12}$
Santiago Garc\'{i}a-Burillo,$^{12}$
Omaira Gonzalez Martin,$^{13}$
Laura Hermosa Mu{\~n}oz,$^{3}$
\newauthor
Erin K. S. Hicks,$^{10, 11, 14}$
Sebastian F. Hoenig,$^{15}$
Nancy A. Levenson,$^{16}$
Chris Packham,$^{14, 17}$
\newauthor
Miguel Pereira-Santaella,$^{18}$
Cristina Ramos Almeida,$^{4, 5}$
Claudio Ricci,$^{19, 20}$
Rogemar A. Riffel,$^{21, 22}$
\newauthor
David Rosario,$^{7}$
Lulu Zhang$^{14}$
\\
$^{1}$Department of Physics, University of Oxford, Keble Road, Oxford, OX1 3RH, UK\\
$^{2}$School of Sciences, European University Cyprus, Diogenes street, Engomi, 1516 Nicosia, Cyprus\\
$^{3}$Centro de Astrobiolog\'{i}a (CAB), CSIC-INTA, Camino Bajo del 497 Castillo s/n, E-28692 Villanueva de la Ca{\~n}ada, Madrid, Spain\\
$^{4}$Instituto de Astrof\'{i}sica de Canarias, Calle V\'{i}a L\'{a}ctea, s/n, E-38205, La Laguna, Tenerife, Spain\\
$^{5}$Departamento de Astrof\'{i}sica, Universidad de La Laguna, E-28206, La Laguna, Tenerife, Spain\\
$^{6}$Departmento de F\'{i}sica de la Tierra y Astrof\'{i}sica, Fac. de CC F\'{i}sicas, Universidad Complutense de Madrid, E-28040 Madrid, Spain\\
$^{7}$School of Mathematics, Statistics and Physics, Newcastle University, Newcastle upon Tyne NE1 7RU, UK\\
$^{8}$Observatoire de Paris, LUX, Collège de France, CNRS, PSL University, Sorbonne University, 75014, Paris, France\\
$^{9}$Max Planck Institute for extraterrestrial Physics, Giessenbachstrasse 1, 85748, Garching, Germany\\
$^{10}$Department of Physics \& Astronomy, University of Alaska Anchorage, Anchorage, AK 99508-4664, USA\\
$^{11}$Department of Physics, University of Alaska Fairbanks, Fairbanks, AK 99775-5920, USA\\
$^{12}$Observatorio Astron\'{o}mico Nacional (OAN-IGN)-Observatorio de Madrid, Alfonso XII, 3, 28014 Madrid, Spain\\
$^{13}$Instituto de Radioastrononom\'{i}a y Astrof´ısica (IRyA), Universidad Nacional Autonoma de Mexico, Mexico\\
$^{14}$Department of Physics and Astronomy, The University of Texas at San Antonio, 1 UTSA Circle, San Antonio, TX 78249, USA\\
$^{15}$School of Physics and Astronomy, University of Southampton, Southampton, SO17 1BJ, UK\\
$^{16}$Space Telescope Science Institute, San Martin Drive, Baltimore, MD 21218, USA\\
$^{17}$National Astronomical Observatory of Japan, National Institutes of Natural Sciences (NINS), 2-21-1 Osawa, Mitaka, Tokyo 181-8588, Japan\\
$^{18}$Instituto de F\'{i}sica Fundamental, CSIC, Calle Serrano 123, 28006 Madrid, Spain\\
$^{19}$Department of Astronomy, University of Geneva, ch. d’Ecogia 16, 1290, Versoix, Switzerland\\
$^{20}$Instituto de Estudios Astrof\'isicos, Facultad de Ingenier\'ia y Ciencias, Universidad Diego Portales, Av. Ej\'ercito Libertador 441, Santiago, Chile\\
$^{21}$Departamento de F\'isica, CCNE, Universidade Federal de Santa Maria, 97105-900 Santa Maria, RS, Brazil\\
$^{22}$Centro de Astrobiolog\'ia (CAB), CSIC-INTA, Ctra. de Ajalvir km 4, Torrejón de Ardoz, E-28850, Madrid, Spain\\
}

\date{Accepted XXX. Received YYY; in original form ZZZ}

\pubyear{\the\year{}}

\begin{document}
\label{firstpage}
\pagerange{\pageref{firstpage}--\pageref{lastpage}}
\maketitle

\begin{abstract}

We present combined JWST NIRSpec and MIRI/MRS integral field spectroscopy data of the nuclear and circumnuclear regions of the highly dust obscured Seyfert 2 galaxy NGC~7582, which is part of the sample of AGN in the Galaxy Activity, Torus and Outflow Survey (GATOS). 
%Mid-infrared 
Spatially resolved analysis of the pure rotational H$_2$ lines (S(1)–S(7)) reveals a characteristic power-law temperature distribution in different apertures, with the two prominent southern star-forming regions exhibiting unexpectedly high molecular gas temperatures, comparable to those in the AGN powered nuclear region. We investigate potential heating mechanisms including direct AGN photoionisation, UV fluorescent excitation from young star clusters, and shock excitation. We find that shock heating gives the most plausible explanation, consistent with multiple near- and mid-IR tracers and diagnostics. Using photoionisation models from the PhotoDissociation Region Toolbox, we quantify the ISM conditions in the different regions, determining that the southern star-forming regions have a high density ($n_H \sim 10^{5}$ cm$^{-3}$) and are irradiated by a moderate UV radiation field ($G_0 \sim 10^{3}$ Habing). Fitting a suite of Paris-Durham shock models to the rotational H$_2$ lines, as well as rovibrational 1-0 S(1), 1-0 S(2), and 2-1 S(1) H$_2$ emission lines, we find that a slow ($v_s \sim 10$ km/s) C-type shock is likely responsible for the elevated temperatures. Our analysis loosely favours local starburst activity as the driver of the shocks and circumnuclear gas dynamics in NGC~7582, though the possibility of an AGN jet contribution cannot be excluded.

\end{abstract}

\begin{keywords}
galaxies: individual: NGC~7582 -- galaxies: active -- galaxies: kinematics and dynamics -- galaxies: nuclei --  galaxies: Seyfert -- infrared: galaxies 
\end{keywords}

%%%%%%%%%%%%%%%%%%%%%%%%%%%%%%%%%%%%%%%%%%%%%%%%%%

%%%%%%%%%%%%%%%%% BODY OF PAPER %%%%%%%%%%%%%%%%%%

\section{Introduction}

At the centres of many galaxies reside Active Galactic Nuclei (AGN), powered by supermassive black hole (SMBH) accretion. AGN play a crucial role in galaxy evolution \citep{2015ARA&A..53...51S}, influencing their host environments through a range of feedback mechanisms, including their gas outflows (forming ionisation cones), powerful relativistic jets that can shock the interstellar medium (ISM), and radiation from their accretion disks to name just a few \citep{fabian2012observational, harrison2017impact, harrison2024observational}. While the source of AGN emission is highly localised (on the scale of less than a parsec), AGN impact extends across the entire galaxy, affecting different regions in diverse ways, from quenching star formation to driving outflows that regulate gas dynamics \citep{martin2021anisotropic, piotrowska2022quenching}.

One key tracer of the physical properties of the ISM is molecular hydrogen (H$_2$), being the most abundant molecule in galaxies \citep{fletcher2021cosmic}. Because the H$_2$ molecule lacks a permanent electric dipole moment it emits lines in the infrared and far-ultraviolet which arise from transitions between its rotational and vibrational energy levels.
%radiation at discrete frequencies due to its lack of a permanent dipole moment. 
%These emission lines arise from transitions involving changes in the rotational, vibrational, or rovibrational states of the two hydrogen nuclei. 
Each of these transitions is characterised by a distinct excitation temperature, making H$_2$ an invaluable diagnostic tool for probing the physical conditions of the molecular gas phase within galaxies. The pure rotational $\nu = 0$–$0$ transitions are particularly valuable as they trace molecular gas with temperatures ranging from cool to warm ($T \sim 100-1000$ K) \citep{roussel2007warm}, with emission lines in the mid-infrared ($ \sim 3 < \lambda < 30 \mu m$). Therefore, H$_2$ rotational lines serve as sensitive tracers of warm molecular gas ($\gtrsim 100$ K), which is abundant in photodissociation regions (PDRs) \citep{hollenbach1997dense, kaplan2021near}.

The analysis of H$_2$ emission lines has become an increasingly valuable tool for probing ISM conditions and excitation mechanisms, particularly over the past few decades with the advent of space-based telescopes. One early study was conducted by \citet{rigopoulou2002iso} who employed Infrared Space Observatory (ISO) \citep{kessler1996infrared} observations to investigate rotational H$_2$ emission in a sample of starburst and Seyfert galaxies, using these data to estimate the mass of warm molecular gas and explore possible excitation mechanisms. Subsequent studies using the Spitzer Space Telescope \citep{werner2004spitzer} further advanced this field. \citet{higdon2006spitzer} observed rotational H$_2$ emission in numerous ultraluminous infrared galaxies (ULIRGs), deriving the temperature and mass of their warm molecular gas and concluding that the emission was consistent with an origin in PDRs associated with star formation. Additionally, \citet{roussel2007warm} utilised Spitzer data from many galaxies to compare H$_2$ rotational emission across different galaxy morphologies, helping to constrain the dominant excitation mechanisms in galaxies of varying types. Collectively, studies such as these have demonstrated the diagnostic power of rotational H$_2$ emission lines in unveiling the physical processes shaping the ISM.

Some of the H$_{2}$ rovibrational emission lines are accessible from the ground and thus have historically been the widely studied. 
%H$_2$ transitions. 
They trace hot molecular gas at temperatures of $\gtrsim 1000$~K, serving as important diagnostics of shocks, UV fluorescence, and X-ray heating in star-forming regions and AGN environments \citep[e.g.][]{davies2003molecular,rodriguez2004molecular,hicks2009role, colina2015understanding, Riffel2021}
. However, they probe only a small fraction of the total molecular gas reservoir, as most of the gas remains at much lower kinetic temperatures, even in AGN-dominated systems. 
%Consequently, rovibrational lines are relatively weak compared to the pure rotational lines, and their interpretation often relies on extrapolating from the hot component to infer the properties of the cooler gas. Ground-based integral field studies \citep{davies2005molecular, emonts2017outflows} reveal that this emission is common in regions influenced by AGN activity, including outflows and photodissociation regions. These lines have further been employed to estimate hot molecular gas outflow rates and constrain the properties of intense ionising sources, offering valuable insights into AGN feedback and ISM conditions. 
In contrast, the pure rotational lines directly probe the dominant warm molecular phase and are more representative of the bulk of the warm gas, though they remain inaccessible from the ground due to atmospheric absorption.

The launch of the James Webb Space Telescope (JWST) \citep{rigby2023science, gardner2023james} has revolutionised the study of AGN and their host galaxies. With its unparalleled sensitivity, JWST’s NIRSpec \citep{Jakobsen_2022, boker2023orbit} and MIRI/MRS \citep{wells2015mid, argyriou2023jwst} instruments enable the spatially resolved spectroscopy of galaxies with unprecedented spatial and spectral resolution compared to other space missions.
%, particularly at the lowest redshifts. 
Thus JWST observations allow for a detailed spatially resolved analysis of H$_2$ emission lines in nearby galaxies
%across localised regions within the same galaxy, 
making it possible to quantify the influence of AGN with far greater precision than ever before.
%, shedding new light on their role in galaxy evolution.

\citet{pereira2022low} analysed JWST/MIRI observations of rotational H$_2$ lines in the nuclear region of nearby galaxy NGC~7319, revealing that interactions between the AGN-driven jet and molecular gas in the galactic disk act to decelerate the jet, producing asymmetric radio hotspots. This provided valuable insights into the complex interplay between AGN outflows and the surrounding ISM through H$_2$ emission line analysis. Similarly, \citet{armus2023goals} utilised MIRI observations of rotational H$_2$ lines to measure the warm molecular gas mass within the central 100 pc of the AGN in NGC~7469. Their analysis revealed a decelerating, stratified AGN driven outflow with broad, blueshifted high-ionisation lines. This emphasised the remarkable capability of JWST to probe multiphase gas dynamics in AGN environments, advancing our understanding of the complex feedback mechanisms in such systems. Furthermore, \citet{davies2024gatos} used MIRI observations of rotational H$_2$ lines from and around the AGN in NGC~5728 to investigate the impact of shocks from its ionisation cone on the surrounding ISM. By combining rovibrational and pure rotational H$_2$ lines, they identified moderate-velocity ($\sim 30$ km s$^{-1}$) shocks in dense gas ($n_H \sim 10^5 \ \mathrm{cm}^{-3}$) irradiated by an external UV field ($G_0 = 10^3$ Habing). This provided insights into the role of shock heating in shaping the molecular gas dynamics within AGN hosting galaxies. Moreover, \citet{esparza2025molecular} investigated warm H$_2$ emission in the active galaxy MCG$-$05$-$23$-$16 using JWST, in comparison with cold molecular gas traced by molecular CO emission lines from ALMA data. They found that the warmer H$_2$ exhibits significantly more complex kinematics than the cold component, highlighting the impact of AGN activity and localised star formation on the molecular gas near the nucleus. Collectively, studies such as these, and others \citep[e.g.][]{costa2024blowing, bohn2024goals, almeida2025jwst}, highlight the unique power of H$_2$ analyses with JWST in spatially resolving molecular gas properties and kinematics in AGN environments, shedding new light on how AGN driven processes influence the ISM.

The paper is structured as follows. Section~\ref{sec:target_section} summarises the properties of the target, 
%reviews previous studies of our target galaxy, 
NGC~7582. 
%outlining its structure and highlighting why it is a compelling target for study. 
Section~\ref{sec:method_section} describes the JWST NIRSpec and MIRI/MRS observations.
%including data collection, reduction, and corrections. 
In Section~\ref{sec:resultsanddiscussion}, we present results and the analysis of the emission lines, the kinematics and   
%and kinematics of warm molecular gas traced by pure rotational H$_2$ lines in the nuclear and circumnuclear regions. Section~\ref{sec:H2_excitation_modelling_subsection} details the 
modelling of the H$_2$ line fluxes.
%followed by excitation diagrams for multiple apertures in Section~\ref{sec:excitation_diagrams_subsection}. Building on 
%Section~\ref{sec:hot_star_forming_regions_subsection} examines additional diagnostic emission line maps and fluxes across different regions, while in 
We then present models to quantify the ISM parameters and fit shock models before contrasting our new results with previous work to assess the origin of shocks within NGC~7582. Finally, in Section~\ref{sec:conclusions_section} we present our conclusions.

\section{NGC~7582}
\label{sec:target_section}

NGC~7582 is a nearby ($z \sim 0.00525$; \citealp{braito2017high}, $D \sim 22$ Mpc) inclined \citep[$i = 58^\circ$;]{wold2006nuclear} barred spiral galaxy that hosts a heavily dust-obscured, Compton-thick AGN, classified as a Seyfert 2 \citep{bianchi2009complex}. The AGN is associated with strong radio emission and a well-defined ionisation cone \citep{2022ApJ...925..203J}, originally deduced through optical [O III] emission and comparison to a prominent H$\alpha$-emitting, H II-rich disk \citep{morris1985velocity}. The central SMBH has been estimated to have a mass of $5.5 \times 10^7 M_\odot$ based on mid-IR [Ne II] emission line analysis \citep{wold2006nuclear2}. 
%There also remains debate regarding 
Of interest are the presence 
of a kinematically distinct core (KDC) in the innermost nuclear region \citep{2022ApJ...925..203J} associated with stars %forming 
in a circumnuclear ring formed from gas in the disk with major axis $\sim$200 pc from the AGN 
%as seen at multiple wavelengths 
\citep{riffel2009agn, Alonso-Herrero2021, garcia2021galaxy}, 
as well as a 
%and a 
possible radio jet emanating almost perpendicular to the ionisation cone \citep{ricci2018optical}.

While the potential presence of a radio AGN jet 
is not the primary focus of this study, we will refer to it when we discuss the origin of the H$_{2}$ rotational lines.
%it remains an important consideration. 
%If present, the jet would pass close to two southern star-forming regions central to our analysis. 
%Establishing whether such a jet exists is therefore relevant, as it could influence the interpretation of our results. 
The primary evidence supporting the presence of the jet comes from extended radio emission \citep[e.g.,][]{forbes1998star}, out to  $\sim$4${\arcsec}$ north and south of the AGN, but mostly concentrated out to  $\sim2$${\arcsec}$. However, such knot-like radio structures can also arise from intense star formation.
%which can also generate shocks that may be misinterpreted as jet-like features.
Nonetheless, the nucleus of NGC~7582 is complex, harbouring an AGN, as well as a star-forming disk, see also \citet{reunanen2010vlt}, presenting an ideal subject for in-depth investigation into the impact of an AGN on its host galaxy. Additionally, NGC~7582 
is highly obscured
%exhibits a high dust content 
along the line of sight to the nucleus with $A_V \sim 3.6$ mag \citep{ward1980new}.

Roughly 1$\arcsec$ south of the nucleus, NGC~7582 hosts two prominent star-forming regions, previously identified and studied by \citet{wold2006nuclear} and \citet{ricci2018optical}, where they were 
referred to as M1 and M2. These regions are highly luminous in hydrogen recombination lines.
%(indicating high star-formation). 
\citet{wold2006nuclear} interpreted them as young star-forming clusters, estimating their ages to be approximately 1 Myr and suggesting that each may contain thousands of O-type stars. \citet{riffel2009agn} further investigated these regions by comparing the equivalent widths (EWs) of their observed Br$\gamma$ lines with those predicted by various evolutionary photoionisation models. Following the methodology outlined in \citet{dors2008ages}, they estimated the clusters' ages to be $\sim 5$ Myr, with each potentially hosting around 500 O-type stars.

\begin{figure}
   \centering
   \includegraphics[width=\columnwidth]{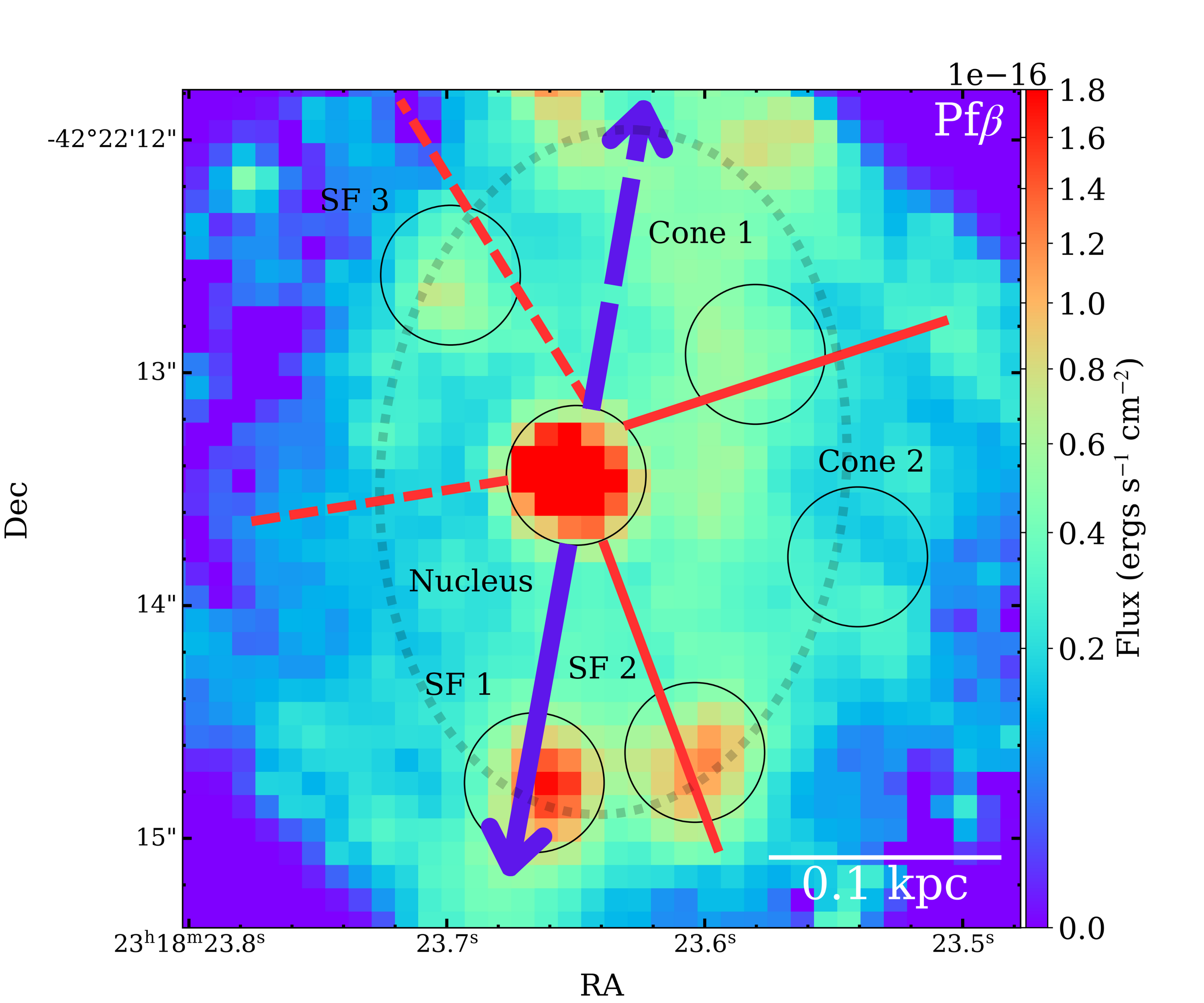}
   \caption{Pf$\beta$ (7-5) hydrogen recombination flux map from NIRSpec-IFS, highlighting regions of strong nuclear activity and/or star-formation. Circles denote apertures for which spectra were extracted. Red lines show the edges of the well established ionisation cones \citep{morris1985velocity}. Dark blue arrows show the proposed direction of the bipolar jets from \citet{ricci2018optical}. Solid lines show orientation towards us and dashed show orientation away from us. The dotted black ellipse shows the approximate position of the star forming ring. North is up, East is to the left.}
    \label{fig:selected_apertures}%
    \end{figure}

In Fig.~\ref{fig:selected_apertures} we present the hydrogen recombination Pf$\beta$ 4.654 $\mu m$ line flux map of the nuclear region of NGC~7582, annotated to show the star-forming disk/ring, the edges of the ionisation cones, and the putative AGN jet directions, based on their relative directions from the nucleus as presented in fig. 17 of \citet{ricci2018optical}. Aperture locations for key regions of interest are also indicated; spectra from these apertures form the basis of our analysis throughout this paper. Details of the observations, data reduction, and aperture extraction are provided in Section~\ref{sec:method_section}. Throughout this paper, all maps of NGC~7582 are shown with the same orientation where North is up and East to the left.

In our nomenclature, the two prominent southern star-forming regions, previously referred to as M1 and M2 \citep{wold2006nuclear, ricci2018optical}, are designated SF~1 and SF~2 to explicitly emphasise their star-forming nature. The selected apertures (Fig.~\ref{fig:selected_apertures}) encompass key regions of interest within the nuclear region, providing a diverse sample for analysis. These include the central AGN-dominated region, the historically well-studied and luminous star-forming regions: SF~1 and SF~2, and two regions within the ionisation cone: Cone~1, located at the northern edge, and Cone~2, a region within the cone itself. 
Additionally, an aperture is placed over SF~3, 
%a smaller star-forming region situated northwest of the nucleus, 
which, unlike SF~1 and SF~2, is not aligned with any potential jet and lies in front of the eastern ionisation cone boundary.
% , compared to the proposed model by \citet{ricci2018optical}, however all three of our selected star forming regions lie on the circumnuclear star forming ring. 
We give the relative RA and Dec of the centre of our apertures to the nucleus in the Appendix, Table~\ref{tab:appendix_table}.
%\bf{Additional star forming clumps, shown as high Pf-$\beta$ flux regions 
%(which are also star forming clumps on the circumnuclear ring) further north} 
%in Fig.~\ref{fig:selected_apertures} were not included 
%as apertures
%in this study, as they are not fully covered by the MIRI/MRS field of view (FOV) in every channel. However, we will occasionally refer to these regions, 
%which have been designated as M4 and M5 in the study by 
%also been studied by 
%\citet{ricci2018optical}}}.
%who designated them as M4 and M5}.
%Overall, our selected apertures provide a diverse and comprehensive dataset for investigating the interactions between AGN components and their host galaxy in the particular case of NGC~7582.

\citet{wold2006nuclear} also identified three additional star-forming clusters in the northern circumnuclear ring, designated M3, M4, and M5. Although less luminous in hydrogen recombination line flux than the southern clusters \citep{ricci2018optical}, they remain noteworthy. These northern clumps are excluded from our analysis due to limited spatial coverage in most MIRI/MRS observations.
% , with M4 and M5 only partially resolved within the NIRSpec FOV (corresponding to the enhanced flux in the northern regions in Fig.~\ref{fig:selected_apertures}).
While not directly analysed, they are referenced throughout this study for comparison with the southern clusters.

The study by \citet{garcia2021galaxy} used ALMA observations of NGC~7582 to probe the cold molecular gas via CO emission, indirectly tracing H$_2$ by assuming a CO-to-H$_2$ conversion factor. They identified a dusty, molecular-rich torus within $\sim 10$ pc of the AGN, a molecular gas deficit across the rest of the inner $\sim 50$ pc, and the molecular-rich star-forming ring at a radius of $\sim 200$ pc. They also traced an outflow originating within the ring, leading to the observed depletion of molecular gas in the central regions — a signature consistent with the presence of the ionisation cone and AGN feedback. Moreover, our selected SF~1, SF~2, and SF~3 apertures all correspond to regions with strong 351 GHz continuum emission (top left panel of fig. 6 in \citealp{garcia2021galaxy}), consistent with strong star formation.

% The orientations of the AGN ionisation cone and putative AGN jet are indicated in Fig.~\ref{fig:selected_apertures} by the red lines and dark blue arrows, respectively, based on their relative directions from the nucleus as presented in fig. 17 of \citet{ricci2018optical}. The red lines mark the approximate outer edges of the cone, with dashed lines and arrow denoting the receding components (eastern cone and northern jet) and solid lines and arrow indicating the approaching components (western cone and southern jet). 
% %NGC~7582 is highly inclined, including its circumnuclear star-forming disk; however, if there were a southern jet, it is believed to be positioned in front of the disk along our line of sight. 
% Throughout this paper, all maps of NGC~7582 are shown with the same orientation where North is up and East to the left.

\section{Observations and data reduction}
\label{sec:method_section}

\subsection{NIRSpec and MIRI/MRS data collection}
\label{sec:datacollection_subsection}

The target studied here is part of the Galaxy Activity, Torus, and Outflow Survey ({\textcolor{blue}{\href{https://gatos.myportfolio.com/}{GATOS}}}; \citealt{garcia2021galaxy, alonso2021galaxy, garcia2024galaxy}). The present study uses near-IR to mid-IR (2.87-28.1~$\mu$m) data collected with the integral-field spectrographs MIRI/MRS with a spectral resolution of R$\sim$1300-3700 (\citealt{labiano2021wavelength}), acquired on October 31st 2023, and NIRSpec with the grating-filter pairs G395H/F290LP (2.87–5.27~$\mu$m) with R$\sim$2700 (\citealt{Jakobsen_2022,boker2022near}), acquired on 7th April 2024. These data are part of the JWST Cycle 2 GO proposal ID\,3535 (PI: I. Garc\'ia-Bernete and D. Rigopoulou).

For the NIRSpec observations, a 4-point medium-size cycling dither pattern optimised for extended sources was adopted, and the NRSRAPID readout mode with 3 groups per integration and 7 integrations per exposure was used. The same configuration was used for the dedicated LEAKCAL observations. For the MIRI/MRS observations, a 4-point extended-source dither pattern and the FASTR1 readout mode were used. Each exposure consisted of 65 groups per integration for the short (A) and medium (B) wavelength settings, and 121 groups per integration for the long (C) setting, with 1 integration per exposure. Background observations were obtained with the same readout settings but using a 2-point dither pattern. Note that for these particular observations, the MIRI/MRS guide star was missing leading to the wrong astrometry and incorrect absolute RA and Dec coordinates (a pointing issue). We corrected these data cubes by shifting the coordinates so that the bright AGN position aligns with its RA and Dec as seen with NIRSpec, where the astrometry was correct.

\subsection{NIRSpec and MIRI/MRS data reduction}
\label{sec:datareduction_subsection}

For the data reduction, we primarily followed the standard pipeline procedure and the same configuration of the pipeline stages described in \citet{garcia2022high} and \citet{pereira2022low} to reduce the data. Some hot and cold pixels are not identified by the current pipeline, so we added some extra steps as described in \citet{pereira2024extended} and \citet{garcia2024galaxy} for NIRSpec and MIRI/MRS, respectively.

One dimensional spectra were extracted from apertures of radius 0.3 arcseconds (corresponding to $\sim$30 pc in radius at the distance of NGC~7582).
%We analyse the spectra of various nuclear regions in NGC~7582 using circular apertures, with 
Their locations and designations are shown in Fig.~\ref{fig:selected_apertures}, showing the hydrogen recombination Pf$\beta$ emission line from NIRSpec G395H/F290LP. 
%This emission line is useful in displaying the locations and morphology of different key components of the circumnuclear region, as it highlights both the central AGN and regions of active star formation. 

%\subsection{PSF Corrections}

\begin{figure*}
   \centering
   \includegraphics[width=\linewidth]{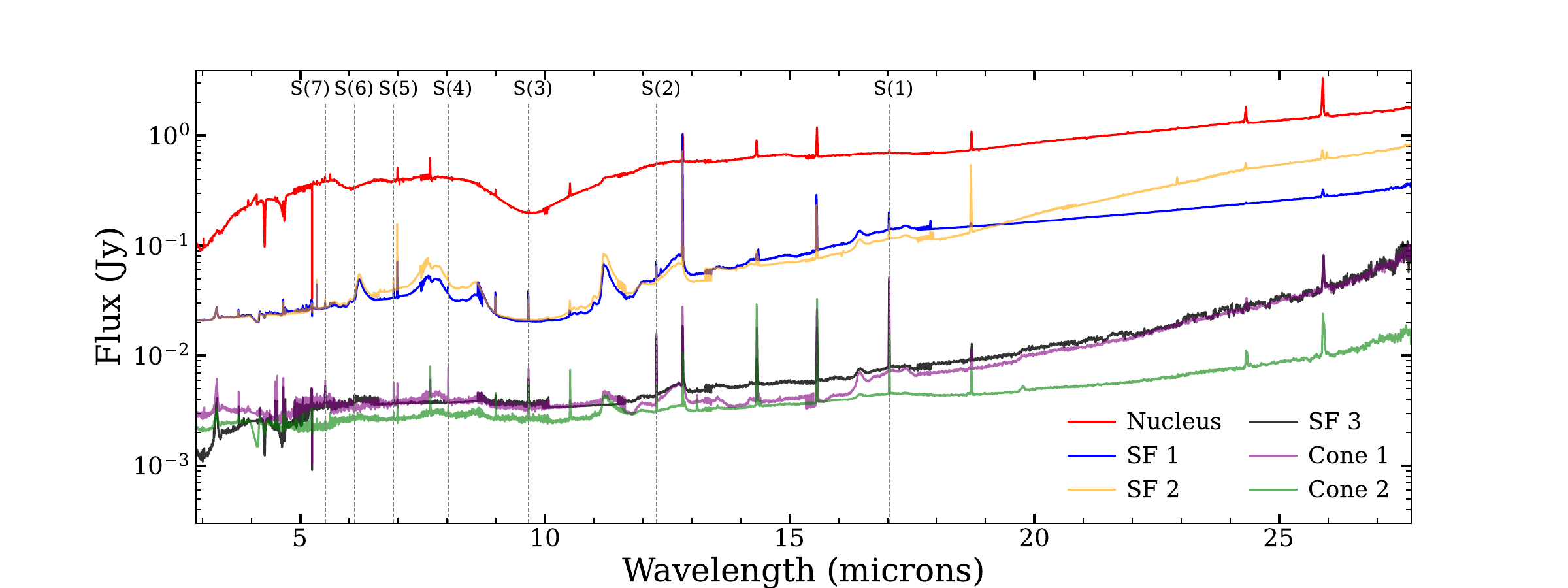}
   \caption{Observed spectrum (rest-frame) for each aperture after applying PSF corrections, but before any extinction corrections. Each line shows the total flux within each aperture at every wavelength bin available across the combined NIRSpec and MIRI/MRS wavelength range. Grey dotted lines show the locations of rotational H$_2$ lines S(1) to S(7). Many atomic emission lines are also clearly visible in each aperture but are not labelled for clarity.}
              \label{fig:uncorrected_fluxes}%
    \end{figure*}

Aperture-specific point spread function (PSF) corrections were then applied to account for the impact of the PSF on the measured fluxes. The PSFs produced by JWST instruments for point sources are well-characterised, and can be modeled using WebbPSF \citep{2014SPIE.9143E..3XP}. For the nuclear aperture, we applied the WebbPSF models directly for NIRSpec G395H/F290LP and MIRI/MRS. Since the nuclear emission is dominated by the AGN, which is effectively a point source, this provides a reliable correction. For the star-forming regions (SF~1, SF~2, and SF~3), the WebbPSF models were convolved with a Gaussian kernel of $\sigma = 0.1$ arcseconds to account for their spatially extended nature, 
%containing stellar clusters rather than single point sources. 
For the remaining apertures (Cone~1 and Cone~2), the flux redistribution caused by the PSF is expected to balance out, as any flux loss from the aperture should be compensated by flux gained from surrounding spaxels 
%as they contain no AGN or significant star forming clumps, 
therefore PSF corrections were not necessary for these apertures. Overall, these corrections were small, of order $\sim 10$\% for each channel. 
The final spectra are presented in 
%We present these `measured' spectra in 
Fig.~\ref{fig:uncorrected_fluxes}.

\subsection{Extinction Corrections}
\label{sec:datacorrections_subsection}

The nucleus of NGC~7582 is highly dust-extinguished, therefore, it is essential to account for the effects of extinction. 
%While dust extinction is significantly more pronounced in the optical and ultraviolet compared to the near- and mid-infrared bands \citep{cardelli1989relationship, salim2020dust}, it still plays a role in reddening the spectra, reprocessing shorter-wavelength light and re-emitting it in the far-infrared. 
Correcting for extragalactic dust extinction is a non-trivial task, with various approaches available \citep{calzetti1996reddening, xue2016precise}. In this work, we employ the differential extinction model developed by \citet{2024MNRAS.529.1386D}. We refer the reader to their paper for a detailed description of the model. In brief, the extinction within each aperture is modeled as a combination of multiple dust screens, each contributing differently to the total attenuation. The extinction of each screen is represented as $e^{-\tau_{9.8}\,\tau(\lambda)}$, where the parameter $\tau_{9.8}$ corresponds to the optical depth at 9.8\,$\mu m$. The model assigns different weights to components with a range of optical depths and dust temperatures. Estimates of the extinction are provided through fitting various features such as the silicate absorption at 9.8$\mu m$, 
%and extra extinction of 
the H$_2$ rotational S$(3)$ line 
%due to it lying in the trough of the silicate absorption just to name a couple, 
as well as the mid-infrared hydrogen recombination lines,
allowing for a more physically-motivated representation of the extinction compared to simpler screen or mixed models. As \citet{2024MNRAS.529.1386D} show, the derived differential extinction model is versatile, and encompasses many different physical ISM conditions such as when the radiation source contains multiple contributions, as is the case here with the presence of an AGN and star-forming regions. The spectra from each aperture are fit using this model, and the resulting extinction correction factors are applied. We give our derived $\tau_{9.8}$ for each region in the Appendix, Table~\ref{tab:appendix_table}. Extinction corrections have been applied to all line fluxes measured in our spectra.

%We correct for PSF effects and dust extinction before measuring any line fluxes. To measure these for each emission line we fit the continuum using a cubic polynomial function with the line region masked. After subtracting the continuum, the emission lines remained as well-defined Gaussian profiles. The flux of each line was calculated by integrating the area under the line pixel-by-pixel.

\subsection{Flux and kinematics maps}

The velocity and velocity dispersion of each spectral line was determined by fitting Gaussian profiles to each continuum subtracted emission line using the CapFit routine \citep{cappellari2023full}. The velocity, $v$, is calculated by comparing the centroid wavelength of the fitted Gaussian profile, $\lambda$, to the systemic wavelength of the line for the galaxy, $\lambda_0$. 
This is given by $v = c \times(\lambda-\lambda_0)/\lambda_0$, where $c$ is the speed of light, and is valid for $v << c$, which is the case here. 
The velocity dispersion, $\sigma_v$, is similarly derived from the Gaussian fit parameters using: $\sigma_v = c \times \sigma / \lambda_0$, where $\sigma$ is the best-fit Gaussian dispersion, expressed in the same units as $\lambda_0$.
%This method provides a robust means of extracting kinematic information from the observed spectral lines, offering insights into the dynamical state of the emitting gas.}}

Line flux, velocity, and velocity dispersion maps were produced following the steps outlined
%by applying the aforementioned procedure 
to individual spaxels for various emission lines within the NIRSpec and MIRI/MRS FOV of NGC~7582. To facilitate a direct comparison between different emission lines, all MIRI maps were reprojected (bilinear interpolation) to match the exact NIRSpec FOV using the \texttt{reproject\_interp} function from \citet{robitaille2020reproject} - this was only done for creating maps, with all quantitative analysis being done using the original data cubes.

\subsection{VLT/SINFONI data}

We supplement our JWST observations with archival near-IR VLT/SINFONI \citep{eisenhauer2003sinfoni} data of NGC~7582, providing additional constraints on shock modelling through rovibrational H$_2$ line measurements in SF~1 and SF~2. The SINFONI data cube (PROGRAM ID 093.B-0057, PI: R. Davies) was reduced using the custom SPRED package \citep{abuter2006sinfoni}, with additional steps to remove the nightsky OH airglow emission \citep{davies2007method}. These data were originally published and discussed in detail by \citet{lin2018llama} (see also references therein). We corrected these data for extinction following \citet{calzetti2000dust}, and comparing the Br$\gamma$ flux to the Pf$\beta$ flux in each aperture under the assumption of case B recombination \citep{hummer1987recombination}.

\section{Results and discussion}
\label{sec:resultsanddiscussion}

\begin{figure*}
   \centering
   \includegraphics[width=0.33\linewidth]{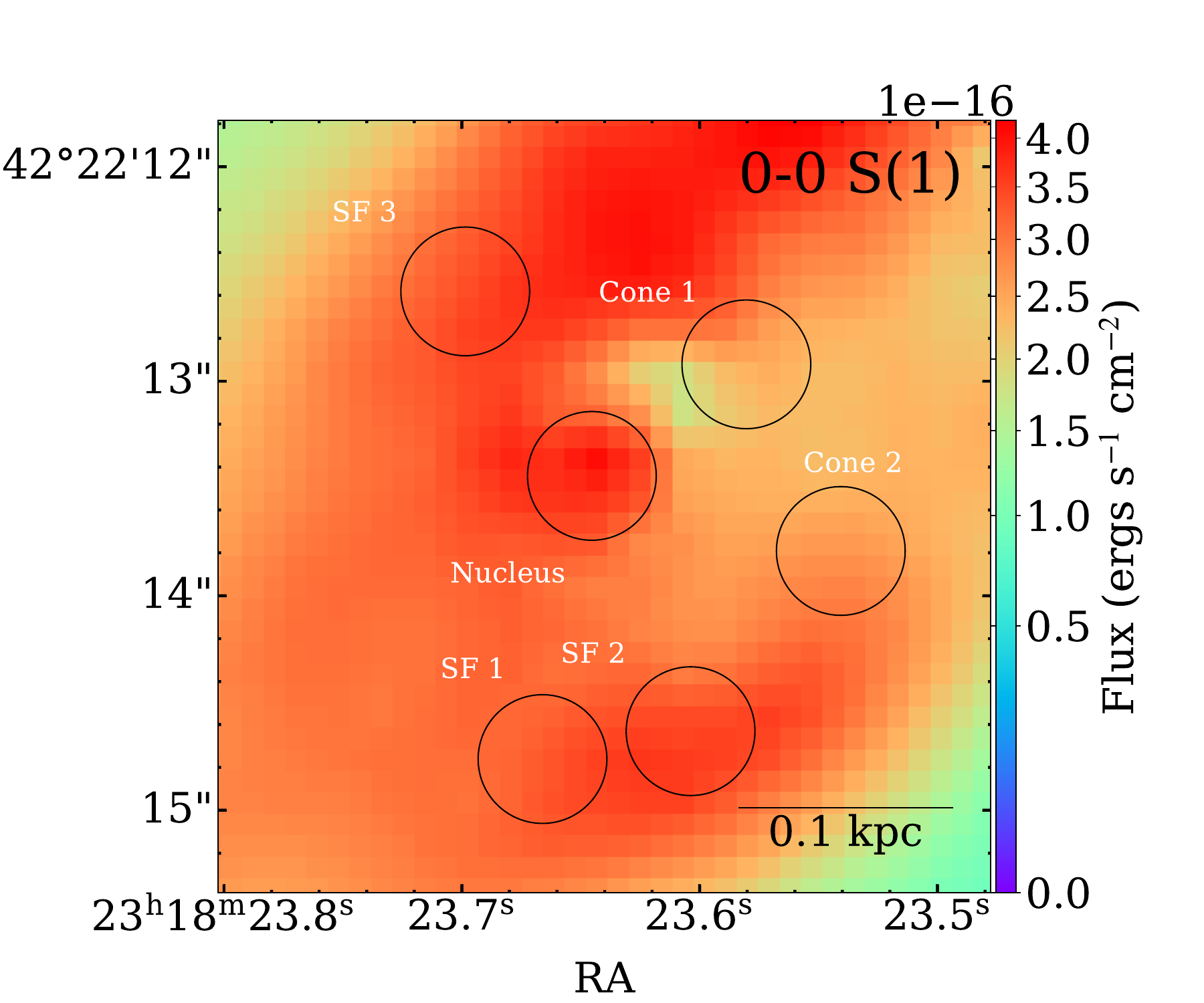}
   \includegraphics[width=0.33\linewidth]{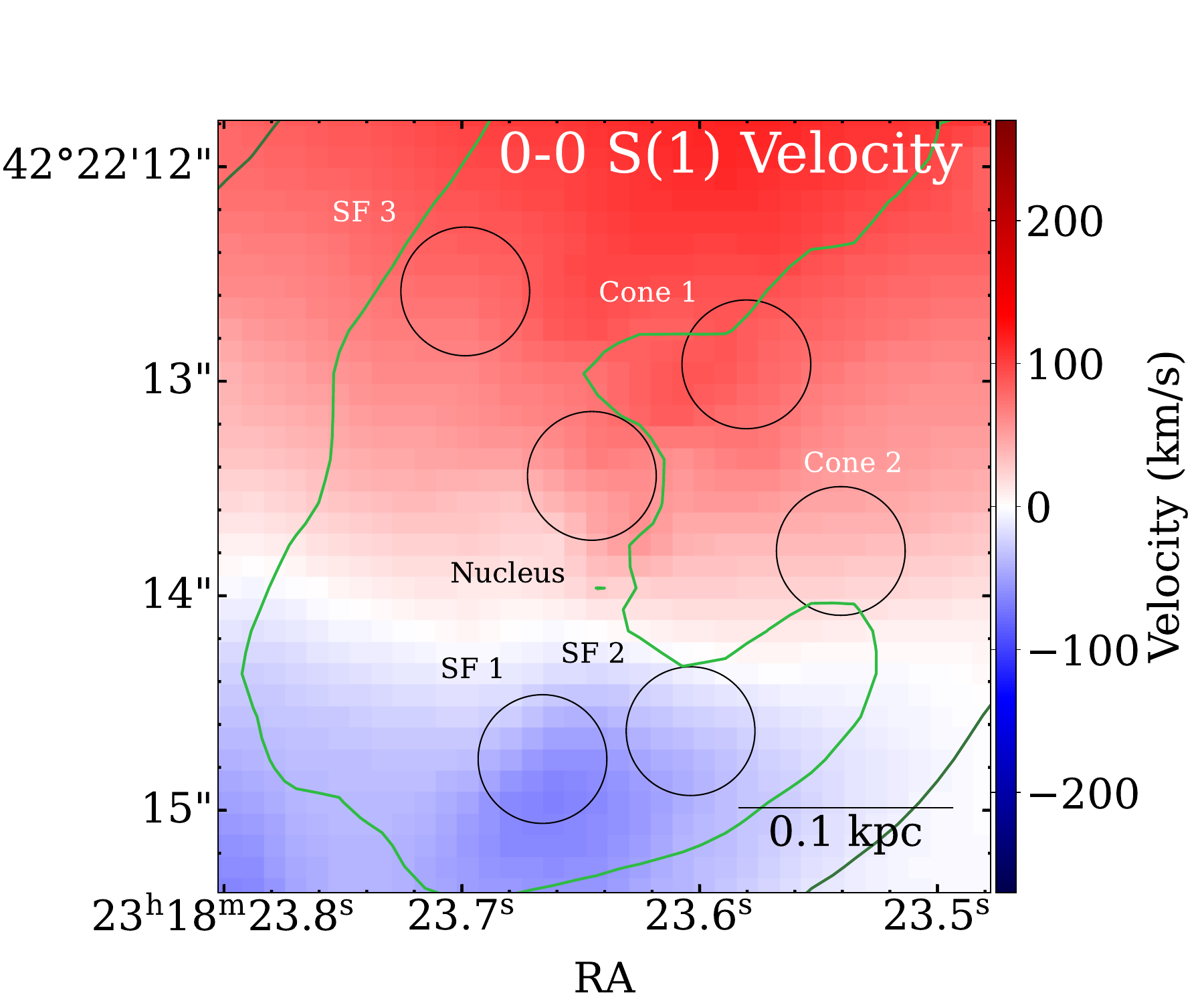}
   \includegraphics[width=0.33\linewidth]{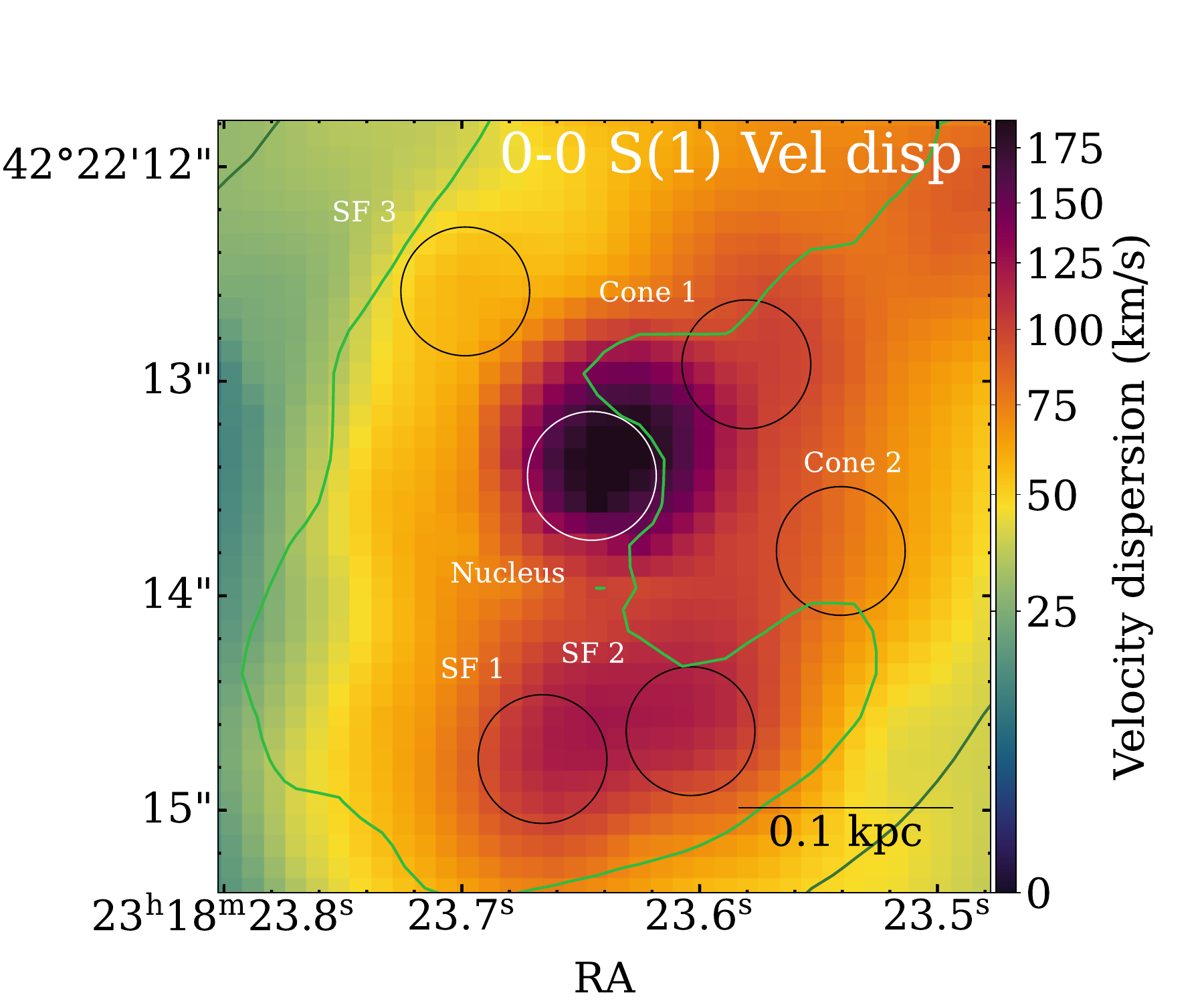}
   \includegraphics[width=0.33\linewidth]{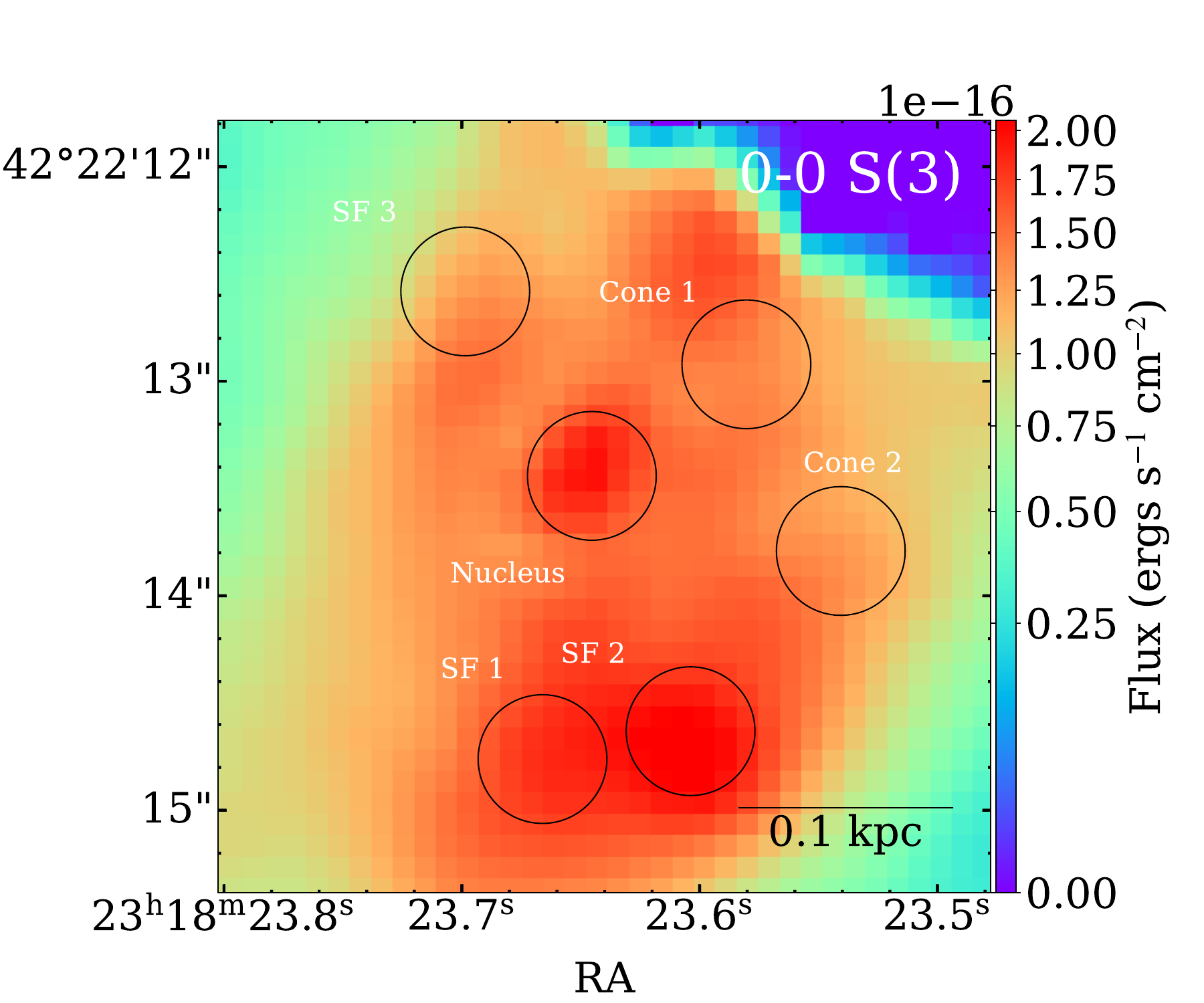}
   \includegraphics[width=0.33\linewidth]{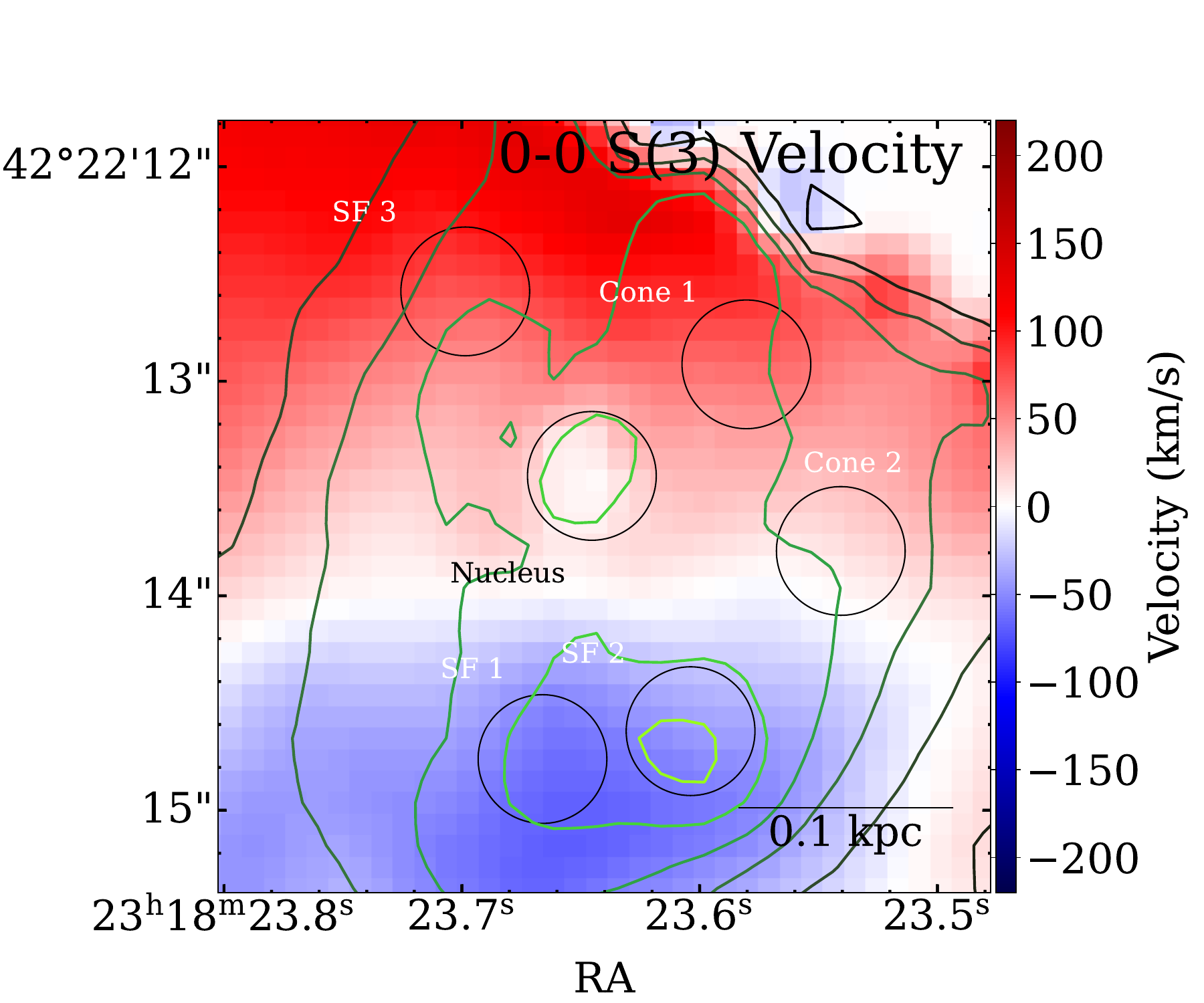}
   \includegraphics[width=0.33\linewidth]{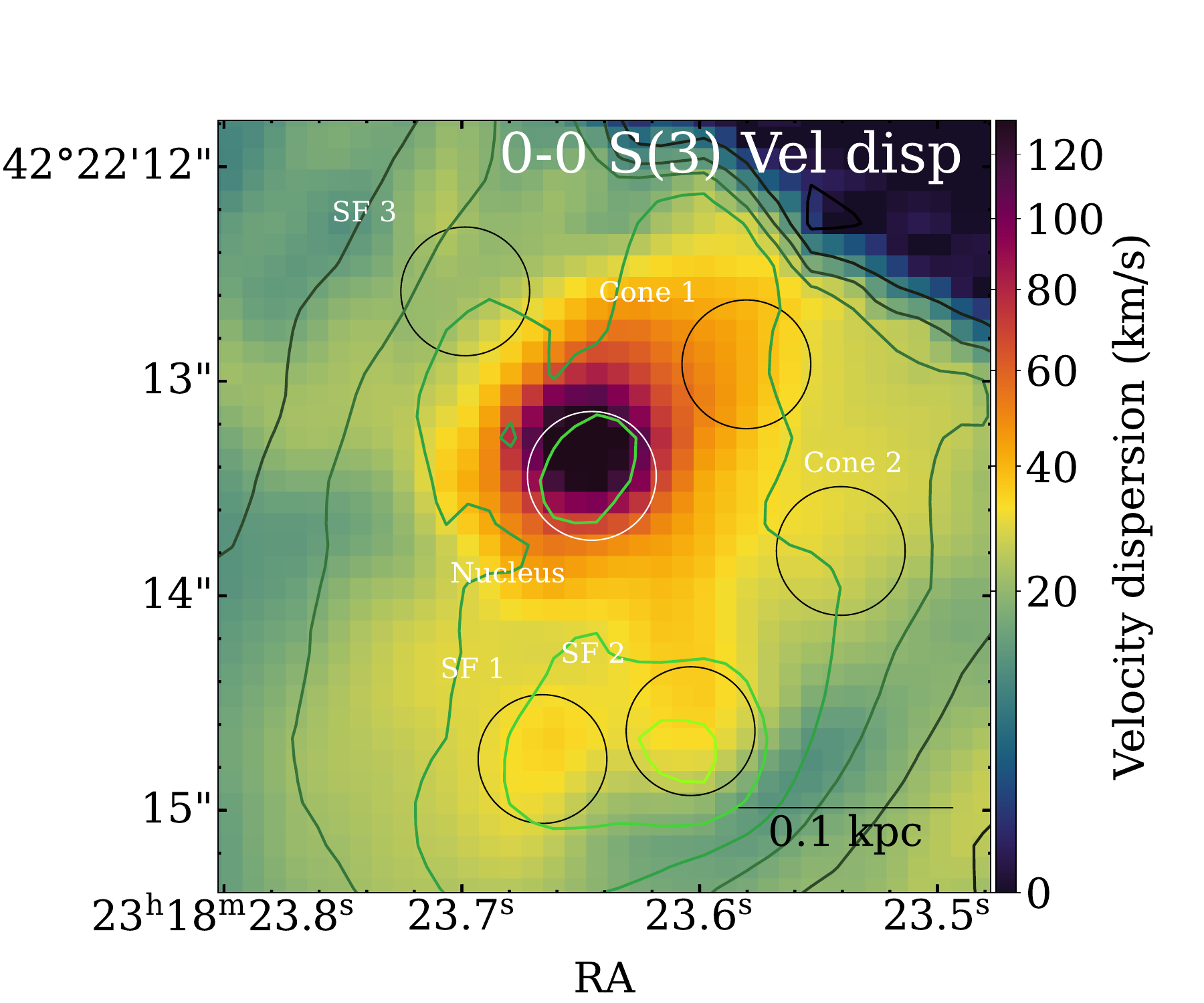}
   \includegraphics[width=0.33\linewidth]{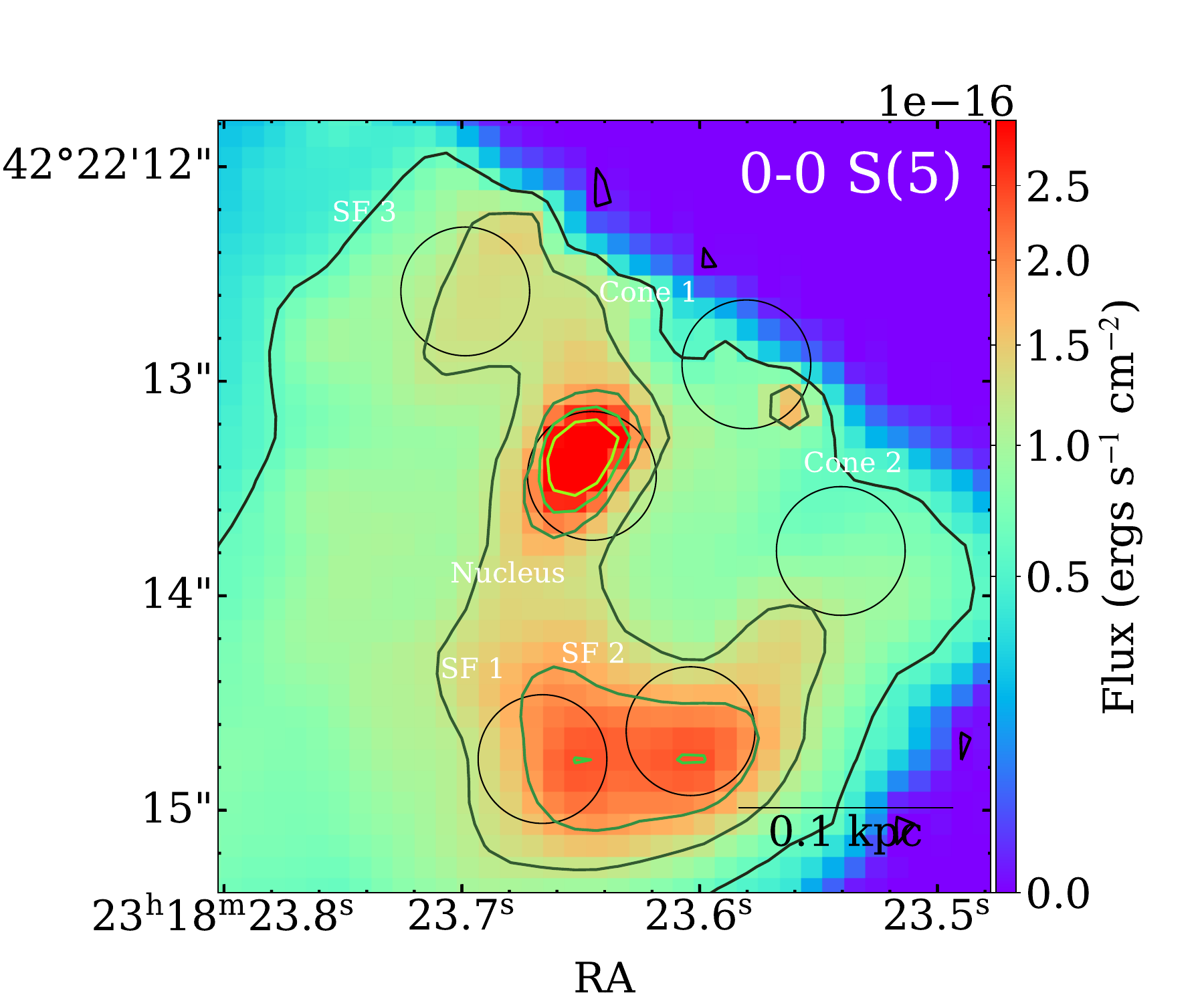}
   \includegraphics[width=0.33\linewidth]{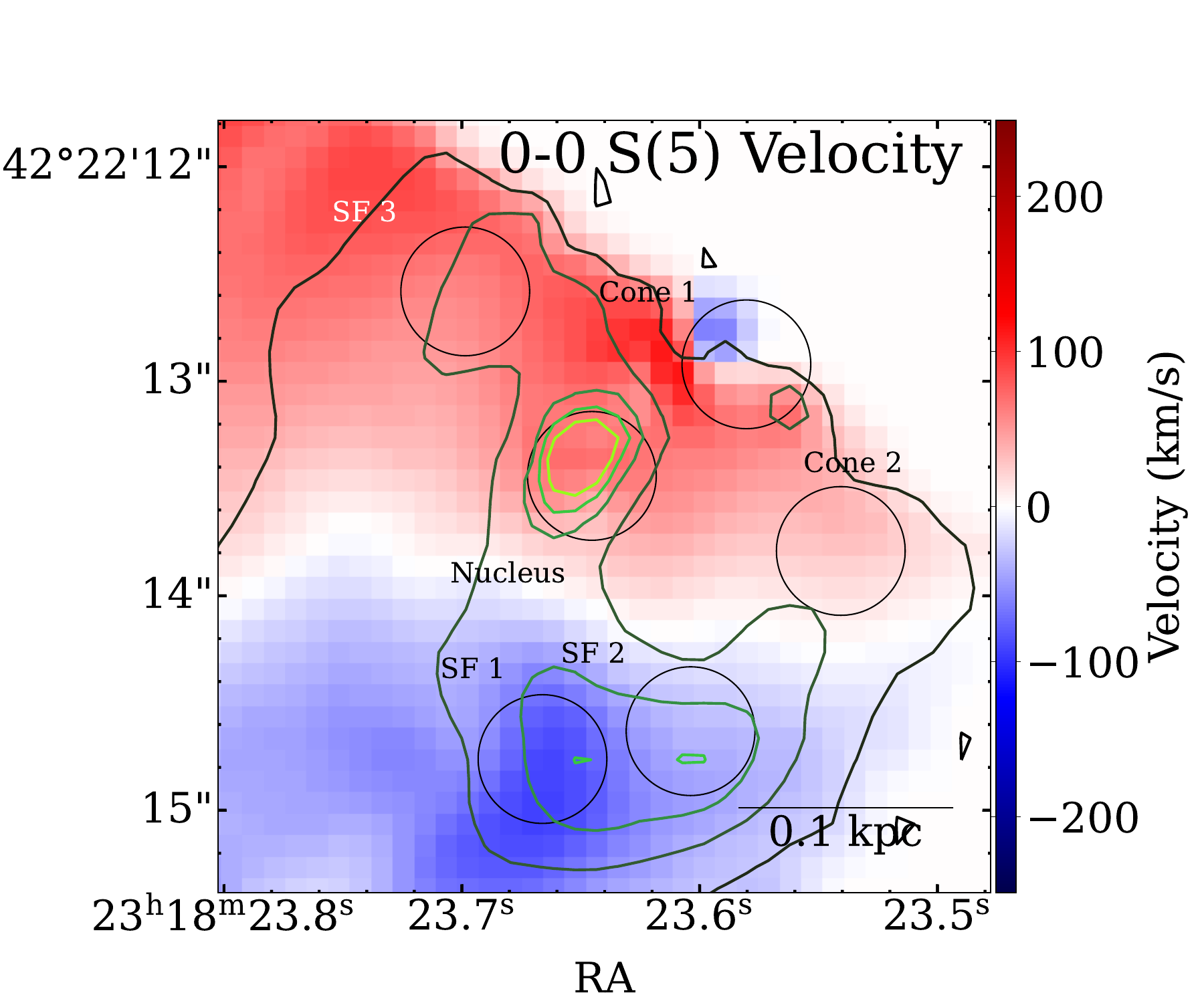}
   \includegraphics[width=0.33\linewidth]{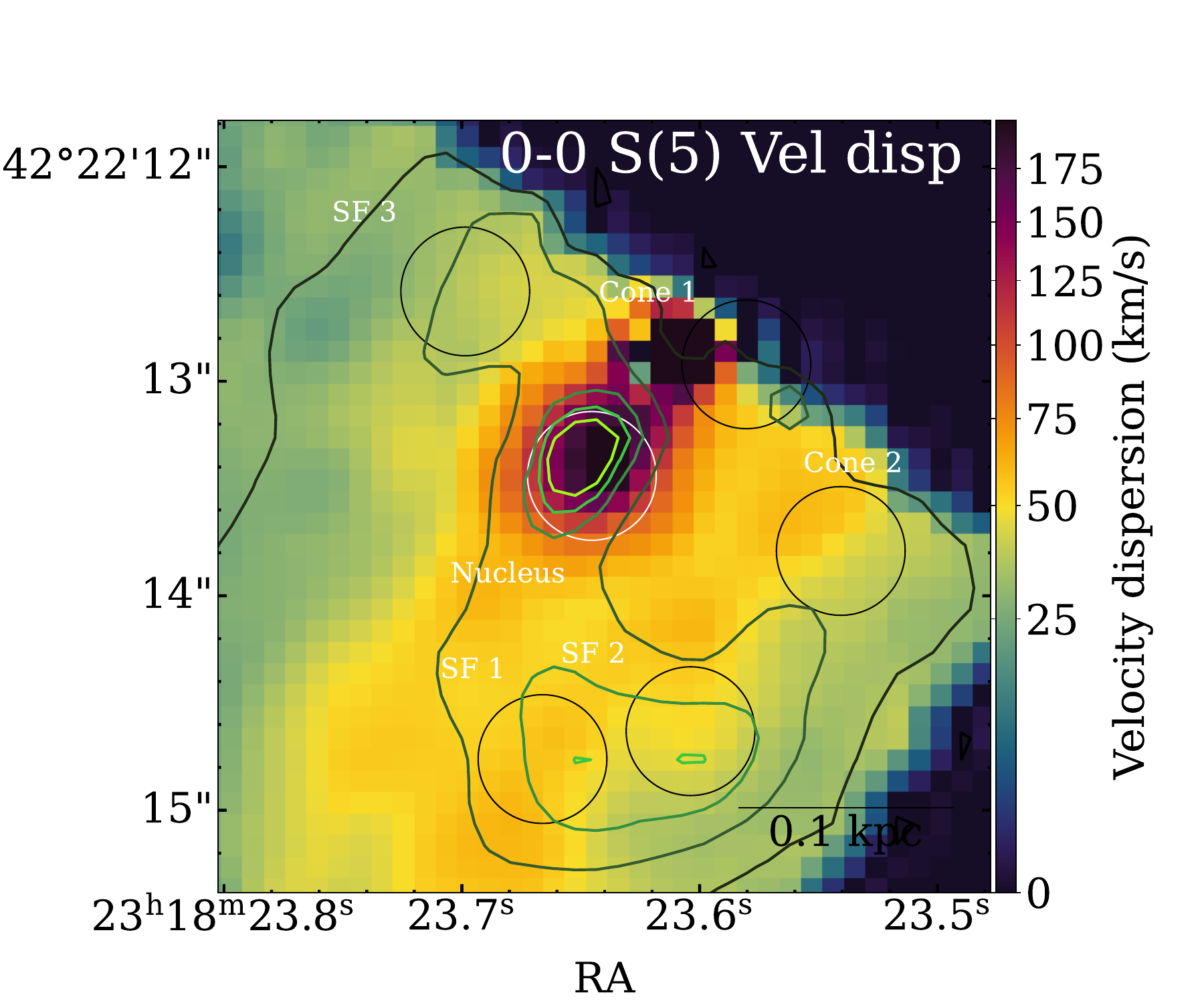}
   \caption{Flux emission, velocity and velocity dispersion maps for H$_2$ 0-0 S(1), S(3) and S(5). Contours show the respective flux line emission. As $J$ increases, the emission lines are tracing warmer molecular gas. This shows that the nucleus, SF~1, and SF~2 contain more warmer molecular gas than any other region.}
    \label{fig:rotational_maps}%
\end{figure*}

\begin{figure}
   \centering
   \includegraphics[width=\columnwidth]{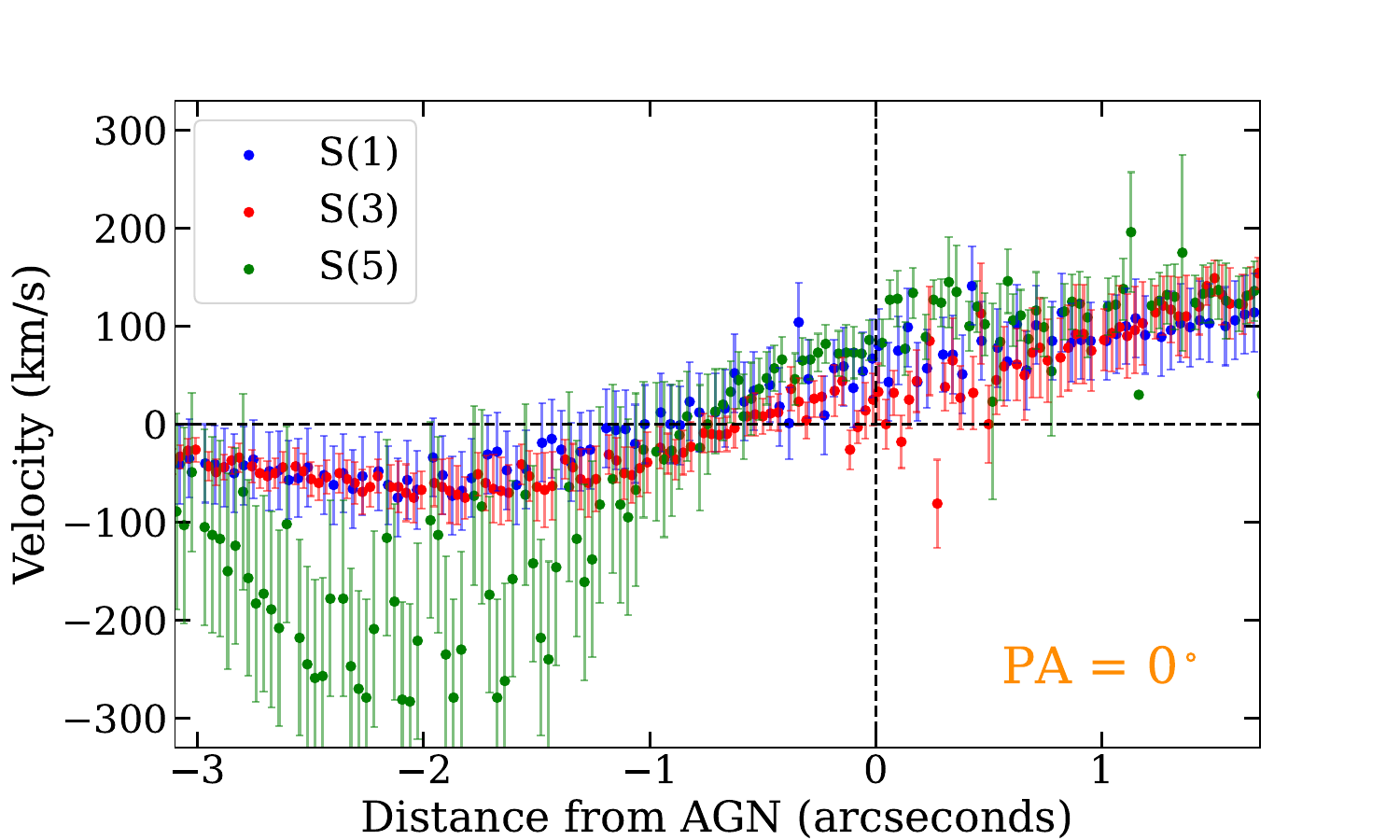}
   \caption{Velocity fit for each spaxel along a pseudoslit of width 3 pixels along PA = $0^\circ$ for the rotational lines S(1), S(3), S(5).}
    \label{fig:PA_vel}%
\end{figure}

\begin{table*}
    \centering
   \caption[]{Measured fluxes for the H$_2$ $\nu = 0$–$0$ S(1) to S(7) lines. Fluxes were corrected for extinction and are stated in units of $10^{-15}$ erg/s/cm$^2$. All apertures were circular with 0.3 arcsecond radii. *: A small portion of Cone~1 lies slightly outside the MIRI/MRS FOV for these lines, leading to some flux not being measured, hence these fluxes are lower limits.}
   \label{tab:H2Fluxes}
   \begin{tabular}{lccccccr}
      \hline
      $\text{Line}$ & $\lambda \, [\mu m]$ & $\text{Nucleus}$ & $\text{SF~1}$ & $\text{SF~2}$ & $\text{SF~3}$ & $\text{Cone~1}$ & $\text{Cone~2}$ \\
      \hline
      S(1) & $17.035$ & $12.8 \pm 0.9$ & $6.1 \pm 0.1 $& $8.10 \pm 0.08$ & $4.34 \pm 0.02$ & $2.941 \pm 0.001$ & $1.70 \pm 0.01$ \\
      S(2) & $12.279$ & $4.8 \pm 1.4$ & $4.01 \pm 0.07$ & $3.94 \pm 0.05$ & $1.60 \pm 0.02$ & $1.09 \pm 0.02$ & $0.52 \pm 0.02$ \\
      S(3) & $9.665$ & $32.5 \pm 0.2$ & $11.01 \pm 0.02$ & $12.42 \pm 0.03$ & $0.27 \pm 0.03$ & $2.83 \pm 0.03$ & $2.12 \pm 0.02$ \\
      S(4) & $8.026$  & $3.0 \pm 1.3$ & $1.2 \pm 0.1$ & $2.18 \pm 0.09$ & $0.06 \pm 0.04$ & $0.39 \pm 0.03$ & $0.08 \pm 0.01$ \\
      S(5) & $6.909$ & $7.4 \pm 1.1$ & $2.75 \pm 0.02$ & $3.80 \pm 0.04$ & $0.05 \pm 0.01$ & $0.29$* & $0.13 \pm 0.01$ \\
      S(6) & $6.109$  & $4.4 \pm 0.8$ & $0.79 \pm 0.03$ & $0.90 \pm 0.03$ & $0.14 \pm 0.04$ & $0.10$* & $0.57 \pm 0.02$ \\
      S(7) & $5.512$  & - & $1.3 \pm 0.2$ & $2.1 \pm 0.1$ & $0.32 \pm 0.08$ & $0.35$* & $0.2 \pm 0.1$ \\
      \hline
   \end{tabular}
\end{table*}

An initial inspection of the combined NIRSpec and MIRI/MRS spectra in the various apertures (Fig.~\ref{fig:uncorrected_fluxes}) shows that SF~1 and SF~2 are remarkably similar, both exhibiting relatively high continuum flux compared to the other non-nuclear apertures. %despite being located furthest from the AGN. 
The main difference between them is that SF~2 displays a more steeply rising MIR continuum beyond $\lambda > 18 \mu m$. This is consistent with our extinction analysis (Table~\ref{tab:appendix_table}), which indicates a slightly higher dust column towards SF~2. This excess mid-IR emission may also be related to the position of SF~2 along the edge of the ionisation cone (Fig.~\ref{fig:selected_apertures}), where enhanced dust heating could further boost its long wavelength continuum. In contrast, SF~3 exhibits a distinctly different spectrum: its continuum flux is much lower, and the shapes of its spectral features differ from those of SF~1 and SF~2. This divergence is notable because SF~3 also lies within the circumnuclear star-forming ring, showing that star-forming regions within the same ring (i.e. likely forming from the same gas) can display markedly different physical conditions.

All H$_2$ $\nu = 0$–$0$ rotational lines from S(1) to S(7) were detected in each aperture with a signal-to-noise ratio (S/N) greater than 5, except in the nucleus where the S(7) 5.511 $\mu m$ line is significantly blended with the [Mg VII] 5.503 $\mu m$ emission line. This metal line is detected only in the nuclear aperture due to having a high ionisation potential (IP) of 187 eV. The measured fluxes (extinction-corrected) and associated uncertainties for all H$_2$ $\nu = 0$–$0$ lines from S(1) to S(7) in each aperture are presented in Table~\ref{tab:H2Fluxes}. Some higher $J$ rotational lines, such as the H$_2$ $\nu = 0$–$0$ S(8) 5.053 $\mu m$, S(9) 4.695 $\mu m$ and S(10) 4.410 $\mu m$ lines lie within the NIRSpec range, but are not detected with adequate S/N in the majority of apertures and so were excluded from our analysis.

We also calculate different mid-IR flux ratios from various other emission lines, these are given for every aperture in Table.~\ref{tab:emission_ratios} and are discussed in context with their utility in Section~\ref{sec:shock_modelling}.

\begin{table*}
    \centering
   \caption[]{Measured log$_{10}$ emission line ratio values in each aperture. The [S III] line could not be fit in the SF~3 and Cone~2 apertures. *: Pf$\alpha$ is poorly detected (low S/N) in these apertures, and so its flux is roughly inferred from Pf$\beta$ assuming Case B recombination is valid, $T_e \sim 10^3$ and $n_e \sim 10^4$cm$^{-3}$, giving a flux ratio of Pf$\alpha$/Pf$\beta$ $\sim 1.81$ \citep{hummer1987recombination}}.
   \label{tab:emission_ratios}
   \begin{tabular}{lccccr}
      \hline
      $\text{Region}$ & $[\text{Ne III}] / [\text{Ne II}]$ & $[\text{Ne V}] / [\text{Ne II}]$ & $[\text{Ar II}] / [\text{Ar III}]$ & $[\text{S IV}] / [\text{S III}]$ & $[\text{Fe II}] / \text{Pf}\alpha$ \\
      \hline
      $\text{Nucleus}$ & 0.43 & -0.21 & -0.35 & 0.01 & 1.37* \\
      $\text{SF~1}$ & -0.75 & -1.84 & -0.36 & -0.32 & 1.32 \\
      $\text{SF~2}$ & -0.34 & -1.41 & -0.62 & -1.00 & 0.75 \\
      $\text{SF~3}$ & -0.10 & -0.57 & -0.07 & - & 0.92* \\
      $\text{Cone~1}$ & -0.19 & -0.36 & -0.44 & -0.35 & 1.00* \\
      $\text{Cone~2}$ & -0.57 & -1.79 & -0.72 & - & 1.14* \\
      \hline
   \end{tabular}
\end{table*}

\subsection{H$_2$ rotational line emission and kinematics maps}
\label{sec:H2_line_kinematics_subsection}

Fig.~\ref{fig:rotational_maps} presents the line emission and kinematic maps for the rotational S(1), S(3), and S(5) transitions. 
%We denote these rotational lines as $S(J)$, where $J$ is the rotational quantum number. 
These lines trace molecular gas at progressively higher kinetic temperatures as $J$ increases, making their comparison particularly useful for identifying regions of elevated gas temperatures. The overall spatial extent of the emission decreases with increasing $J$, indicating that the cooler phase molecular gas has a more extended distribution. Notably, only the nucleus, SF~1, and SF~2 exhibit substantial emission in the higher excitation S(5) line, strongly indicating that these regions contain a significant fraction of the warmer molecular gas. The S(5) map also reveals quite striking extended emission both north and south of the nucleus, but notably not aligning with the ionisation cone as shown in Fig.~\ref{fig:selected_apertures}. However, the S(5) emission almost forms a reverse `S'-shape that encompasses SF~1 and SF~2 in the south and SF~3 in the north, with the inner part of the `S' loosely following the direction of the proposed jet (Fig.~\ref{fig:selected_apertures}).
% This distinct morphology is absent in the S(1) and S(3) maps, suggesting that this extended feature lacks a substantial cooler gas component and is instead dominated by warmer molecular gas.
% These maps may provide initial evidence of a link between the AGN and the heating of molecular gas within the circumnuclear star-forming regions.
%We also see evidence for the star forming ring shown by \citet{riffel2009agn} (who traced it using the 7-4 hydrogen recombination line 2.1661$\mu m$ emission), traced by the increased H$_2$ flux out to a radius of $\sim 200$ pc.

The kinematic maps for the rotational lines presented in Fig.~\ref{fig:rotational_maps} also reveal several intriguing features. The velocity and dispersion maps for the S(1) and S(3) lines display a notably similar morphology, with SF~1 and SF~2 exhibiting significantly larger velocity dispersions compared to most other circumnuclear spaxels. Additionally, SF~1 and SF~2 (particularly SF~1) appear slightly more blueshifted in the S(1), S(3), and S(5) lines relative to their surrounding spaxels on the approaching side of the galaxy. This additional blueshifting around SF~1 and SF~2 was also detected by \citet{garcia2021galaxy} in the CO kinematics with ALMA, who linked it to density-wave driven non-circular outflows from the AGN intermediate line region. This combination of enhanced dispersion and blueshifting suggests that SF~1 and SF~2 are considerably more kinematically disturbed in molecular H$_2$ than other regions. An additional feature is seen uniquely in the S(3) velocity map, where the nucleus shows enhanced blueshifting relative to its immediate surroundings. This blueshift extends southwards toward SF~1 and SF~2, potentially tracing an outflow, although it is interesting we do not see such a strong feature in the S(1) or S(5) kinematics.

The S(5) velocity dispersion map shows significantly elevated dispersion across the nuclear region, with a clear asymmetry between the northern and southern sides. Notably, the dispersion is enhanced around the nucleus compared to throughout the ring, suggesting additional turbulent contributions from shocks or outflows acting in this region. 

%Such turbulence could arise from circumnuclear star formation, which is known to be prevalent in the southern regions and just north of our MIRI/MRS FOV, but may also reflect the impact of AGN driven disk winds. In this scenario, winds interacting with the inner edge of the ring could compress and shock the molecular gas, producing the observed pattern of elevated S(5) velocity dispersions.}

Collectively, the S(5) kinematics follows a similar pattern to those of the cooler S(1) and S(3) lines, albeit with extended enhanced velocity dispersion south of the nucleus. These results reinforce the conclusions drawn from the emission maps: the warmer molecular gas exhibits a more complex and disturbed morphology, particularly around and south of the AGN, as evidenced by distinct kinematic signatures, which could suggest the presence of shocks.

We also present the velocity profiles of the S(1), S(3), and S(5) lines along PA = $0^\circ$ (approximately aligned with the major velocity axis) in Fig.~\ref{fig:PA_vel}. We see the clear redshifted offset, particularly in S(1) and S(5) around the nucleus. In contrast, S(3) displays increased scatter near the nucleus and a velocity trend closer to the systemic velocity, consistent with the additional nuclear blueshifting seen in Fig.~\ref{fig:rotational_maps}. Between approximately $-3$ and $-1\arcsec$, S(5) also shows enhanced blueshifting that aligns with the blueshifted structure over SF~1 observed in Fig.~\ref{fig:rotational_maps}.

% However, we do not see direct evidence unambiguously indicating the definite presence of a jet. While this makes it challenging to definitively confirm or rule out the existence of a jet based on these diagnostics alone, it does confirm that NGC~7582 exhibits turbulent molecular gas features and extended warm molecular gas near the AGN, which could suggest the presence of shocks.

% Outflow features have been seen around other AGN, such as NGC~5728, where an outflow is causing additional blueshifting of the gas immediately within and around the nucleus \citep{schommer1988ionized, shimizu2019multiphase, durre2019agn}. Interestingly, the signature is absent in both the cooler S(1) and the warmer S(5) velocity maps, which is not straightforward to explain. If driven by an AGN outflow or jet, one might expect a corresponding imprint in the S(1) kinematics (affecting the cooler gas) or in S(5) (affecting the hotter gas). 
%Its presence only in S(3) suggests that the effect may be linked to shocks or heating processes preferentially exciting the intermediate temperature molecular gas, sufficient to enhance S(3) emission but not strong enough to significantly affect the S(5) line, although this cannot be confirmed from the velocity maps alone, motivating a more quantitative analysis of the H$_2$ emission.}

\subsection{H$_2$ excitation modelling}
\label{sec:H2_excitation_modelling_subsection}

A possible H$_2$ excitation mechanism is shocks as outlined by \citet{burton1992mid}, who demonstrated that H$_2$ rotational transitions can be efficiently excited by slower shocks compared to those responsible for the near-IR vibrational H$_2$ emissions. Their work also highlights the dependence of the S(1) line intensity on the temperature of the photodissociation region (PDR), with significant excitation occurring only when $T \gtrsim 100$ K, assuming a fixed ortho-to-para H$_2$ ratio (OPR) of 3:1 and local thermal equilibrium (LTE).

Since the higher-order S($J$) transitions require greater energy, and hence hotter PDRs, to be appreciably excited compared to S(1), we adopt this lower cutoff of $T \gtrsim 100$ K as the baseline for the temperature range probed in our analysis. The wavelength coverage of our NIRSpec G395H/F290LP observations precludes the observation of vibrational H$_2$ lines, which are better tracers of hotter gas ($T \gtrsim 1000$ K) due to their higher excitation temperatures. Consequently, the kinetic temperature range of the molecular gas that we can reliably constrain using the mid-IR S(1) to S(7) lines is approximately $100 \lesssim T \lesssim 1000$ K, with only a small fraction at hotter temperatures probed only by vibrational lines \citep{2016ApJ...830...18T}.

% We do nonetheless report the detection of the higher-temperature rovibrational lines $\nu = $ $1$-$0$ O$(8)$ 4.163 $\mu m$, O$(7)$ 3.808 $\mu m$, and O$(6)$ 3.501 $\mu m$ in the nucleus as well as SF~1 and SF~2 regions. These lines, typically excited at temperatures $T$ > $1000$ K, are not observed with sufficient signal-to-noise (S/N) for a robust quantitative analysis (S/N > 3).

To model the fluxes of the rotational H$_2$ lines, we adopt the formalism outlined in \citet{burton1992mid}, \citet{pereira2014warm}, \citet{jones2024jwst}, and \citet{2016ApJ...830...18T} (hereafter TS16). We refer the reader to these works for a comprehensive discussion of the theoretical framework. In essence, we assume that the populations of the different rotational energy levels — characterised by their column densities, $N$, follow a continuous power-law distribution determined by the molecular gas temperature, under the assumption of local thermal equilibrium (LTE). This distribution is given by:

\begin{equation}
    \label{eq:power-law-temp}
    dN = mT^{-n} dT.
\end{equation}

Here, $m$ is a scaling coefficient which can be easily found by integrating Equation~(\ref{eq:power-law-temp}) over the total H$_2$ column density, $N_{\text{tot}}$, between upper and lower bounds of the temperature distribution. Importantly, $n$ represents the power-law index. Physically, a larger $n$ (so a more negative exponent overall) results in a steeper decline in the population of higher energy rotational levels, indicating that a greater fraction of the molecular H$_2$ resides in the lower energy states. This manifests as relatively higher observed fluxes in the lower rotational transitions, such as S(1) and S(2). Conversely, a smaller $n$ (less negative exponent) leads to enhanced fluxes in the higher transitions, such as S(6) and S(7). The power-law distribution is constrained by the lower and upper temperature limits, $T_l$ and $T_u$, respectively. These do not correspond to physical kinetic temperatures of the molecular gas, rather the integration limits that make the power-law distribution finite and physically interpretable. For rotational line analyses, it is standard practice to adopt a fixed upper limit of $T_u = 2000$ K, as populations at higher temperatures contribute negligibly to the observed flux distribution. This assumption is well-supported in the literature (TS16) and is confirmed by our own calculations of the total H$_2$ column density. We find that varying $T_u$ beyond 2000 K has an insignificant impact on the derived level populations and the resulting flux.

The lower temperature cutoff parameter, $T_l$, has a more significant influence on the derived results, as the bulk of the H$_2$ mass occupies lower energy states associated with these rotational transitions, thus are predominantly excited at relatively low temperatures. An important question, however, becomes: just how low can the lower temperature cutoff be before the S(1) emission becomes undetectable? Consequently, the choice of $T_l$ significantly affects the inferred H$_2$ column density and must be handled with care. \citet{burton1992mid} demonstrated that assuming a fixed ortho-to-para ratio of 3:1 is reasonable for temperatures $\gtrsim 300$ K. Additionally, since the lowest detectable line in our MIRI/MRS observations is the S(1) transition— which itself is poorly excited at temperatures $\lesssim 150$ K— we feel that adopting a fixed $T_l$ within this range is reasonable. Therefore, we also investigate applying a fixed $T_l = 300$ K when fitting the power-law model, which would be close to the expected $T_l$ for an ortho-to-para ratio of 3:1 for these lines. To assess the robustness of this assumption, we also perform fits without fixing the ortho-to-para ratio, allowing $T_l$ to vary freely. As we demonstrate later, most regions naturally converge toward a best-fit $T_l$ close to 300 K, supporting the validity of our assumption. A detailed theoretical analysis of potential deviations from this ratio, while potentially insightful, lies beyond the scope of this study.

Unfortunately, the lowest-energy rotational H$_2$ line, S(0) at 28.221 $\mu m$, lies beyond the wavelength coverage of MIRI/MRS (28.1 $\mu m$) and therefore cannot be included in our analysis.

% Measuring S(0) would have provided a stronger constraint on the cooler molecular gas component, probing molecular temperatures $T \lesssim 100$ K, which previous studies have shown to be prevalent for the molecular gas content in the nuclear regions of galaxies \citep{falgarone2005warm, 2016ApJ...830...18T}. 

\subsection{Excitation diagrams}
\label{sec:excitation_diagrams_subsection}

\begin{figure}
    \centering
    \includegraphics[width=0.79\columnwidth]{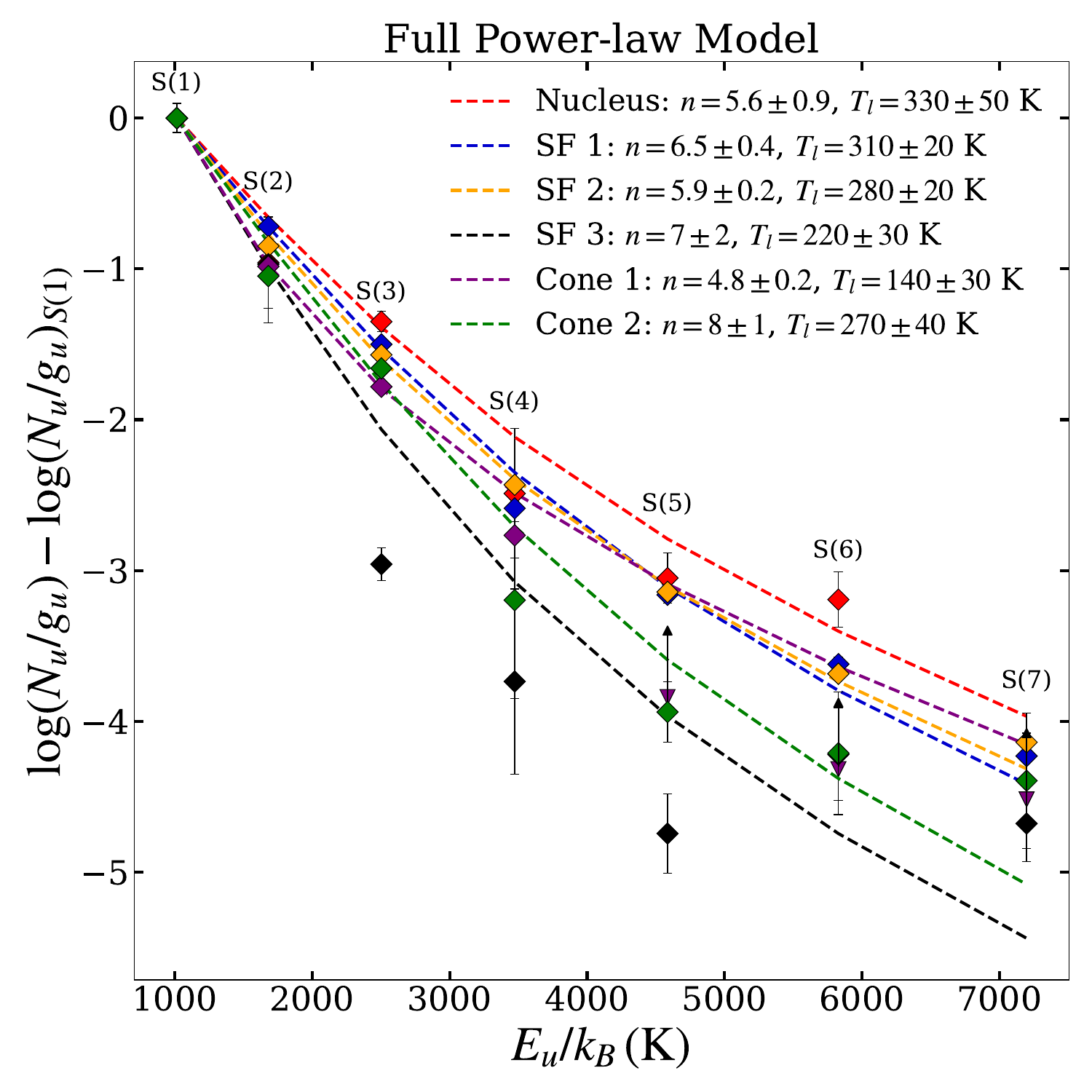}
    \includegraphics[width=0.79\columnwidth]{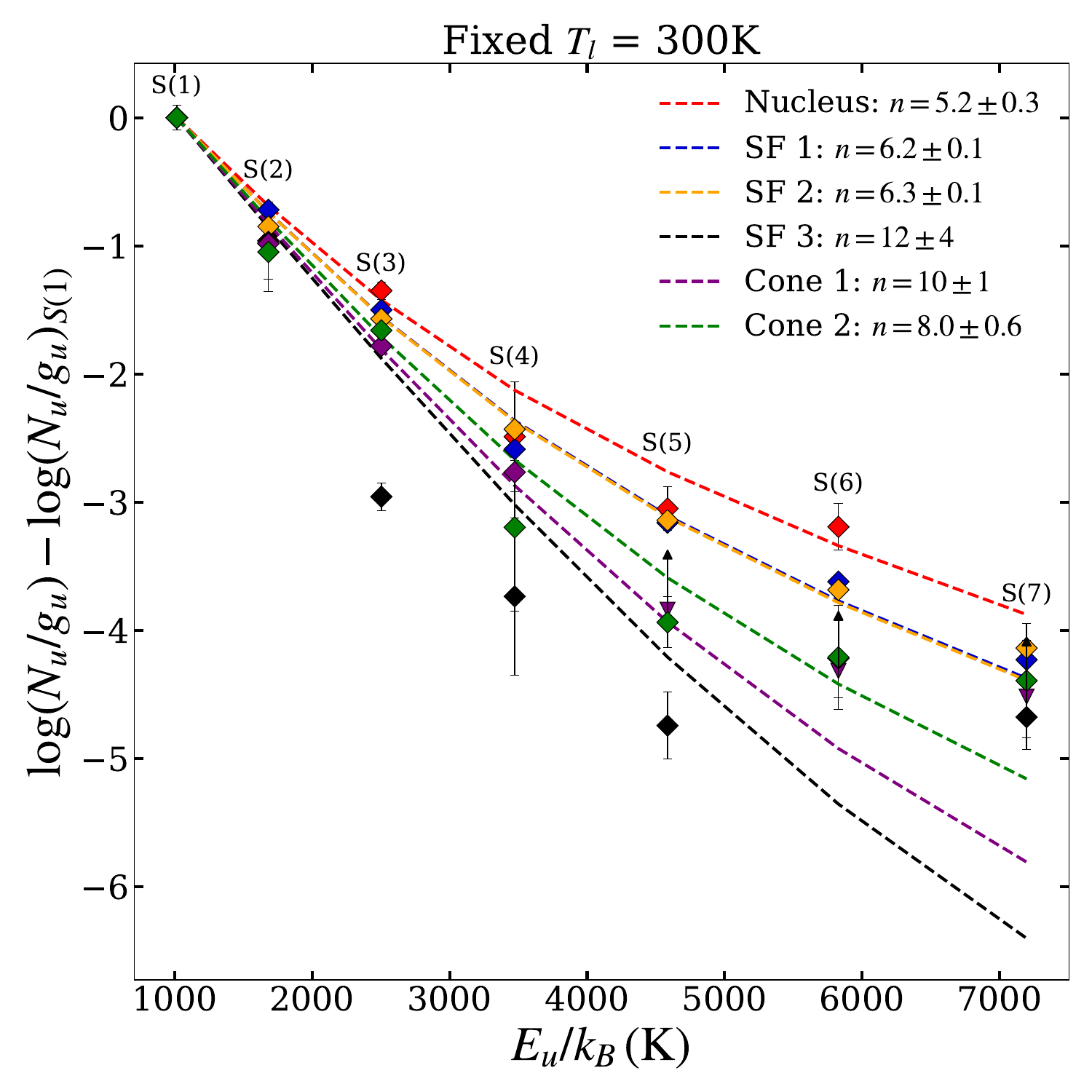}
    \includegraphics[width=0.79\columnwidth]{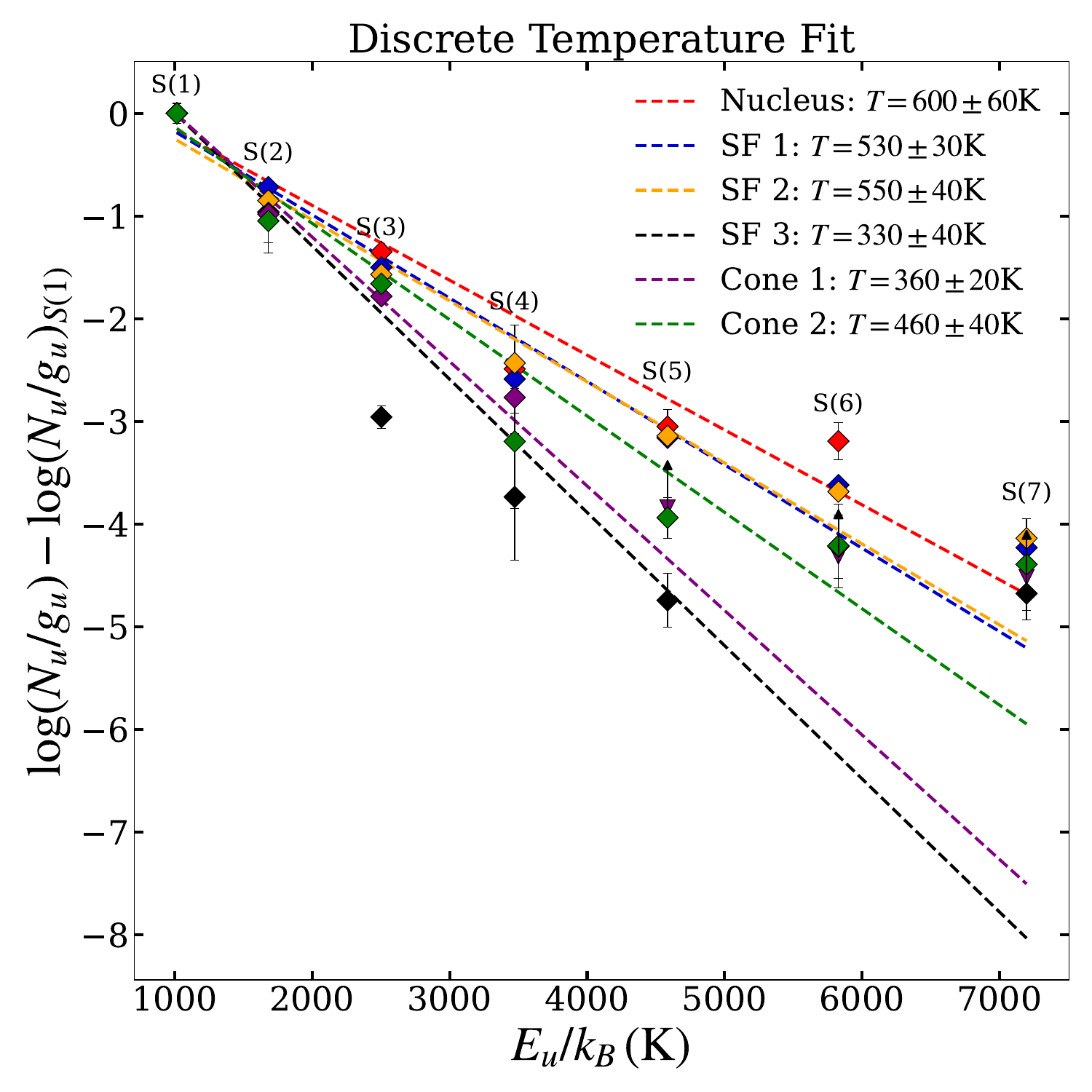}
    \caption{H$_2$ excitation diagrams for each aperture with three model fits. Top: Power-law temperature distribution with a free OPR. When including rotational lines down to S(1), a best-fit $T_l \sim 300$ K approximately corresponds to an OPR of 3:1. Middle: Power-law temperature distribution with the OPR fixed to 3:1, leaving $n$ as the sole free parameter. Bottom: Single-temperature model, where the only free parameters are the excitation temperature, $T$, and the total H$_2$ column density. A small portion of the Cone 1 aperture extends beyond the MIRI/MRS FOV for S(5), S(6), and S(7), hence these line fluxes are lower limits which we show as triangular markers.}
    \label{fig:TS16_fits}
\end{figure}

\begin{table}
    \centering
    \begin{tabular}{lcccc}
    \hline
    \textbf{Region} & $n$ & $T_l$ (K) & log N$_{H_2}$ (cm$^{-2}$) & log M$_{H_2}$ (M$_\odot$) \\
    \hline
    Nucleus     & $5.6 \pm 0.9$  & $330 \pm 50$ & $21.45 \pm 0.17$ & $5.18 \pm 0.17$ \\
    SF 1     & $6.5 \pm 0.4$  & $310 \pm 20$ & $21.14 \pm 0.07$ & $4.87 \pm 0.07$ \\
    SF 2     & $5.9 \pm 0.2$  & $280 \pm 20$ & $21.32 \pm 0.06$ & $5.01 \pm 0.06$ \\
    SF 3     & $7 \pm 2$  & $220 \pm 30$ & $21.09 \pm 0.33$ & $4.82 \pm 0.33$ \\
    Cone 1   & $4.8 \pm 0.2$  & $140 \pm 40$ & $21.34 \pm 0.34$ & $5.07 \pm 0.34$ \\
    Cone 2   & $8 \pm 1$  & $270 \pm 40$ & $20.64 \pm 0.17$ & $4.37 \pm 0.17$ \\
    \hline
    \end{tabular}
    \caption{Best fit $n$ and $T_l$ from our TS16 modelling, as well as the derived H$_2$ column density and mass for $T>300$ K for each region. Hotter regions exhibit overall lower $n$ combined with higher $T_l$.}
    \label{tab:H2_params}
\end{table}

All of our TS16 power-law models are generated using the \texttt{H2Powerlaw} Python implementation\footnote{\url{https://github.com/astrolojo/H2Powerlaw}}. We construct excitation diagrams, whereby measured fluxes of H$_2$ rotational lines are converted into column densities and are plotted against the energy of the upper state of the line transition, allowing us to analyse the temperature distribution of the molecular gas in each aperture region. The theory behind modelling excitation diagrams from H$_2$ rotational line flux observations is discussed for completeness in the Appendix~\ref{sec:appendix_H2}.

We fit the excitation diagrams for every aperture region using both the power-law distribution model and an alternative fit that assumes the molecular H$_2$ resides at a single, fixed temperature. The latter, 
%while 
provides a convenient linear fit on the excitation diagram and allows the column density to be directly inferred from the intercept with the normalised column density axis. 
%lacks a physically intuitive basis, as it assumes a constant temperature across the region. Nonetheless, we include 
These fixed-temperature models 
%for completeness, as it is 
have also been employed in previous studies \citep{rigopoulou2002iso, roussel2007warm, rosenberg2013excitation, davies2024gatos, esparza2025molecular, almeida2025jwst}. 
%and conveniently provides an intuitive way to interpret the temperature of each region. 
The excitation diagrams, constructed from the measured fluxes listed in Table~\ref{tab:H2Fluxes} and normalised by the S(1) flux for each aperture, along with the corresponding fits for the three different models, are presented in Fig.~\ref{fig:TS16_fits}. 
%We also include 
The individual rotational diagrams for each aperture 
with every model plotted 
separately are shown in the Appendix 
%for clarity, 
Fig.~\ref{fig:separate_rotation_fits}. Finally, the best-fit $T_l$ and $n$, along with the inferred H$_2$ column densities and H$_2$ mass of each region are given in Table~\ref{tab:H2_params}.

The TS16 power-law model 
%without a fixed 
with a varying ortho-to-para ratio, shown in the top panel of Fig.~\ref{fig:TS16_fits}, provides the best fit among all tested models.
%having the lowest $\chi^2$ in all apertures.
%This is perhaps not surprising, given that this model imposes the fewest assumptions and includes more free parameters. 
However, the physical implications of such fits are complicated as variable ortho-to-para ratios remain poorly understood theoretically \citep{burton1992mid, sternberg1999ratio, takahashi2001ortho}. 
%Furthermore, comparing the best-fit power-law indices, $n$, across different regions becomes less straightforward. %For instance, although the nucleus exhibits a larger $n$ than Cone~1, it also has a significantly higher $T_l$. Consequently, while the nucleus contains a higher proportion of `colder' gas compared to Cone~1, the temperature of this `colder' gas is substantially higher than that of Cone~1 (roughly given by the best-fit $T_l$), complicating comparison.
For every model 
the best fits 
%for every model 
are those for 
the nuclear, SF~1, and SF~2 regions, which is expected given that these exhibit the strongest line fluxes and the highest S/N. As shown in the middle panel of Fig.~\ref{fig:TS16_fits}, models with 
%assuming 
a fixed lower temperature limit of $T_l = 300$ K provide a good fit within one standard deviation for all three of these regions, suggesting that an ortho-to-para ratio of 3:1 is a reasonable assumption. This allows us to confidently fix $T_l$ and focus on comparing the resulting power-law indices, $n$, instead.
%as presented in Fig.~\ref{fig:TS16_fits}. 
%With $T_l$ held constant, variations in $n$ become more directly comparable, facilitating a clearer interpretation of the relative differences between regions.

Overall, when keeping $T_l$ fixed, the nucleus has the lowest $n$, (middle panel of Fig.~\ref{fig:TS16_fits}), corresponding %suggesting that it has 
to a larger proportion of hot molecular H$_2$ gas, which is perhaps not unexpected. More intriguing is the fact that SF~1 and SF~2 have very similar $n$ values, which are only slightly higher than that of the nucleus but significantly smaller than those of Cone~1, Cone~2 and SF~3. This suggests that SF~1 and SF~2 have 
%only a slightly higher proportion of cooler molecular H$_2$ than that found in the nucleus, 
%i.e. that these regions have 
a similar proportion of hot molecular gas as the nucleus
which is somewhat puzzling. 
% Generally, one would expect star-forming regions to be colder
%as colder gas is necessary for gravitational collapse. 
This observation is further supported by the results in the top panel of Fig.~\ref{fig:TS16_fits}, where the best-fit $T_l$ values for the nucleus, SF~1, and SF~2 are virtually identical within the error bars, as are the corresponding $n$ values. When allowing the ortho-to-para ratio to vary within the power-law distribution, the inferred temperature distributions for these three regions are also indistinguishable within the error margin. This is particularly remarkable given that the star-forming regions are clearly spatially resolved and located about 100 pc away from the nucleus (i.e. not strongly within the PSF even if our corrections were not applied), making it unlikely that nuclear contamination is responsible for their heating.

Furthermore, the bottom panel of Fig.~\ref{fig:TS16_fits} presents the best-fit discrete temperature model for each aperture, showing the same trend where the nucleus, SF~1, and SF~2 are much hotter than the other regions, as shown by having the least steep gradients. Once again, the results from fitting the discrete temperature model suggest that SF~1 and SF~2 are at nearly the same temperature as the nucleus, further supporting the conclusions drawn from the power-law fits.

Cone~1, Cone~2, and SF~3 exhibit lower S/N in the higher $J$ lines, resulting in larger scatter of the S(3)-S(7), data points that deviate from the line fit trends of the lower-$J$ transitions. The excitation diagram for SF~3, in particular, is not well fit by any of the models considered here.
%(power-law distributed or discrete temperature). 
This is because SF~3 is fainter and has comparatively poor S/N in rotational line fluxes.
SF~3 was included as an additional star-forming region for comparison with the more pronounced and intriguing SF~1 and SF~2. Although a comparison of the excitation diagrams of the northern star forming clumps M3, M4, or M5 would have been informative, we note that these regions are only partially covered by the MIRI FOV, hence the study of their H$_{2}$ properties is not possible.
%{Also interesting could have been one of the northern star-forming regions}, such as M3, M4, or M5 from \citet{ricci2018optical}, which exhibit higher fluxes and reduced AGN contamination. \textcolor{red}{However, these regions are not fully covered by the NIRSpec and MIRI/MRS FOVs of our observations, with small parts of M4 and M5 only covered in some. 
%While this limits direct comparisons between SF~1 and SF~2 and other more prominent star-forming regions in the ring}, 
Nonetheless, 
the H$_2$ column density derived for SF~3 reveals a significantly cooler temperature distribution, despite its close proximity to the AGN, supporting our conclusion that the temperature of the molecular gas in SF~1 and SF~2 is unexpectedly elevated. Such H$_2$ excitation temperatures are not just higher compared to other regions in this galaxy, but also compared to studies of other nearby galaxies, e.g., \citet{2016ApJ...830...18T, hernandez2023dissecting, riffel2025blowing}.
%Regardless of the model chosen to fit the H$_2$ excitation diagram, the same quantitative conclusion emerges: the southern circumnuclear star-forming regions of NGC~7582 contain hotter molecular gas than expected for typical star-forming regions, with temperature distribution comparable to that found in the AGN-heated nucleus. 
In the following subsection, we delve deeper into these two regions to further investigate their properties and explore the mechanisms that might be responsible for the gas heating.

\subsection{Analysis of the properties of the circumnuclear ISM through mid-IR lines}
\label{sec:hot_star_forming_regions_subsection}

To investigate the origin of the heating mechanism, we extend our analysis beyond just the H$_2$ lines, considering additional mid-IR lines and line ratios as diagnostic tracers.

\begin{figure}
    \centering
    \includegraphics[width=\columnwidth]{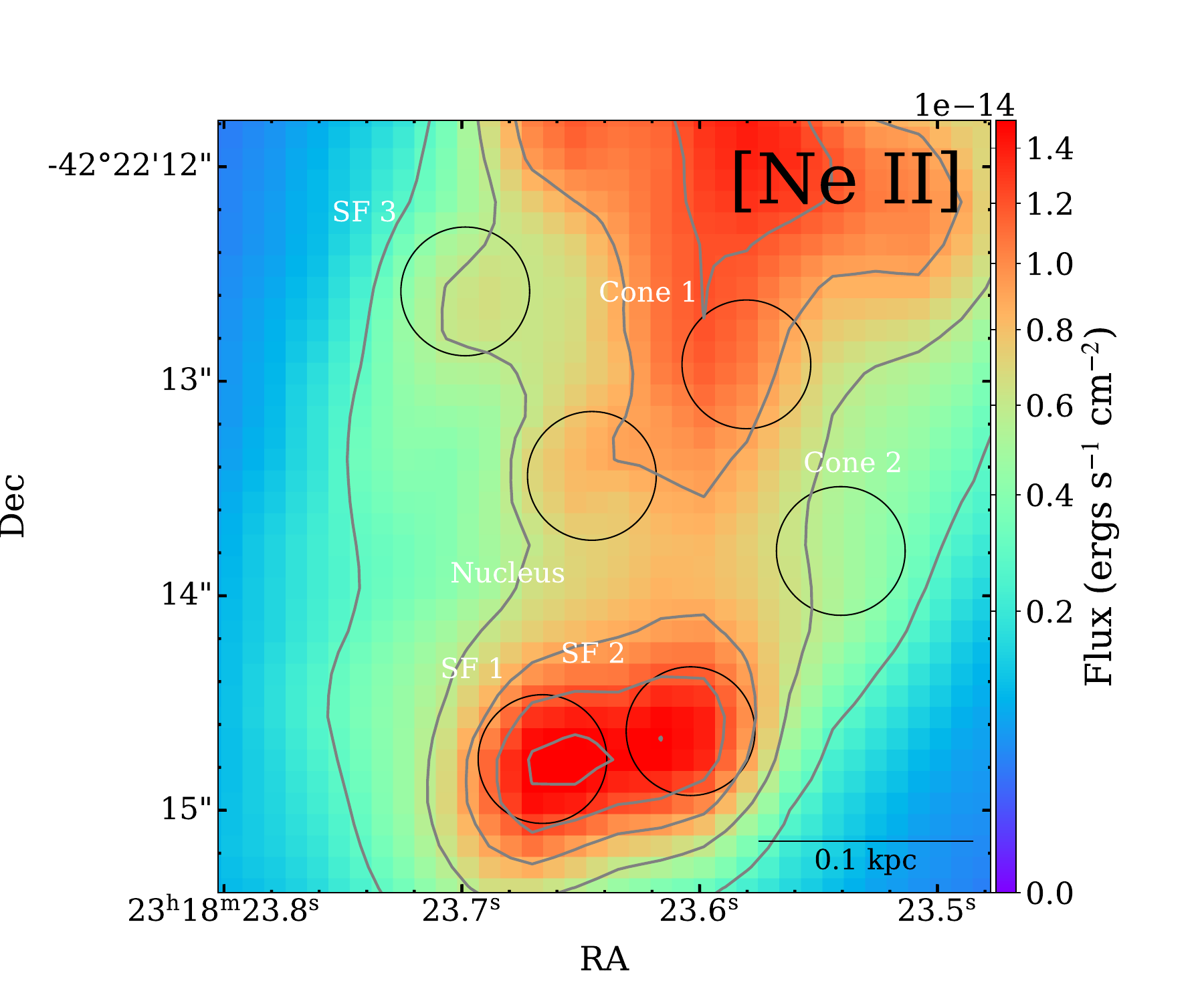}
    \includegraphics[width=\columnwidth]{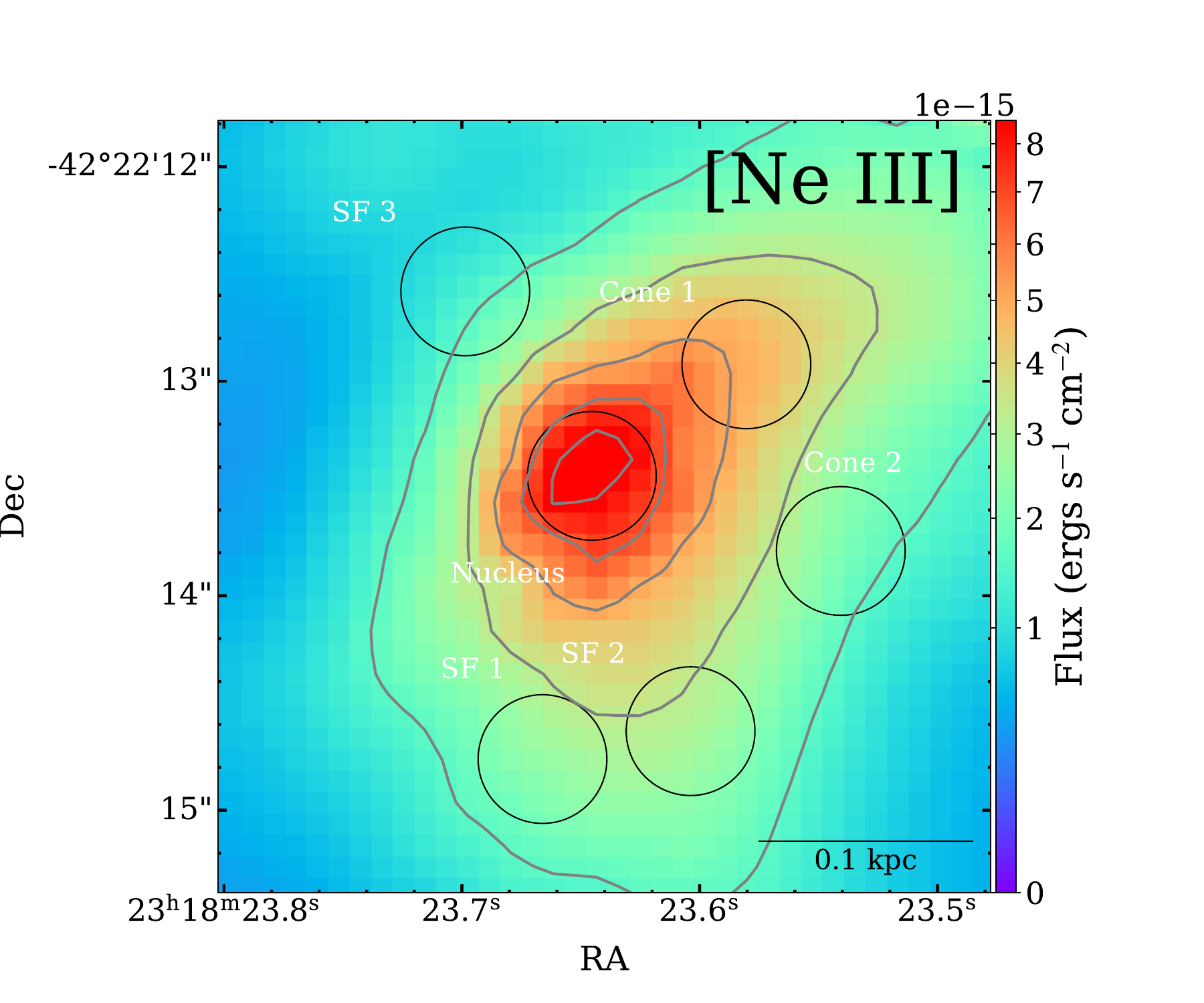}
    \includegraphics[width=\columnwidth]{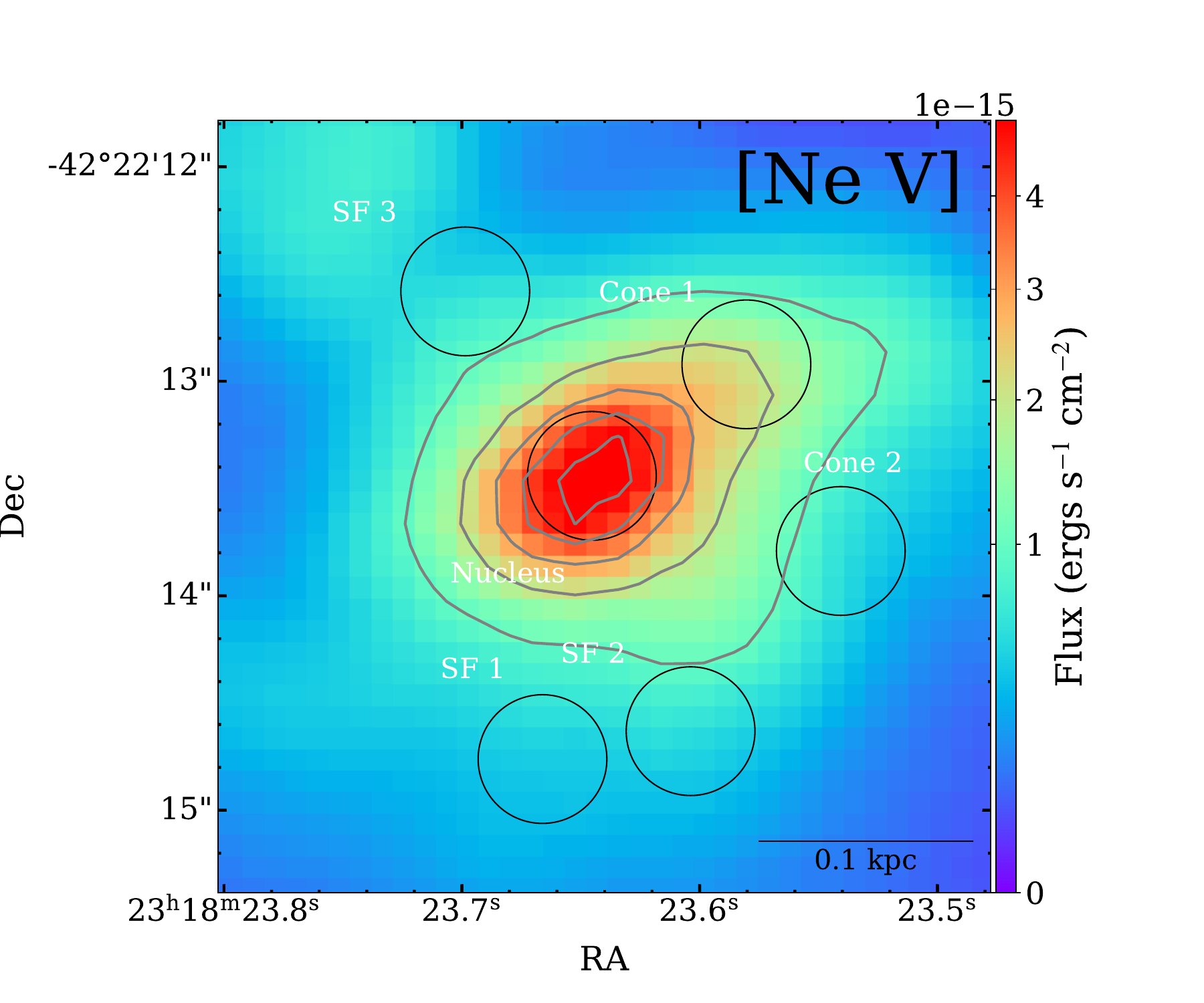}
    \caption{Top to bottom: [Ne II] 12.8 $\mu m$, [Ne III] 15.8 $\mu m$, [Ne V] 14.3 $\mu m$ flux maps. [Ne V] is a high ionisation line, IP = 97 eV map: there is no increase in its emission in SF~1 or SF~2, supporting the notion that they are not photoionised by the AGN.}
    \label{fig:Ne_ratio}
\end{figure}

It is well established that [Ne II] line emission traces star formation,
%(when its fluxes are relatively high compared to other lines with higher IPs), 
[Ne III] can arise from both stellar photoionisation and AGN photoionisation, whereas [Ne V] requires highly energetic ionising photons, and is only excited by AGN.
%and so cannot be tied to star formation. 
%To more rigorously rule out AGN photoionisation as the dominant heating mechanism for the warm molecular H$_2$ in the circumnuclear regions, 
Fig.~\ref{fig:Ne_ratio} shows maps of the [Ne II], [Ne III] and [Ne V] lines for NGC~7582. It is evident that 
%we closely examine the [Ne V] 14.32\,$\mu$m emission. We find that 
strong [Ne V] emission is entirely confined to the nucleus, with much weaker, morphologically distinct detections in the western side of the ionisation cone, clearly visible in the contours shown in the same figure %Fig.~\ref{fig:Ne_ratio} 
as an extension westward from the nucleus.
In SF~1, the [Ne V] line appears as a weak, single-Gaussian component with a flux of order $\sim$1\% that of the nucleus and, critically, with an identical central wavelength to the nucleus. Given that SF~1 is kinematically offset from the nucleus (typically blueshifted), any intrinsic [Ne V] emission would be expected to appear at a different wavelength. The identical line centres therefore indicate this signal is due entirely to PSF contamination from the nucleus.
SF~2 shows a similar nuclear PSF feature, but with an additional, weaker blueshifted [Ne V] component. This results in a subtle double-Gaussian line shape and suggests a small amount of genuine [Ne V] emission, possibly originating from SF~2 itself. This is consistent with its location on the southern edge of the western ionisation cone (Fig.~\ref{fig:selected_apertures}). However, this component is minor and overall the [Ne V] flux is greatly outshone by the nearby [Cl II] 14.36\,$\mu$m line, a reliable tracer of star formation, which is much stronger in both SF~1 and SF~2, and almost absent in the nucleus.
These findings are consistent with the 
%the nucleus hosts 
[Ne V] emission being powered by the AGN.
% \sout{In addition, [Ne V] emission is found to originate from within the ionisation cone, while SF~1 and SF~2 are \textcolor{red}{predominantly} low ionisation regions. Any trace of [Ne V] emission from the southern star-forming regions is either due to slight PSF contamination from the AGN or, in the case of SF~2, a very minor intrinsic component due to \textcolor{red}{it being located at the edge of the ionisation cone}.}
The analysis of these lines suggests that the molecular gas in SF~1 and SF~2 is not being heated through direct AGN photoionisation.
%This distinction is also reflected in their nearly identical H$_2$ temperatures (Fig.~\ref{fig:TS16_fits}), suggesting that AGN photoionisation must not significantly influence the molecular gas in these regions.

% \begin{figure*}
%     \centering
%     \includegraphics[width=0.33\linewidth]{graphics/NeII_map_2.pdf}
%     \includegraphics[width=0.33\linewidth]{graphics/NeIII_map_2.pdf}
%     \includegraphics[width=0.33\linewidth]{graphics/new_NeV_map_2.pdf}
%     \includegraphics[width=0.33\linewidth]{graphics/new_Ne_III_II_ratio_map_2.pdf}
%     \includegraphics[width=0.33\linewidth]{graphics/new_Ne_V_II_ratio_map_2.pdf}
%     \includegraphics[width=0.33\linewidth]{graphics/new_Ne_V_III_ratio_map_2.pdf}
%     \caption{Top: [Ne II] 12.8 $\mu m$, [Ne III] 15.8 $\mu m$, [Ne V] 14.3 $\mu m$ flux maps. [Ne V] 14.32 $\mu m$ is a high ionisation line, IP = 97 eV map: there is no increase in its emission in SF~1 or SF~2, supporting the notion that they are not high ionisation regions. Bottom: ratios of the different Neon line fluxes. [Ne III] / [Ne II] and [Ne V] / [Ne II] highlights regions of higher ionisation, confirming that SF~1 and SF~2 do not display high temperature molecular H$_2$ likely due to having high ionisation. Higher ratio regions by the north east and south west corners of the ratio maps are caused by higher noise in the spaxels by the edge of the MIRI/MRS FOV and do not correspond to physical structures.}
%     \label{fig:Ne_ratio}
% \end{figure*}

% We next examine the [S IV] 10.5 $\mu m$ line map shown in Fig.~\ref{fig:S_map}. 
% %The [S IV] line traces the outward facing ionisation cone whereas 
% \textcolor{red}{
% %SF~2 lies partially within the edge of the ionisation cone. 
% The outward-facing cone west of the nucleus is clearly visible, with SF~2 located along its line of sight rather than being merely affected by diffuse contamination by the PSF.}

% \begin{figure}
%     \centering
%     \includegraphics[width=\columnwidth]{graphics/SIV_map_2.pdf}
%     \caption{[S IV] 10.5 $\mu m$ flux map, tracing the nucleus and the edges of the outward facing ionisation cone.}
%     \label{fig:S_map}
% \end{figure}

\begin{figure}
    \centering
    \includegraphics[width=\columnwidth]{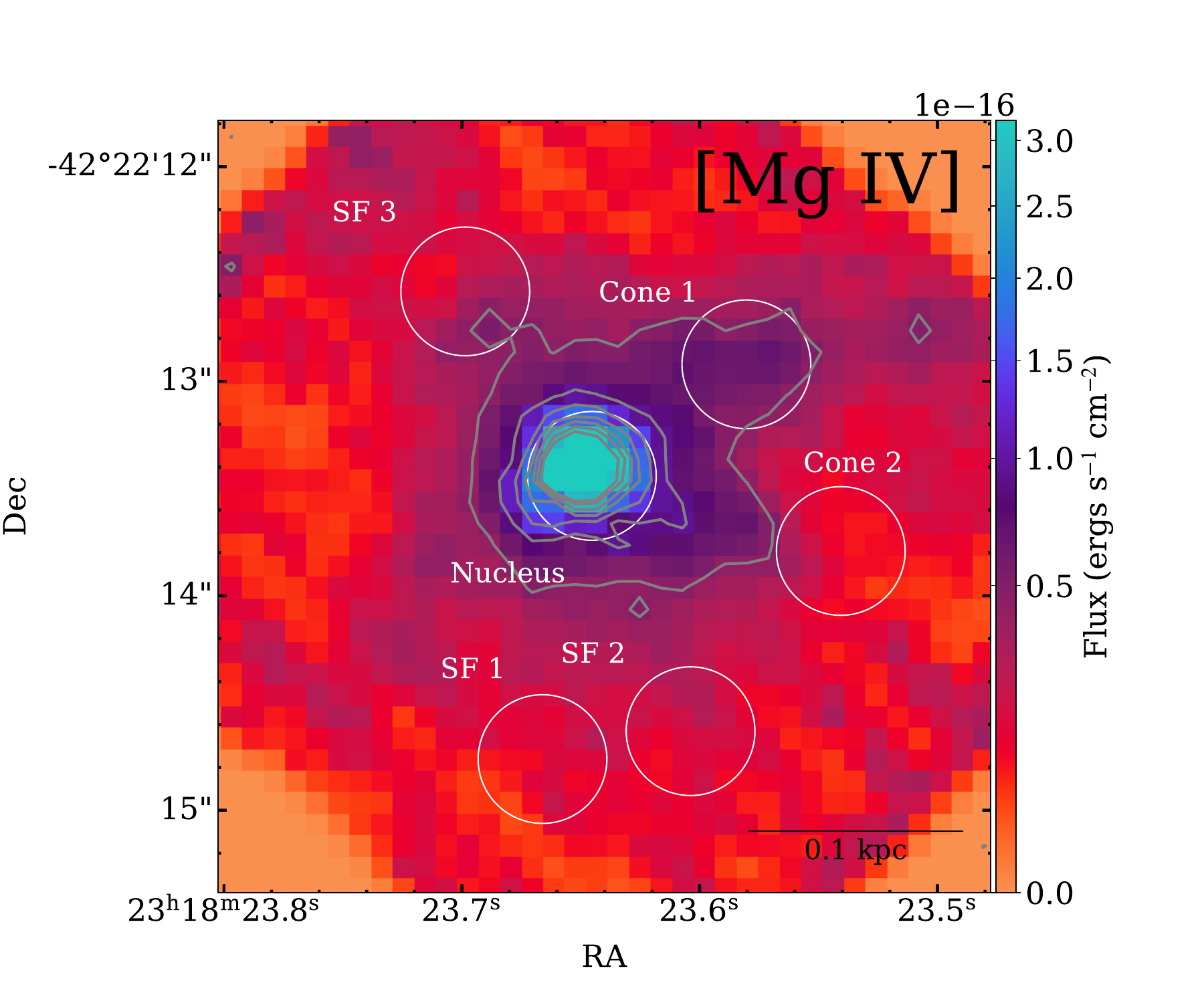}
    \caption{[Mg IV] 4.487 $\mu m$ flux map tracing gas that is highly ionised. IP = 80 eV. The emission is concentrated in the nucleus and around the edges of the ionisation cone.}
    \label{fig:Mg_map}
\end{figure}

[Mg IV] 4.487 $\mu m$ is also a tracer of high ionisation (IP = 80 eV). We present the [Mg IV] emission map in Fig.~\ref{fig:Mg_map}, with unsurprisingly the absence of strong [Mg IV] emission in SF~1 and SF~2 or anywhere south of the nucleus providing further evidence against direct ionisation by the AGN, 
%as such a scenario would ionise the regions and lead to detectable [Mg IV] emission. 
Notably, [Mg IV] emission does appear to trace the edges of the ionisation cone, as indicated by the contours in Fig.~\ref{fig:Mg_map}, aligning well with previous literature on the western-facing cone, confirming that it contains strongly ionised gas as expected. Moreover, the [Mg IV] distribution closely resembles that of [Ne V], being confined to the AGN and the western cone edges. The lack of any additional [Mg IV] extension beyond [Ne V] suggests that both lines trace the same ionised gas and are excited by the same mechanism —photoionisation by the AGN, as is well established for [Ne V].
%This resemblance indicates that [Mg IV] does not provide any new information or evidence for additional high-ionisation shocks, since such shocks would produce a more extended or distinct [Mg IV] distribution relative to [Ne V].}

\begin{figure}
    \centering
    \includegraphics[width=\columnwidth]{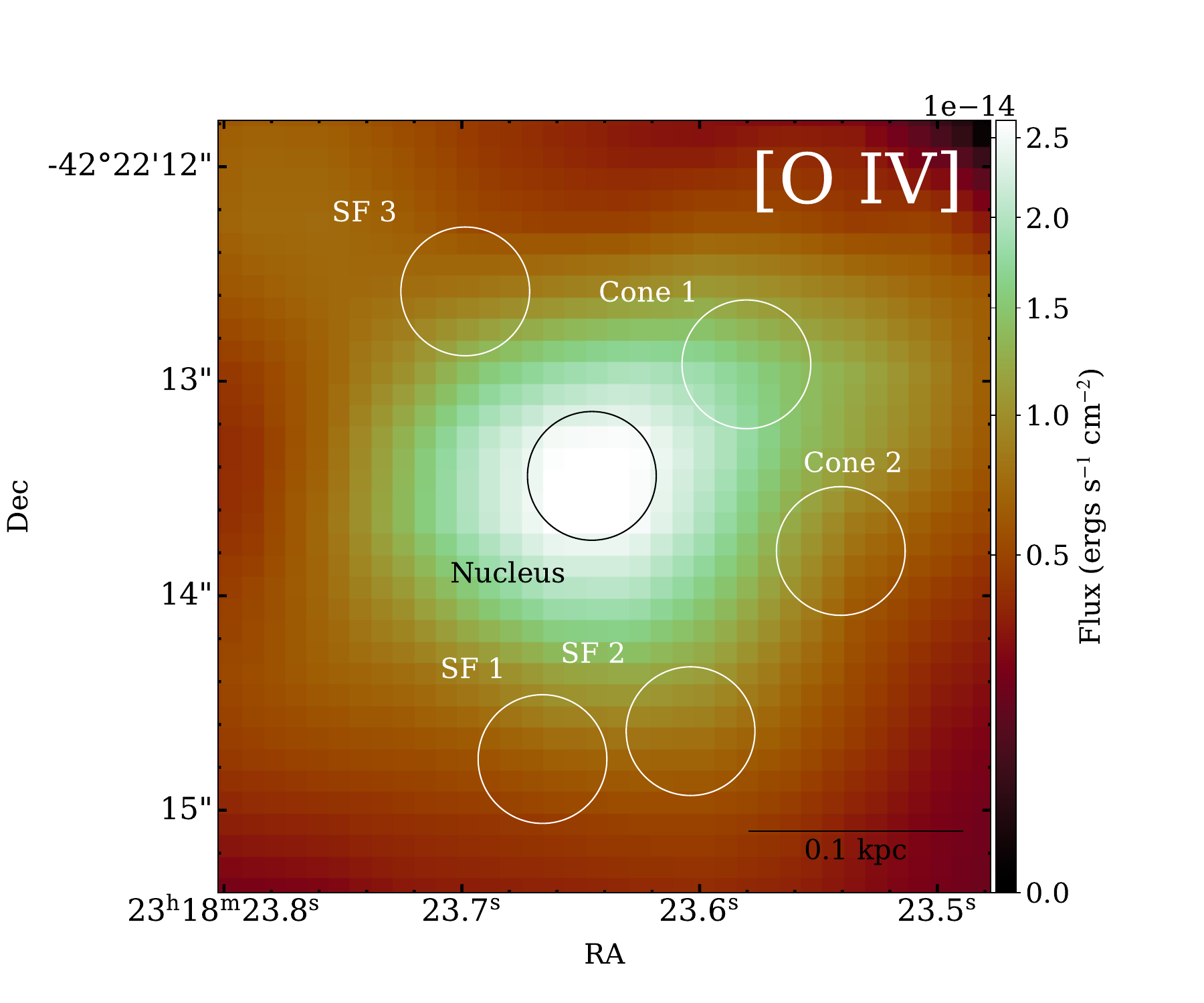}
    \caption{[O IV] 25.89 $\mu m$ flux map tracing the AGN region and potentially O-type star clusters, showing little evidence that SF~1 and 2 could be being heated by either of these scenarios. We note that the nucleus produces a large PSF, as this line falls within the longest MIRI/MRS wavelength channel.}
    \label{fig:OIV_map}
\end{figure}

The [O IV] 25.89 $\mu m$ line (IP = 54.9 eV) is a well-established tracer of AGN activity as well as the presence of young, hot, massive O-type stars \citep{thornley2000massive, pereira2010mid, alonso2011local}. 
In NGC~7582, this line is detected by MIRI/MRS with a high S/N ratio. We present its spatial distribution in Fig.~\ref{fig:OIV_map}. We see no morphologically pertinent [O IV] emission beyond the nucleus, suggesting that SF~1 and SF~2 are not influenced by the AGN.
%and no evidence in favour of them containing significant O-type star clusters, although we cannot rule this out based on the emission of [O IV] alone \citep{thornley2000massive}. 
The spatial extent of [O IV] flux around the nucleus appears large due to the broad PSF of MIRI/MRS channel 4, where this line is detected. While SF~3, Cone~1, and Cone~2 appear contaminated by nuclear [O IV] flux, SF~1 and SF~2 are much less affected, reinforcing the conclusion that their elevated temperatures are not due to AGN photoionisation or a result of PSF contamination from the nucleus.
%Therefore, we cannot conclude from the [O IV] emission that these regions contain significant numbers of O-type stars, contrary to the suggestion by \citet{wold2006nuclear} and \citet{riffel2009agn}.
%\textcolor{red}{The [O IV] line is primarily concentrated in the nuclear region and not the ring}.

\begin{figure}
    \centering
    \includegraphics[width=\columnwidth]{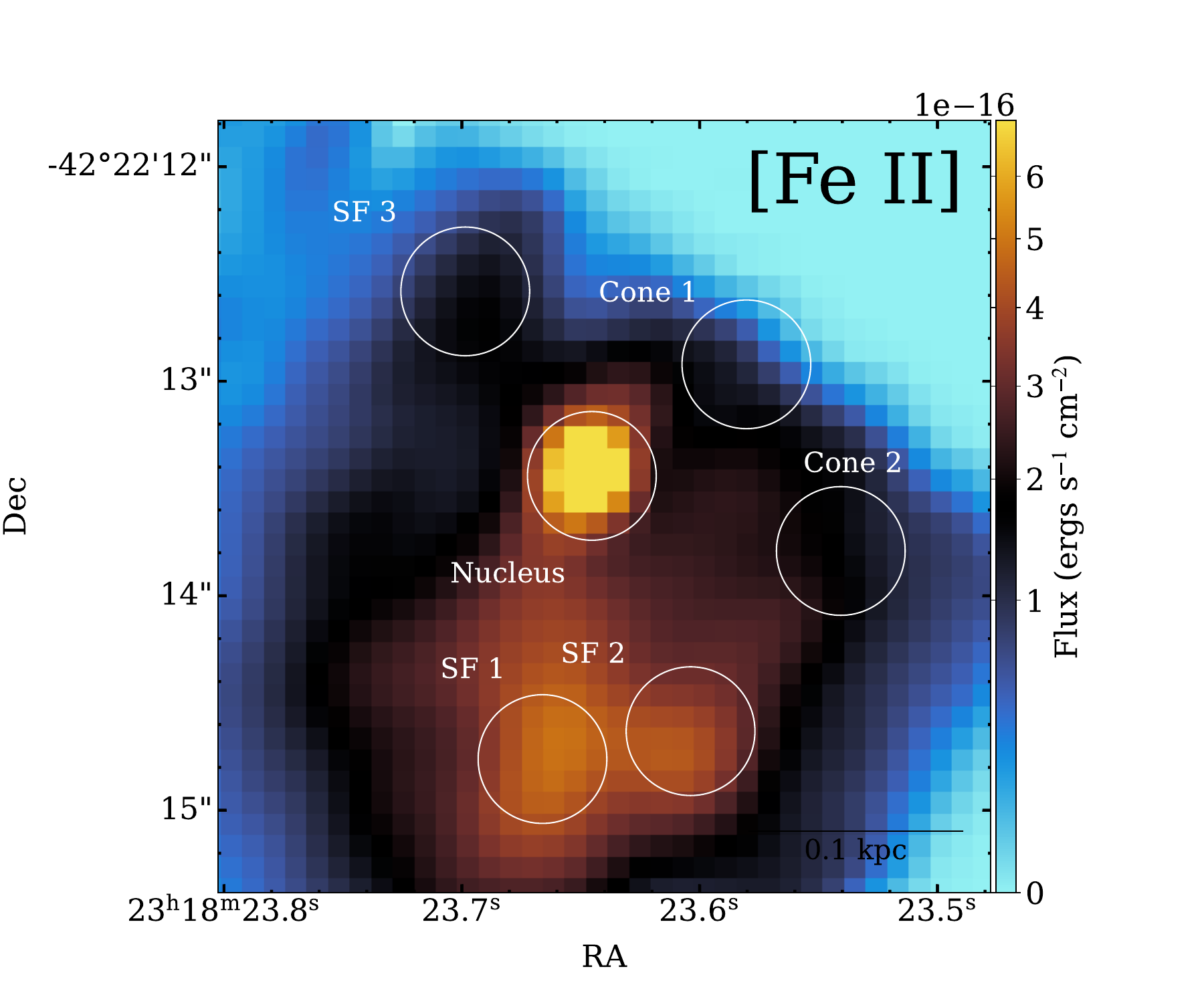}
    \caption{[Fe II] 5.34 $\mu m$ flux map tracing low-IP shocked gas (IP = 7.9 eV). We see clear extended emission around SF~1 and SF~2, suggesting then that shocks could be heating them.}
    \label{fig:Fe_map}
\end{figure}

The [Fe II] 5.34 $\mu m$ line, which has a low ionisation potential of only 7.9 eV, is useful as a well-established diagnostic for low-ionisation shocked gas \citep{kawara1988forbidden, forbes1993radio}. The corresponding map, along with the aperture locations, is presented in Fig.~\ref{fig:Fe_map}. The nucleus exhibits significant [Fe II] emission, indicative of strong AGN activity \citep{baldwin2004origin}. More notably, there is a strikingly clear extension of [Fe II] flux south by southeast of the nucleus, encompassing both SF~1 and SF~2, while no similar directional increase is observed elsewhere, however, the MIRI/MRS FOV for this line does not extend to the northern regions of the star-forming ring, preventing us from determining whether the northern star-forming clumps, M3, M4, M5, also show enhanced [Fe~II] 5.34 $\mu m$ emission. Nonetheless, this provides evidence that SF~1 and SF~2 may contain substantial shocked gas, which could be the cause of their observed higher than expected H$_2$ temperatures.

\citet{ricci2018optical} reported strong near-IR [Fe II] 1.64~$\mu$m emission in the nucleus, also extending along a position angle of $\sim20^\circ$, aligning with Fig.~\ref{fig:Fe_map}. The southern [Fe II] extension across SF~1 and SF~2 may arise from an AGN jet interacting with the surrounding ISM. As the jet disrupts, it could lose collimation and disperse into a broader outflow, driving shocks that compress, fragment, and entrain ambient gas, thereby enhancing [Fe II] emission. \citet{ricci2018optical} further showed that any such jet would be nearly orthogonal to the ionisation cone, traced by the Br$\gamma$ hydrogen recombination line in the near-IR. However, we find no clear evidence of this jet in our Pf$\beta$ recombination map (Fig.~\ref{fig:selected_apertures}), nor in other lines such as Pf$\alpha$ or Br$\alpha$.

Furthermore, \citet{ricci2018optical} proposed that SF~2 may be photoionised by the ionisation cone, given its position along the southern edge of the western-facing cone. We conclude that although SF~2 likely experiences some small additional photoionisation from the cone, the molecular gas in both regions is primarily shock heated. (SF~2 shows weak high-ionisation features (e.g. [Ne V]), weak Br$\alpha$ emission, and identical molecular gas temperature and excitation to SF~1.)
%Our results lend partial support: SF~2 exhibits weak high-ionisation features (e.g. [Ne V]) and a slightly different spectrum with enhanced mid-IR continuum compared to SF~1 (Fig.~\ref{fig:uncorrected_fluxes}). However, multiple diagnostics instead point to shocks as the dominant heating mechanism in both regions. SF~1 and SF~2 show nearly identical molecular gas temperatures, H$_2$ excitation diagrams, and [Fe II] emission, along with similarly low ionisation levels as indicated by their Neon ratios (bottom row of Fig.~\ref{fig:Ne_ratio}). If photoionisation were dominant in SF~2, we would expect stronger Br$\alpha$ emission and more pronounced differences in thermal and excitation properties relative to SF~1. }

\begin{figure*}
    \centering
    \includegraphics[width=0.49\linewidth]{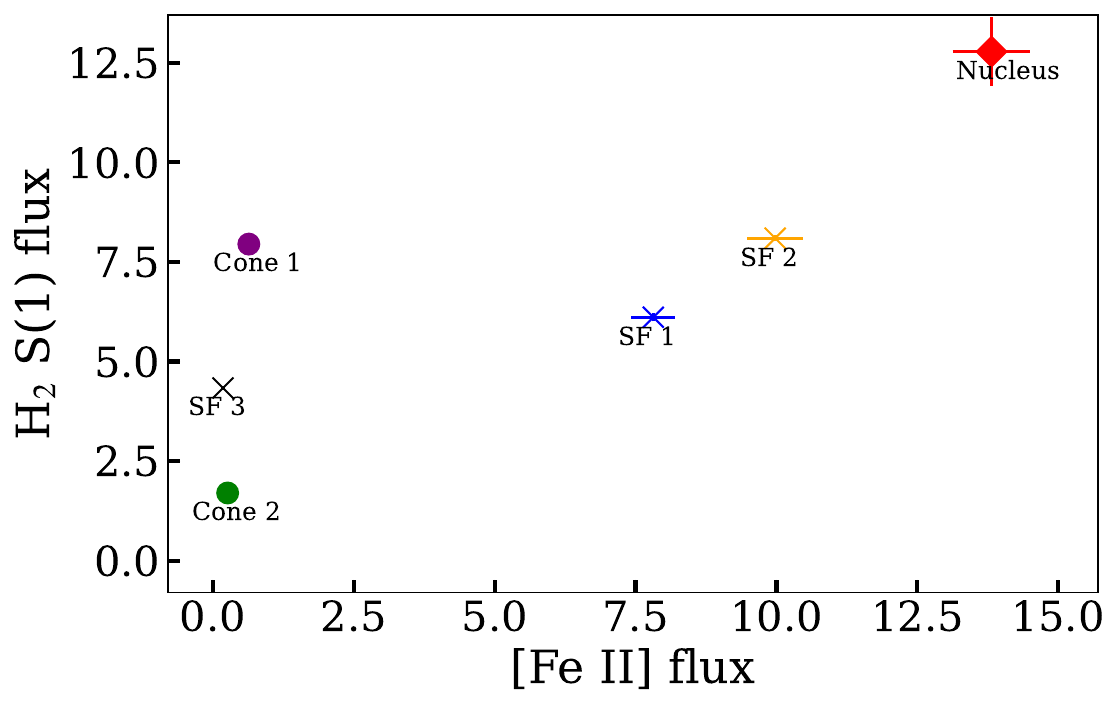}
    \includegraphics[width=0.49\linewidth]{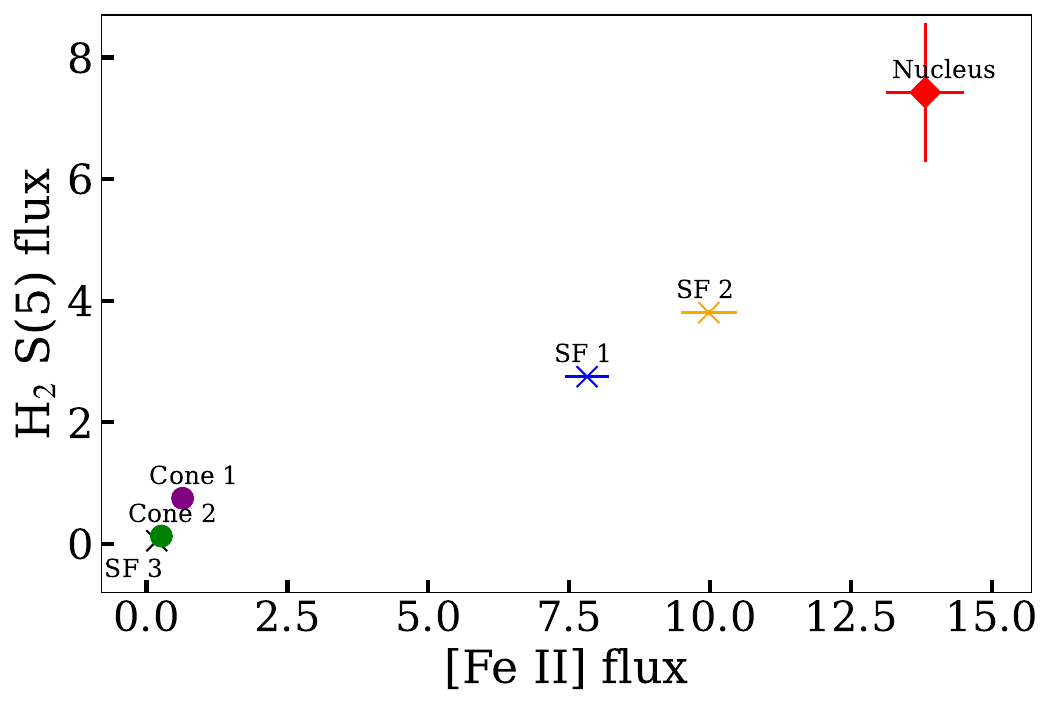}
    \caption{H$_2$ rotational line vs [Fe II] 5.34 $\mu$m flux (in units of 10$^{-15}$ erg/s/cm$^2$). Left: S(1) transition, right: S(5) transition.}
    \label{fig:Fe_H2}
\end{figure*}

We also compare the amount of [Fe II] to H$_2$ flux in Fig.~\ref{fig:Fe_H2}. Both [Fe II] vs S(1) (left) and [Fe II] vs S(5) (right) display linear correlations across the selected aperture regions (but we note that in the left panel the correlation appears to be driven in part by the nucleus, and without this point the trend is less pronounced). This indicates a physical connection between shocked gas traced by [Fe II] and the warm molecular gas. The correlation is strongest for the S(5) line, which probes higher excitation temperatures, suggesting that the warmest H$_2$ is most closely linked to [Fe II] emission, thus further supporting shock excitation as the dominant heating mechanism in these regions.

Significant [Fe II] emission can also arise from photoionisation by star formation and AGN activity. For example, NGC~4303 exhibits a star-forming ring bright in [Fe II] emission \citep{riffel2016sinfoni}, as does NGC~7469 \citep{bianchin2024goals}, and numerous other galaxies with circumnuclear star formation. Photoionisation models are capable of reproducing [Fe II] emission even in the absence of coronal line emission \citep{dors2012x}, as the ionisation potential of [Fe II] is relatively low, allowing diffuse AGN photons to contribute significantly. A common method to distinguish [Fe II] emission caused by shocks from that due to photoionisation is through diagnostic line ratios, particularly comparing [Fe II] to hydrogen recombination lines in near-IR diagrams \citep{colina2015understanding}. For this purpose we consider the [Fe II] 5.34 $\mu$m / Pf$\alpha$ ratio as it serves as an established shock sensitive tracer in the mid-IR \citep{hollenbach1989molecule, vivian2022goals}. We provide this ratio in Table~\ref{tab:emission_ratios}. \citet{herrero2025miconic} compared this ratio in many different regions of Seyfert nuclei and low luminosity AGNs and demonstrated that shocked regions typically exhibit log$_{10} \text{[Fe II]} / \text{Pf}\alpha \gtrsim 0.7$. While all regions show ratios consistent with potential shocks, we caution that the large uncertainties, due to the low S/N of Pf$\alpha$, limit the robustness of this diagnostic outside SF~1 and SF~2. However, for SF~1 and SF~2, where Pf$\alpha$ is measured with high S/N, this ratio is consistent with the presence of shocks in these regions. 

[Mg IV] has also been shown to be a tracer of shocked gas in highly ionised regions \citep{pereira2024extended}, and as we see no clear extension south of the nucleus compared to [Fe II], this suggests that these shocks would likely be low ionisation and low velocity.

\subsection{Shock models}
\label{sec:shock_modelling}

%We explore the possibility of shock heating further 
In this subsection we employ 
the well-established Paris-Durham models (\citealp{flower1985theoretical, flower2003influence}) to investigate a shock origin for the observed H$_{2}$ emission. The input energy flux, both radiative and kinetic, from a potential shock can be reprocessed and emitted as H$_2$ rovibrational and/or rotational line emission. Different types of shocks produce distinct ratios and fluxes of rotational and vibrational H$_2$ lines \citep{kristensen2023shock}. Shocks with velocities exceeding 30 km/s significantly dissociate H$_2$, causing its abundance to decrease sharply.
% As a result, if a shock is present, it is likely to have a velocity lower than 30 km/s, as higher velocities would rapidly destroy H$_2$ and prevent the detection of strong rotational lines.
This does not by itself rule out the shock origin; for instance, a lower velocity shock could still arise from an AGN jet propagating through a clumpy medium, where interactions cause the jet to split and redirect across a range of velocities \citep{mukherjee2016relativistic, mukherjee2018relativistic}.

\citet{kristensen2023shock} discuss the application of shock models using JWST data and highlight various diagnostics for Paris-Durham models, as outlined by \citet{godard2019models}, including the use of H$_2$ lines. They suggest that the primary parameters influencing the type of shock present and the resulting distribution of H$_2$ line fluxes are the proton number density, $n_H$, the hardness of the radiation field, characterised by the UV field strength $G_0$, the shock velocity, $v_s$, and the strength of any potential magnetic field, quantified by the scaling factor $b$, where $B = b \times \sqrt{n_H(\text{cm}^{-3})}$ $\mu$G. Additionally, the models have minor dependencies on the cosmic H$_2$ ionisation rate, $\xi_{H_2}$, and the fractional abundance of PAHs relative to hydrogen, $X(\text{PAH})$.

The models predict that higher velocity shocks, or shocks in stronger magnetic fields, typically correspond to C-type (continuous) shocks, whereas lower velocity shocks, or shocks in weaker magnetic fields, are more likely to be J-type (jump) shocks. %(although this is because classifying as a jump shock requires the shock be supersonic, but the presence of a magnetic field alleviates this, so that even fast shocks can become continuous rather than jump when a magnetic field is present). 
This distinction between shocks leads to a larger proportion of the shock energy being dissipated via pure rotational transitions of excited H$_2$ in C-type shocks. They suggest that in a C-type shock, with the same input energy as a J-type shock, a stronger magnetic field causes the neutral and ionised fluids to decouple. This results in the sound speed in the medium being lower than the ion-magnetosonic speed, thus creating a magnetic precursor ahead of the shock front. 
%The precursor causes the input kinetic energy to be deposited into exciting H$_2$ over a larger physical distance. Consequently, each H$_2$ molecule excited by the shock receives less energy on average, leading to less overall heating of the H$_2$, making it more likely to be excited to lower energy levels, such as those corresponding to pure rotational transitions. As a result, for C-type shocks, the model predicts a significantly larger flux of rotational lines compared to vibrational lines, with the opposite being true for J-type shocks. 
Overall, C-type shocks excite H$_2$ to kinetic temperatures of $\sim 10^3$ K, whereas J-type shocks excite it to $\sim 10^4$ K, hence leading to larger vibrational line fluxes in J-type shocks compared to rotational.

\citet{kristensen2023shock} emphasise that analysing rotational lines alone often leave shock models degenerate. To address this, several studies (e.g. \citealt{davies2024gatos}; Delaney et al., submitted) combine JWST data with ground based near-IR observations of higher energy H$_2$ vibrational lines (e.g. 1-0 S(1), S(2), S(3)) to constrain plausible shock models. We adopt a similar approach, incorporating archival VLT/SINFONI rovibrational line fluxes together with additional ISM parameter modelling of the PDR, enabling us to select a non-degenerate and more robust shock model.

\begin{figure}
    \centering
    \includegraphics[width=0.85\linewidth]{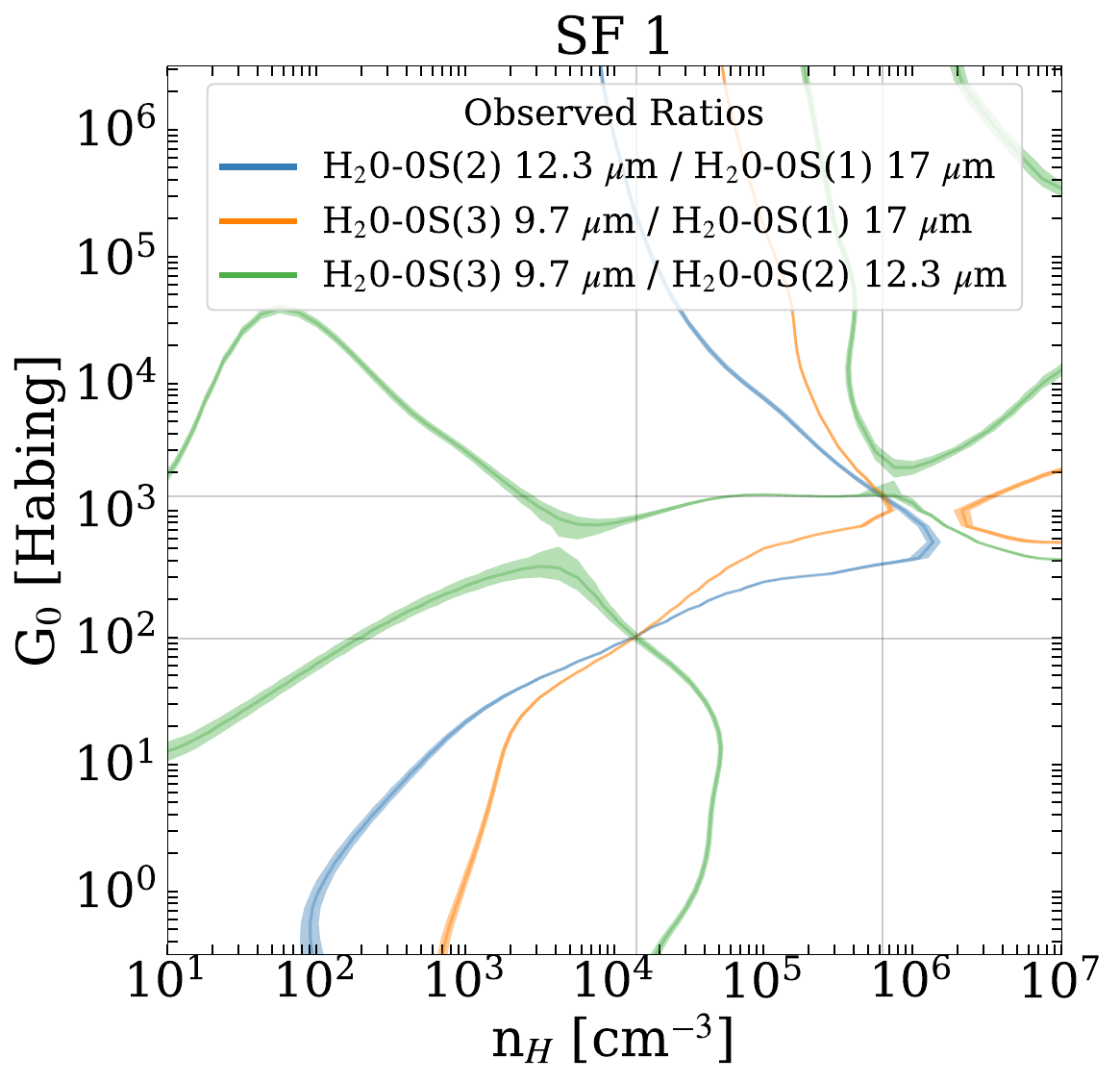}
    \includegraphics[width=0.85\linewidth]{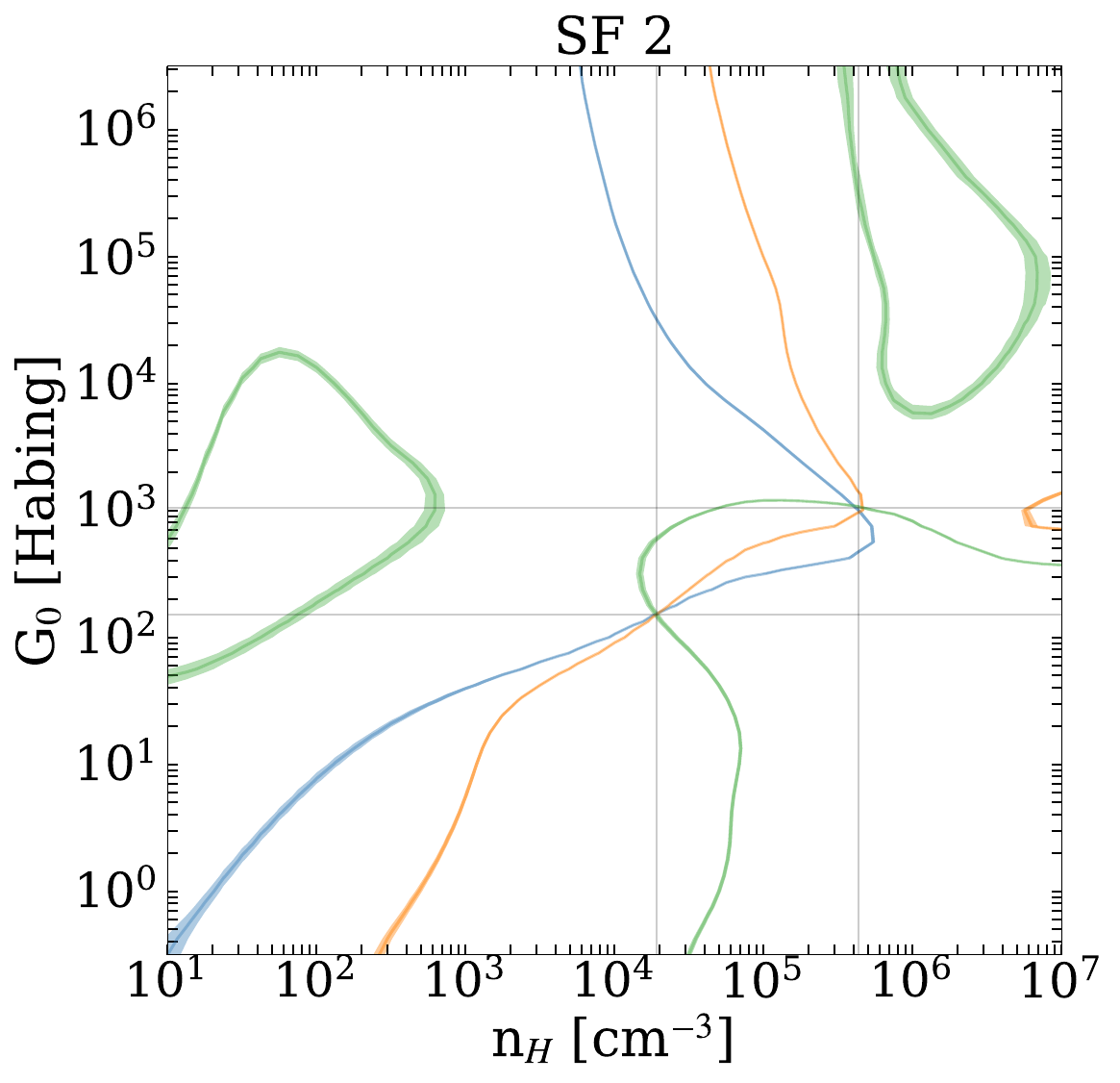}
    \includegraphics[width=0.85\linewidth]{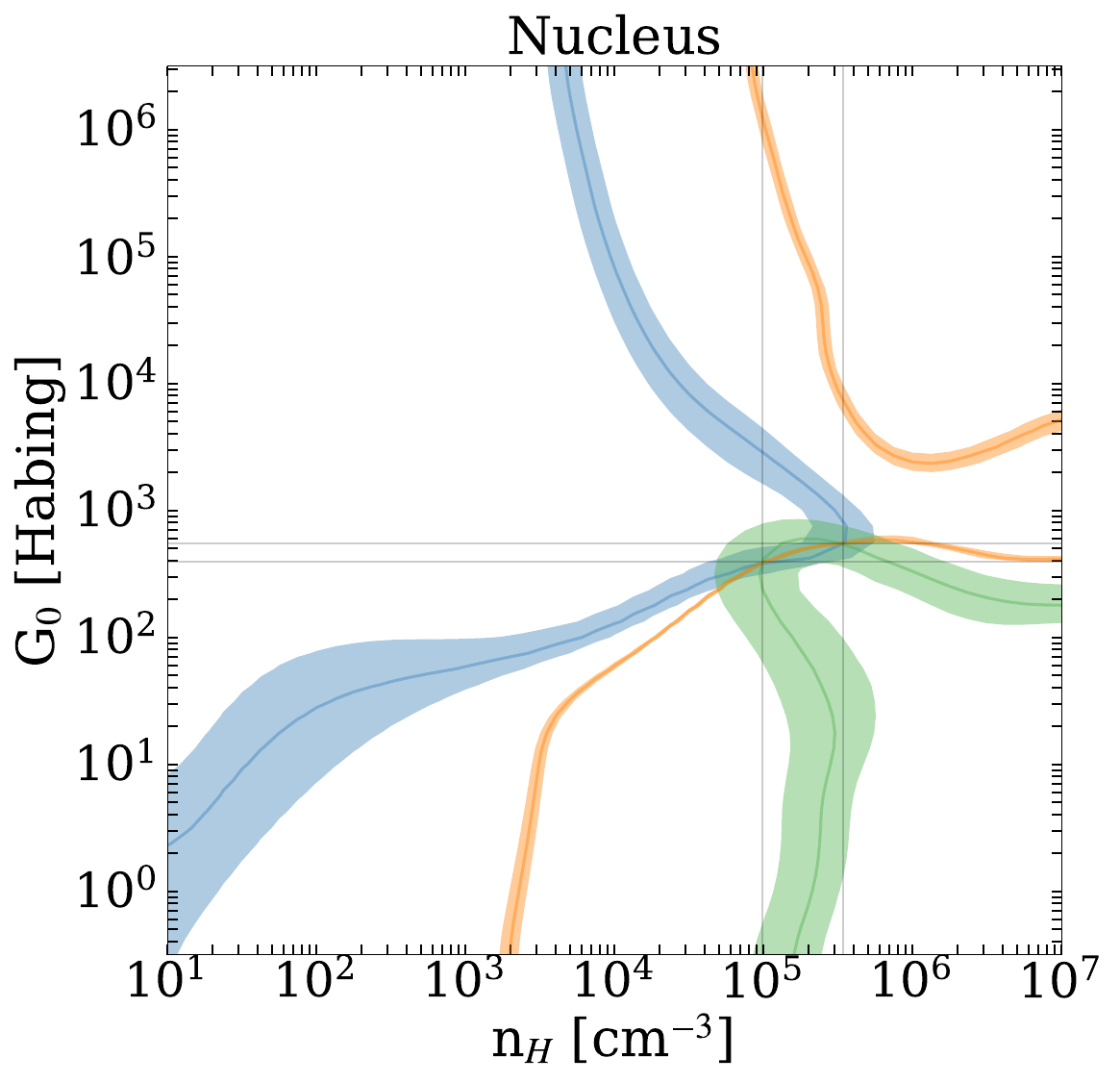}
    \caption{Inferred $G_0$ and $n_H$ for the nucleus, SF~1, and SF~2 from applying The PhotoDissociation Region Toolbox \citep{Pound_2022} with the observed three lowest rotational H$_2$ line ratios (found from inspecting where all three curves intersect). The widths vary due to uncertainties in the measured fluxes for each region. For SF~1 and SF~2, this gives two independent solutions, which only differ by roughly an order of magnitude in their predicted ISM parameters.}
    \label{fig:pdrtpy_plots}
\end{figure}

To better constrain the properties of the ISM we use the PhotoDissociation Region Toolbox, \texttt{pdrtpy}\footnote{\url{https://github.com/mpound/pdrtpy-nb}} \citep{Pound_2022}, which constrains physical conditions in PDRs. The PDR models span a range of $n_H$, and $G_0$, assuming no magnetic fields. By inputting at least three line ratios, the intersection of model contours provides the best-fitting values of $n_H$ and $G_0$. This is shown graphically in Fig.~\ref{fig:pdrtpy_plots} using line fluxes for 0-0 S(1), S(2), and S(3) in the nuclear, SF~1, and SF~2 apertures. This analysis yields the same two distinct solutions of $n_H \sim 10^{4}$ or $\sim 10^{5-5.5}$ cm$^{-3}$ and $G_0 \sim 10^{2}$ or $\sim 10^{3}$ Habing for both SF~1 and SF~2.

Our aim here is not to constrain ISM parameters with high precision, but rather with sufficient accuracy to resolve otherwise degenerate shock models. Although uncertainties of up to an order of magnitude remain, this precision is adequate since different shock models that reproduce similar H$_2$ excitation diagrams can typically predict ISM parameters differing by more than an order of magnitude. By incorporating independent constraints from PDRToolbox, we can exclude physically unrealistic scenarios and isolate the most plausible shock solution. This step is crucial, as degeneracies can persist even when both rotational and rovibrational lines are considered \citep{davies2024gatos}. The addition of PDRToolbox analysis therefore provides the necessary constraints to break all remaining degeneracies, yielding the most reliable shock solution and parameterisation.

We also limit our shock models to those with stronger magnetic fields, $b \gtrsim 1$. NGC~7582 shows prominent radio emission \citep{ward1980new, orienti2010radio} (with SF~2 being identified as region R3 in the latter, with SF~1 the clear extension eastwards of R3), which may drive AGN outflows. Previous radio studies, such as \citet{farnes2014spectropolarimetry}, revealed that the central region exhibits a high degree of linear polarisation, indicative of a significant Faraday depth. The presence of linear polarisation suggests that the synchrotron emitting plasma in the nucleus contains a well-ordered magnetic field component, as an entirely random field would result in depolarisation of the emission. This polarisation is therefore a direct consequence of a strong magnetic field \citep{frick2011faraday}. Quantitatively, these radio studies suggest that magnetic field strengths in the nuclear regions of NGC~7582 are typically $\gtrsim$1 mG, supporting the assumption that $b \gtrsim 1$ even at the inferred densities of $n_H \sim 10^{6}$ cm$^{-3}$. Consequently, low magnetic field shock models can be reasonably excluded from our analysis. The combination of these constraints, along with the ISM conditions inferred from the PDRToolbox, and the inclusion of rovibrational line fluxes, provides sufficient information to resolve the degeneracy between shock models.

We apply these constraints on our selection of appropriate shock models. Specifically, we require parameters to be roughly within $n_H \sim 10^{4}-10^{5.5}$ cm$^{-3}$, $G_0 \sim 10^2-10^4$ Habing, and restricted to $b \gtrsim 1$. We examine the model grids from the Paris-Durham Shock Code 1.1.0\footnote{\url{https://app.ism.obspm.fr/ismdb/}}, and compare the predicted H$_2$ fluxes with those measured in SF~1 and SF~2 for each shock model. We include all well detected rotational lines from JWST, 0-0 S(1) to S(7), along with the 1-0 S(1), 1-0 S(2), and 2-1 S(1) rovibrational lines from SINFONI.

\begin{figure}
   \centering
   \includegraphics[width=\columnwidth]{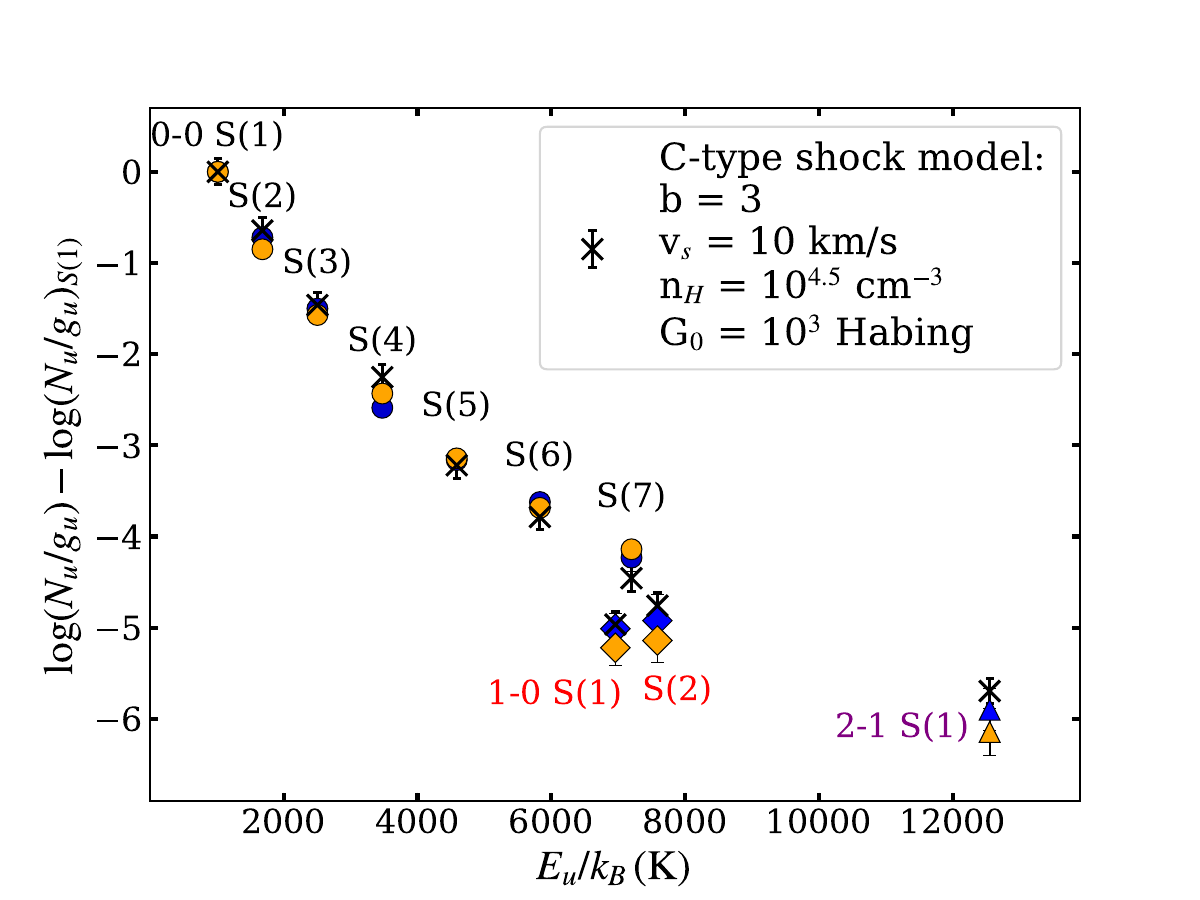}
   \caption{Excitation diagram of SF~1 (blue) and SF~2 (orange) JWST MIRI/MRS H$_2$ 0-0 S(1) to S(7) (circles), and VLT/SINFONI 1-0 S(1), 1-0 S(2) (diamonds), and 2-1 S(1) (triangle) fluxes with the best-fit Paris-Durham Shock Code 1.1.0 model prediction (crosses) with parameter values given in the legend.}
    \label{fig:shock_model_fits}
\end{figure}

We identify only two good fits to all available rotational and rovibrational line fluxes for SF~1 and SF~2, which happen to have nearly identical parameters, differing only in $n_H$ by one order of magnitude. These models both correspond to a C-type shock with $b = 3$ and $v_s = 10$ km/s, assuming $G_0 = 10^3$ Habing, $n_H = 10^4$ or $10^5$ cm$^{-3}$ respectively. We acquire our best overall fit model by linearly interpolating between them (as they differ only in $n_H$), this is presented in Fig.~\ref{fig:shock_model_fits}. These results align closely with the ISM conditions inferred from PDRToolbox, showing consistency between all of our analyses, while also predicting a low-velocity shock ($v_s < 30$ km/s), necessary for negligible H$_2$ dissociation. Furthermore, the presence of a strong magnetic field in this region, as inferred from \citet{farnes2014spectropolarimetry}, supports the plausibility of this shock scenario. The H$_2$ kinematics in Fig.~\ref{fig:rotational_maps} show elevated velocity dispersions in SF~1 and SF~2, exceeding those in surrounding ISM regions by approximately 10–20 km~s$^{-1}$. This enhancement supports the presence of the best fitting low-velocity shock model, which is responsible for the increased H$_2$ rotational temperatures observed in these regions.

The shock is of relatively low velocity and does not possess sufficient energy to dissociate H$_2$ or disrupt the star forming disk. Instead, it primarily acts to elevate the gas temperature, resulting in significantly enhanced H$_2$ rotational line fluxes. Moreover, the strong nuclear magnetic field plays a crucial role in shaping the shock structure. The presence of a magnetic precursor extends the shock front upstream, distributing kinetic energy over a larger physical scale. This process reduces the energy imparted to individual H$_2$ molecules, thereby suppressing dissociation while still maintaining sufficient excitation to produce the observed strong rotational transitions.

These results demonstrate that even low energy shocks could significantly influence the thermal and dynamical properties of the surrounding gas. The source of this shock is the focus of the final subsection.

%\subsection{Is there a jet in NGC~7582?}
\subsection{Is there a jet-ISM interaction in NGC~7582?}
\label{sec:Is_there_a_jet_subsection}

Throughout this paper, we have explored various scenarios to explain the observed properties of SF~1 and SF~2, considering whether the observed shock heating is driven by the AGN; either through direct radiation or feedback mechanisms such as the jet proposed by \citet{ricci2018optical}, or by local starburst activity. In this final subsection, we summarise our findings within the context of the ongoing debate about the presence of a jet in NGC~7582.

We emphasise that confirming or excluding a jet ultimately requires radio observations. IR data only allow us to infer its presence through its impact on the ISM rather than direct detection. Hence, the strongest evidence for the AGN jet comes from radio studies \citep{forbes1998star, orienti2010radio}, which reveal extended emission south of the nucleus. This feature has been interpreted as a possible jet by \citet{ricci2018optical}, who showed that the extended [Fe~II] and Br$\gamma$ emission in the southern regions cannot be explained by the established ionisation cone. They identified similar spectral signatures not only in SF~1, SF~2, but also in the northern star-forming clumps (M3, M4, M5), suggesting that these bright IR clusters trace a Seyfert jet rather than simply being dust-embedded. The extended [Fe~II] emission further indicates shocks south of the nucleus, consistent with a jet propagating through the surrounding ISM.

Furthermore, in our kinematic analysis of the warm molecular gas (Figs.~\ref{fig:rotational_maps} and~\ref{fig:PA_vel}),
% and particularly of the [Fe II] kinematics (Fig.~\ref{fig:FeII_velocities_and_dispersions}),
we observe slight additional blueshifting over the AGN position, especially in S(3), extending southwards across SF~1, roughly aligned with the jet direction proposed by \citet{ricci2018optical}. This blueshift spans a few spaxels, which could result from the PSF smoothing a more narrow kinematic feature produced by an AGN jet. The blueshift also roughly follows the enhanced southern [Fe II] emission (Fig.~\ref{fig:Fe_map}), tracing the shocked gas. Taken together, these features may indicate an AGN jet propagating through the ISM towards the observer and over SF~1 and SF~2, shock-heating the surrounding gas. We note that in a clumpy medium, such as NGC~7582 with its circumnuclear ring hosting multiple star-forming clumps, the AGN jet need not be collimated. While the radio emission may remain collimated, optical or IR counterparts may be absent, as the jet could fragment and redirect multiple times at varying velocities \citep{mukherjee2016relativistic, mukherjee2018relativistic}.

However, these shocks could instead be triggered by cloud–cloud collisions within the intermediate line region that hosts the starburst ring. Such collisions may also account for the non-circular motions seen in CO~(3–2) emission (Appendix D.1 of \citealt{garcia2021galaxy}). To reconcile our findings with other previous work, we argue that many of the `jet' signatures identified by \citet{ricci2018optical} are not unique to AGN jets but could also arise from intense starburst activity. Specifically, the observed asymmetries between the northern and southern emission regions pose a significant challenge to the jet hypothesis. For example, our [Fe II] emission map exhibits strong emission only in the south, a trend also noted by \citet{ricci2018optical} in their near-IR [N I] 5198 and [O I] 6300 line maps. As they discussed, attributing this asymmetry solely to high dust attenuation in the north due to the obscuring disk is insufficient, as the asymmetry is also evident in radio observations \citep{forbes1998star}, which we also argue appears more diffuse and knotty and not jet like in morphology. The radio maps presented in \citet{orienti2010radio} also show clumps of strong radio emission eastwards and northwestwards of the nucleus, which they also attribute to star formation instead of a jet. Hence, the absence of emission extending northwards in many IR lines and in the radio makes it difficult to justify the presence of an AGN jet. However, some of our JWST observations do not fully cover the northern edge of the star forming ring. Consequently, while we do not detect [Fe II] emission north of the nucleus in the coverage we do have, this absence does not imply that shocks or AGN jet impacts are not present further north in the ring.

Emission line flux maps of high IP lines such as [Mg IV] (Fig.~\ref{fig:Mg_map}) and [Ne V] (Fig.~\ref{fig:Ne_ratio}) reveal emission tracing the edges of the ionisation cone but no indication of interaction with a high ionisation jet. Furthermore, our low ionisation jet tracers; ([Fe II], H$_2$ S(5), and Pf$\beta$), show enhanced flux over SF~1 and SF~2, but again, no northern counterpart (even when northern FOV is available).
% A genuine AGN jet would require a corresponding northern component, which we see little evidence for from any of our emission line maps. Instead, the morphology of these emission lines appears extended and diffuse, which could be due to local starburst activity within the circumnuclear ring.
Additionally, our PDR modelling indicate that SF~1 and SF~2 exhibit nearly identical ISM conditions. If a jet with the morphology proposed by \citet{ricci2018optical} were penetrating SF~1 but not SF~2, it would be impossible for both regions to have strikingly similar ISM conditions across multiple tracers.

%\textcolor{red}{We also see evidence against the jet hypothesis from the distribution of polycyclic aromatic hydrocarbons (PAHs). We fit the 3.3 $\mu m$ feature using a Drude profile \citep{draine2007infrared}. The 3.3 $\mu m$ PAH flux distribution map (Fig.~\ref{fig:3p3PAHMAP}) reveals that PAHs are not only present but exhibit enhanced flux in \textcolor{red}{the circumnuclear ring, including} SF~1 and SF~2. It is likely that most PAHs would be significantly disrupted by jets \citep{zhang2022evidence}, making it impossible for them to be detected so strongly if these regions were impacted by an AGN jet. Analysing PAHs in NGC~7582 warrant a dedicated study of their own, which is the work of Garcia-Bernete et al. (in prep.). We refer the reader to their work for a more in depth PAH analysis. Moreover, increased PAH emission has previously been observed at shock fronts \citep{donnan2023obscured}, supporting the idea that slow shocks can shield PAH molecules, allowing their emission to persist. This is consistent with the enhanced PAH flux observed in SF~1 and SF~2 (Fig.~\ref{fig:3p3PAHMAP}).}

%\begin{figure}
%    \centering
%    \includegraphics[width=\linewidth]{graphics/3p3_PAH_map_3.pdf}
%    \caption{3.3 $\mu m$ PAH feature flux map showing enhanced PAH flux in SF~1 and SF~2. \textcolor{red}{It is likely that PAH molecules would be majorly dissociated if they were in line of an AGN jet.} The square region in the south west is an artefact caused by poor fits in those spaxels and does not correspond to a physical region.}
    %\label{fig:3p3PAHMAP}
%\end{figure}

We propose that the bright, clumpy emission in the southern circumnuclear regions could arise from intense starburst activity. Such vigorous star formation could naturally produce low velocity shocks, without invoking a specific fragmented or redirected AGN jet to explain the similar flux and velocity signatures in SF~1 and SF~2 seen across many emission lines. As these clumps originate from the same gas which forms the circumnuclear ring, it is plausible that they experience comparable star formation histories, consistent with previous studies \citep{wold2006nuclear, riffel2009agn}, which reasonably accounts for the similar ISM conditions observed. This scenario could also explain the observed north-south asymmetries in emission line distributions, the absence of a northern jet counterpart (within our coverage), 
% and the survival of PAHs,
all of which argue against a jet-driven origin.

% While the simplified, collimated AGN jet proposed by \citet{ricci2018optical} cannot fully account for the detailed observations presented here, a jet propagating through a clumpy ISM could plausibly split, redirect, and shock the gas at a range of velocities and morphologies. Such a scenario could contribute to the shock heating observed in the southern star forming regions.
% Supporting this possibility, \citet{ricci2018optical} reported enhanced [He II] emission within and around SF~2, which cannot be explained by starburst activity alone due to its high ionisation potential (54.4eV). 
% Our line maps further indicate that SF~2 is associated with slightly more highly ionised or dissociated gas, consistent with its location along the cone edge, somewhat elevated [Ne V]/[Ne III] ratios (Fig.~\ref{fig:Ne_ratio}), and marginally increased [S IV] emission (Fig.~\ref{fig:S_map}). Together, these observations suggest that SF~2 may be more strongly affected by AGN-driven processes than SF~1.

\section{Conclusions}
\label{sec:conclusions_section}

We have presented combined JWST NIRSpec and MIRI/MRS 3.87-28.1 $\mu m$ spectra of the inner nuclear region of the Seyfert 2 galaxy NGC~7582, focusing on the analysis of pure rotational H$_2$ $\nu = 0$-$0$ emission lines across multiple apertures. Our main findings are summarised as follows:

\begin{enumerate}[i]
    
\item Fitting three different LTE models to H$_2$ rotational lines S(1) to S(7) across six aperture regions consistently demonstrates that the two southern star-forming regions, SF~1 and SF~2, are significantly hotter than the smaller north-eastern star-forming region, SF~3, as well as the regions within the western-facing ionisation cone, Cone~1 and Cone~2. The temperature distribution of the warm molecular gas ($\sim 100 < T < 1000$ K) in SF~1 and SF~2 more closely resembles that of the AGN-powered nucleus. The best-fitting model, a power-law temperature distribution without a fixed ortho-to-para ratio, indicates that SF~1, SF~2, and the nucleus share a common lower cutoff temperature ($T_l \sim 300$ K, consistent with an ortho-to-para ratio of 3) and a similar power-law index within one standard error.

% Additionally, mapping the flux of the hot H$_2$ molecular gas via the vibrational $\nu = 1$-$0$ O$(8)$ line highlights that SF~1 and SF~2 contain a substantial amount of hot ($T > 1000$ K) molecular gas.

\item The ionisation state of the ISM in the nuclear regions, traced via comparing the [Ne II] 12.8 $\mu m$, [Ne III] 15.6 $\mu m$, and [Ne V] 14.3 $\mu m$ flux emission, indicates that only the nucleus is highly ionised, ruling out direct photoionisation as a dominant heating mechanism for SF~1 and SF~2. We also see no intrinsic [O IV] 25.9 $\mu m$ emission in these regions, suggesting that SF~1 and SF~2 are not photoionised by the AGN or any evidence in favour of them being heated by embedded clusters of hot O-type stars either.

\item By analysing mid-IR line ratios and H$_2$ rotational line fluxes with the PDRToolbox, we loosely constrain the ISM conditions in each aperture. The nucleus exhibits the highest ionisation, consistent with direct AGN heating, whereas the surrounding star forming regions have lower ionisation as expected for star forming regions.  

\item We look for evidence of shocked gas through spatial mapping of [Fe II] 5.34 $\mu m$ and [Mg IV] 4.49 $\mu m$. The [Fe II] emission exhibits a substantial spatial extension south of the nucleus, covering SF~1 and SF~2, suggesting that low-ionisation shocks may be responsible for their increased molecular temperature. We also see a positive correlation between [Fe II] and rotational H$_2$ flux in each aperture, supporting the idea the two are associated. In contrast, [Mg IV] emission is concentrated within and slightly west of the nucleus, tracing the position of the AGN ionisation cone, further supporting the notion that SF~1 and SF~2 are not highly ionised.

\item Kinematic evidence for shocked gas south of the nucleus shows increased velocity dispersions in rotational H$_2$ S(1), S(3), and S(5). The S(5) emission shows distinct spatial extent north and south of the nucleus, revealing a mini-spiral of warm molecular gas out to $\sim 200$ pc. 

\item A low velocity ($v_s \sim 10$ km/s) C-type shock model from the Paris-Durham Shock Code, assuming ISM conditions with a high magnetic field ($b \sim 3$), proton number density ($n_H \sim 10^{4.5}$ cm$^{-3}$), and UV radiation field strength ($G_0 \sim 10^3$ Habing), provides an excellent fit to the observed H$_2$ rotational and rovibrational line fluxes in SF~1 and SF~2. These ISM conditions are consistent with our PDRToolbox modelling and previous literature studies.

\item We investigated whether the observed shocks originate from the AGN jet proposed by \citet{ricci2018optical} interacting with the ISM or from intense starburst activity. The additional blueshifting around and south of the AGN, as well as elevated velocity dispersions extending over the southern star forming regions of the warm molecular gas kinematics, are consistent with a jet interacting with a clumpy medium. However, the absence of a clear northern counterpart (possibly due to limited spatial coverage), the knotty, diffuse appearance of the radio emission resembling star-forming clumps, the similar ISM conditions in SF~1 and SF~2,
% and the survival of PAHs
all favour starburst driven shocks from gas in the circumnuclear ring. Nevertheless, we cannot rule out either an AGN jet or starburst scenario.

\end{enumerate}

%García-Bernete et al. (in prep) aim to expand on our analysis by examining PAH-derived properties of the ISM in NGC~7582 \citep{2024MNRAS.532.1598R}. Early results also indicate unusually high PAH temperatures in the southern circumnuclear star-forming regions, supporting our evidence for enhanced heating. They also plan to extend this study to a larger sample of Seyfert galaxies, particularly those showing similarly elevated temperatures in NIRSpec and MIRI/MRS data, to assess whether the extreme heating in NGC~7582 is common and to better understand its physical drivers.

To conclude, this study finds that the southern circumnuclear star-forming regions in NGC~7582 show unexpectedly high molecular H$_2$ temperatures. Our analysis favours the presence of low-velocity C-type shocks, although the dominant heating mechanism remains unclear. This sheds light on the complex dynamics of circumnuclear disks and underscores the need for multi-faceted approaches to disentangle the roles of shocks, radiation, and outflows in AGN environments. We also show that star formation can significantly impact nearby photodissociation regions, altering H$_2$ excitation and diagnostic line ratios. These results demonstrate JWST’s ability to probe feedback processes in unprecedented detail, opening new avenues for studying AGN-impacted galaxy evolution.

\section*{Acknowledgements}

OV is supported by a Science and Technology Facilities Council (STFC) studentship No. ST/Y509474/1. SGB and FE acknowledge support from the Spanish grant PID2022-138560NB-I00, funded by
MCIN/AEI/10.13039/501100011033/FEDER, EU. MPS acknowledges support under grants RYC2021-033094-I, CNS2023-145506 and PID2023-146667NB-I00 funded by MCIN/AEI/10.13039/501100011033 and the European Union NextGenerationEU/PRTR. RAR acknowledges the support from Conselho Nacional de Desenvolvimento Cient\'ifico e Tecnol\'ogico (CNPq; Proj. 303450/2022-3, 403398/2023-1, \& 441722/2023-7), Funda\c c\~ao de Amparo \`a pesquisa do Estado do Rio Grande do Sul (FAPERGS; Proj. 21/2551-0002018-0), and Coordena\c c\~ao de Aperfei\c coamento de Pessoal de N\'ivel Superior (CAPES; Proj. 88887.894973/2023-00). AJB acknowledges funding from the "FirstGalaxies" Advanced Grant from the European Research Council (ERC) under the European Union's Horizon 2020 research and innovation program (Grant agreement No. 789056). SC and DJR acknowledge the support of the UK's Science and Technology Facilities Council (STFC) through grant ST/X001105/1. EB acknowledges support from the Spanish grants PID2022-138621NB-I00 and PID2021-123417OB-I00, funded by MCIN/AEI/10.13039/501100011033/FEDER, EU. CRA and AA acknowledge support from Agencia Estatal de Investigaci\AA Lon of the Ministerio de Ciencia, Innovaci \AA Lon y Universidades (MCIU/AEI) under the grant "Tracking active galactic nuclei feedback from parsec to kiloparsec scales", with reference PID2022-141105NB-I00 and the European Regional Development Fund (ERDF). AA acknowledges funding from the European Union grant WIDERA ExGal-Twin, GA 101158446. EKSH and DD acknowledge grant support from the Space Telescope Science Institute (ID JWST-GO-03535). SFH acknowledges support through UK Research and Innovation (UKRI) under the UK government’s Horizon Europe funding Guarantee (EP/Z533920/1) and an STFC Small Award (ST/Y001656/1). AAH and LHM acknowledge support from grant PID2021-124665NB-I00 funded by the Spanish Ministry of Science and Innovation and the State Agency of Research MCIN/AEI/10.13039/501100011033 and ERDF A way of making Europe.

%%%%%%%%%%%%%%%%%%%%%%%%%%%%%%%%%%%%%%%%%%%%%%%%%%
\section*{Data Availability}

The data cubes used here are publicly available from the Mikulski Archive for Space Telescopes (MAST)\footnote{\url{https://mast.stsci.edu/portal/Mashup/Clients/Mast/Portal.html}}, with DOI: 10.17909/9sc5-q436.



\bibliographystyle{mnras}
\bibliography{references} % if your bibtex file is called example.bib

%\begin{thebibliography}{99}
%\end{thebibliography}

%%%%%%%%%%%%%%%%%%%%%%%%%%%%%%%%%%%%%%%%%%%%%%%%%%

%%%%%%%%%%%%%%%%% APPENDICES %%%%%%%%%%%%%%%%%%%%%

\appendix

\section{Supporting material}

In order to facilitate reproducibility of our results, Table~\ref{tab:appendix_table} provides the relative RA and Dec of the centre of each of our 0.3 arcsecond apertures from the nucleus (which was centred on the spaxel with the highest amount of overall flux across the entire NIRSpec datacube).

\begin{table*}
    \centering
    \begin{tabular}{lccc}
    \hline
    \textbf{Aperture} & $\Delta$RA ($^{\prime\prime}$) & $\Delta$Dec ($^{\prime\prime}$) & $\tau_{9.8}$ \\
    \hline
    Nucleus     & 0  & 0 & 2.67 \\
    SF 1     & +0.127  & –1.30 & 1.29 \\
    SF 2     & –0.565  & –1.20 & 1.61 \\
    SF 3     & +0.642  & +0.90 & 1.23 \\
    Cone 1   & –0.863  & +0.60 & 2.39 \\
    Cone 2   & –1.173  & –0.40 & 1.29 \\
    \hline
    \end{tabular}
    \caption{Relative RA and Dec offsets for each aperture from the nucleus, as well as the derived $\tau_{9.8}$ for each regions after fitting for differential extinction.}
    \label{tab:appendix_table}
\end{table*}

Table~\ref{tab:appendix_table} also provides our derived $\tau_{9.8}$ parameter for each region from applying the differential extinction tool, which we used for correcting for the effects of dust extinction on our measured spectra shown in Fig.~\ref{fig:uncorrected_fluxes}.

% \textcolor{red}{We also provide kinematics maps for the Pf-$\beta$ hydrogen recombination line in Fig.~\ref{fig:velocities_and)dispersions}. Overall this line kinematics follows the same disk rotation as seen in the H$_2$ rotational line kinematics, Fig.~\ref{fig:rotational_maps}.}

% \begin{figure*}
%    \centering
%    \includegraphics[width=0.49\linewidth]{graphics/Pfbeta_vel_2.pdf}
%    \includegraphics[width=0.49\linewidth]{graphics/Pfbeta_vel_disp_2.pdf}
%    %\includegraphics[width=0.49\linewidth]{graphics/SIV_vel.png}
%    %\includegraphics[width=0.49\linewidth]{graphics/SIV_vel_disp.png}
%    %%%\includegraphics{empty.eps}
%    %%%\includegraphics{empty.eps}
%    \caption{Velocity and velocity dispersion maps for Pf$\beta$. Contours show the line flux emission.}
%     \label{fig:velocities_and)dispersions}%
% \end{figure*}

\subsection{Modelling H$_2$ excitation diagrams}
\label{sec:appendix_H2}

To understand the observed flux of H$_2$ rotational lines, we begin by considering the Boltzmann equation, which describes the population distribution across molecular energy states in thermodynamic equilibrium. This can be adapted for a power-law distribution of temperatures, giving the column density, $N_i$, in a given state $i$ in the following form:

\begin{equation}
    N_i = \int^{T_u} _{T_l} \frac{g_i e^{-E_i / k_B T}}{Z(T)}  m T^{-n} dT .
\end{equation}

This is simply the Boltzmann equation when assuming our power-law temperature distribution, Equation~(\ref{eq:power-law-temp}), as integrating over just the second part within the integral ($m T^{-n}$) with respect to temperature would give the total column density. Here $g_i$ is the statistical weight of state, $E_i$ is the energy of the state, $ k_B$ is the Boltzmann constant, $T$ is the excitation temperature, and $Z(T)$ is the partition function; a temperature dependent sum over all possible states:

\begin{equation}
    Z(T) = \sum_i g_i e^{-E_i / k_B T}.
\end{equation}

The relative population of these states is determined by the exponential term, meaning that higher energy states become increasingly populated as the temperature rises. The observed flux of a given rotational line is directly proportional to the column density in the corresponding upper energy state. For an optically thin transition between levels $J+2 \rightarrow J$, the observed flux can then be expressed as:

\begin{equation}
    \label{eq:column_dens}
    F_J = \frac{h \nu A_J N_{J+2} \Omega}{4 \pi},
\end{equation}

where $h$ is Planck's constant, $\nu$ is the frequency of the transition, $A_J$ is the Einstein spontaneous transition rate, and $\Omega$ is the solid angle of the aperture. Here, the column density $N_{J+2}$ represents the population in the upper state of the transition.

To construct the H$_2$ excitation diagram, we plot the inferred column density for each state, $N_{J+2}$, normalised by its statistical weight, $g_{J+2}$. Assuming a fixed ortho-to-para ratio of 3:1 simplifies the statistical weights to $g_J = 2J + 1$ for even (para-) H$_2$ and $g_J = 3(2J + 1)$ for odd (ortho-) H$_2$.

By plotting these column densities logarithmically (we use log$_{10}$ here) against the upper state energy divided by Boltzmann's constant, $E_u / k_B$, we obtain the excitation diagram. This representation allows the excitation temperature to be inferred from the slope of the data points. While the upper state energy often exceeds the actual gas temperature (e.g., $E_u / k_B \sim 1050$ K for the S(1) transition), recall that significant emission can still arise from cooler regions (e.g., down to $T \sim 150$ K).

\begin{figure*}
    \centering
    \includegraphics[width=0.42\linewidth]{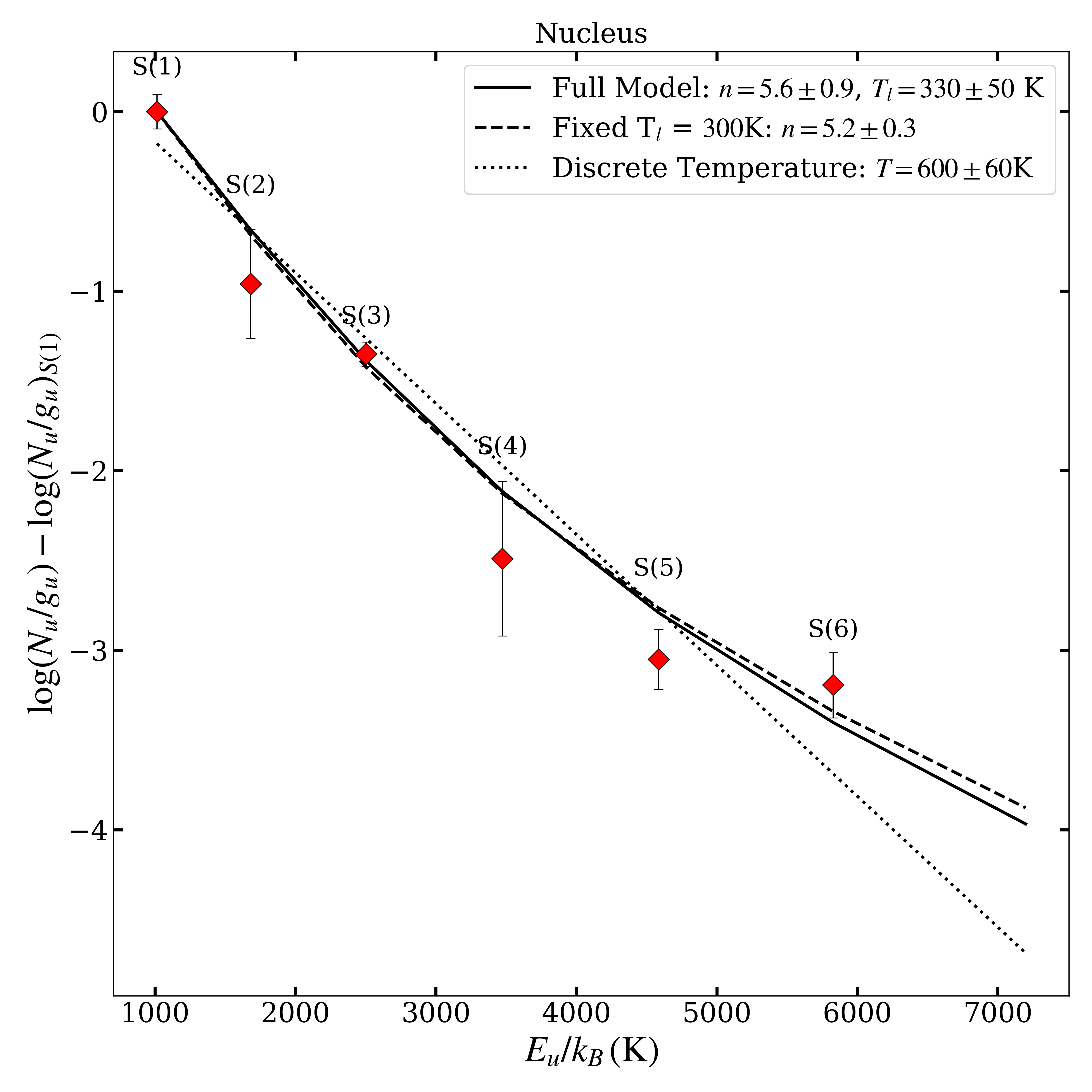}
    \includegraphics[width=0.42\linewidth]{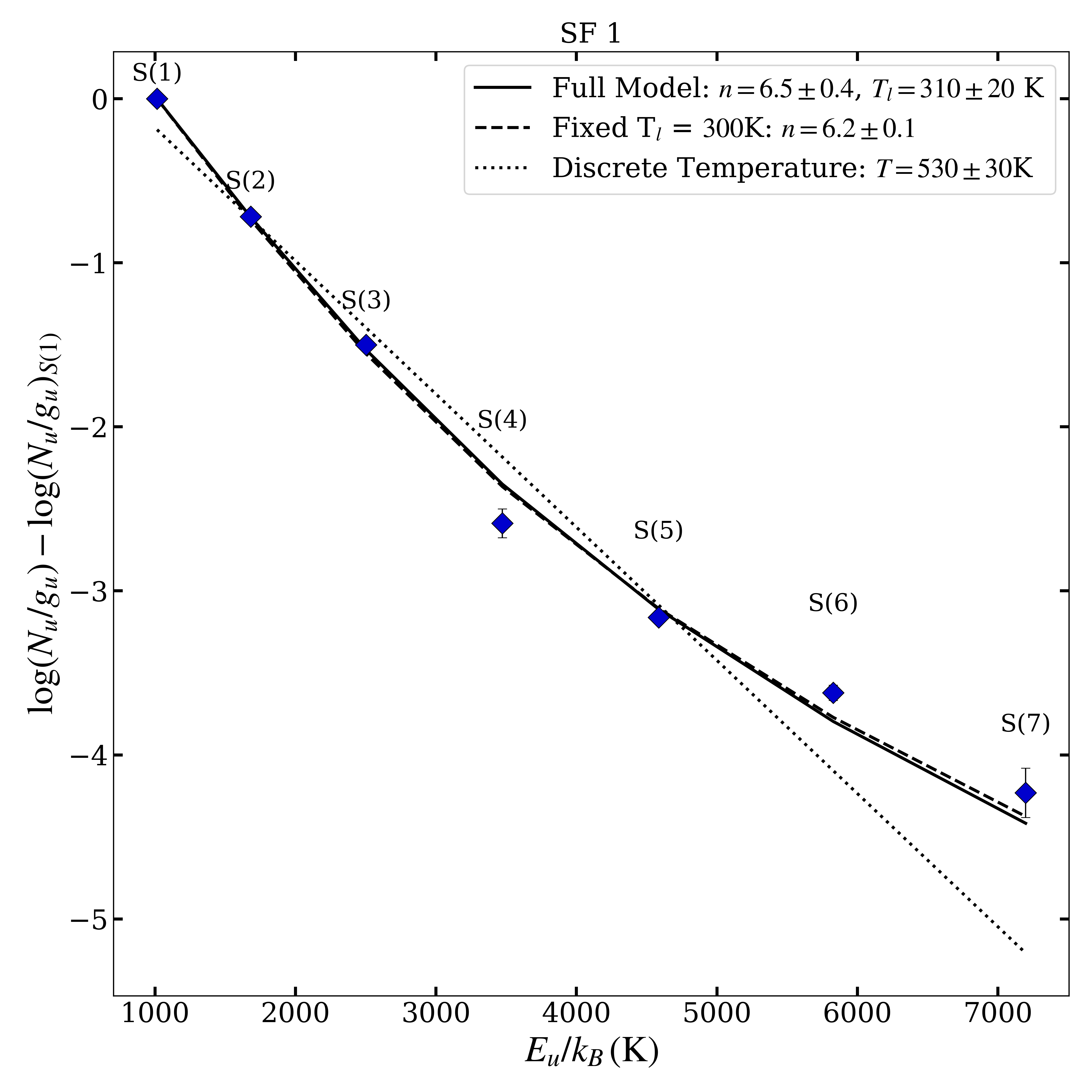}
    \includegraphics[width=0.42\linewidth]{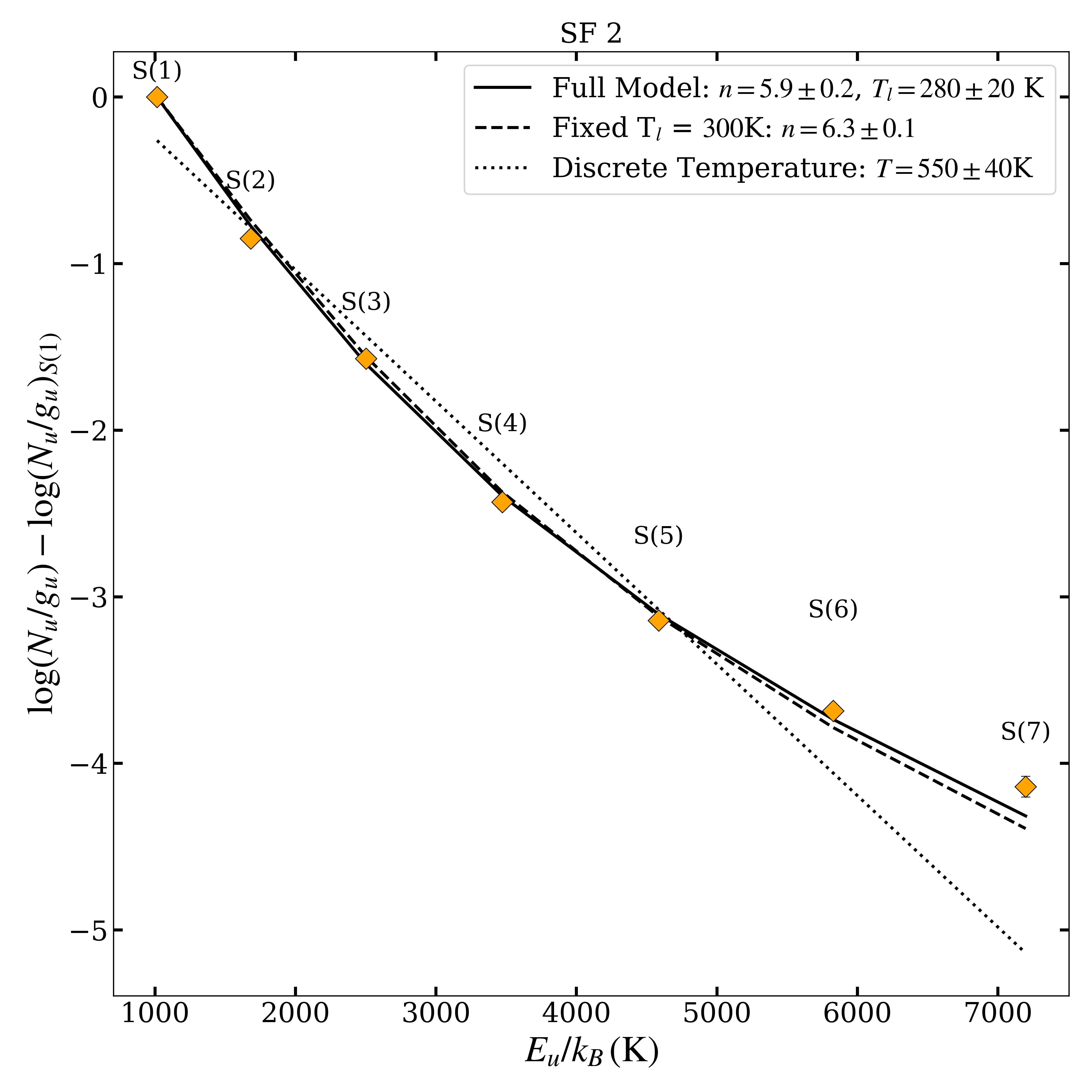}
    \includegraphics[width=0.42\linewidth]{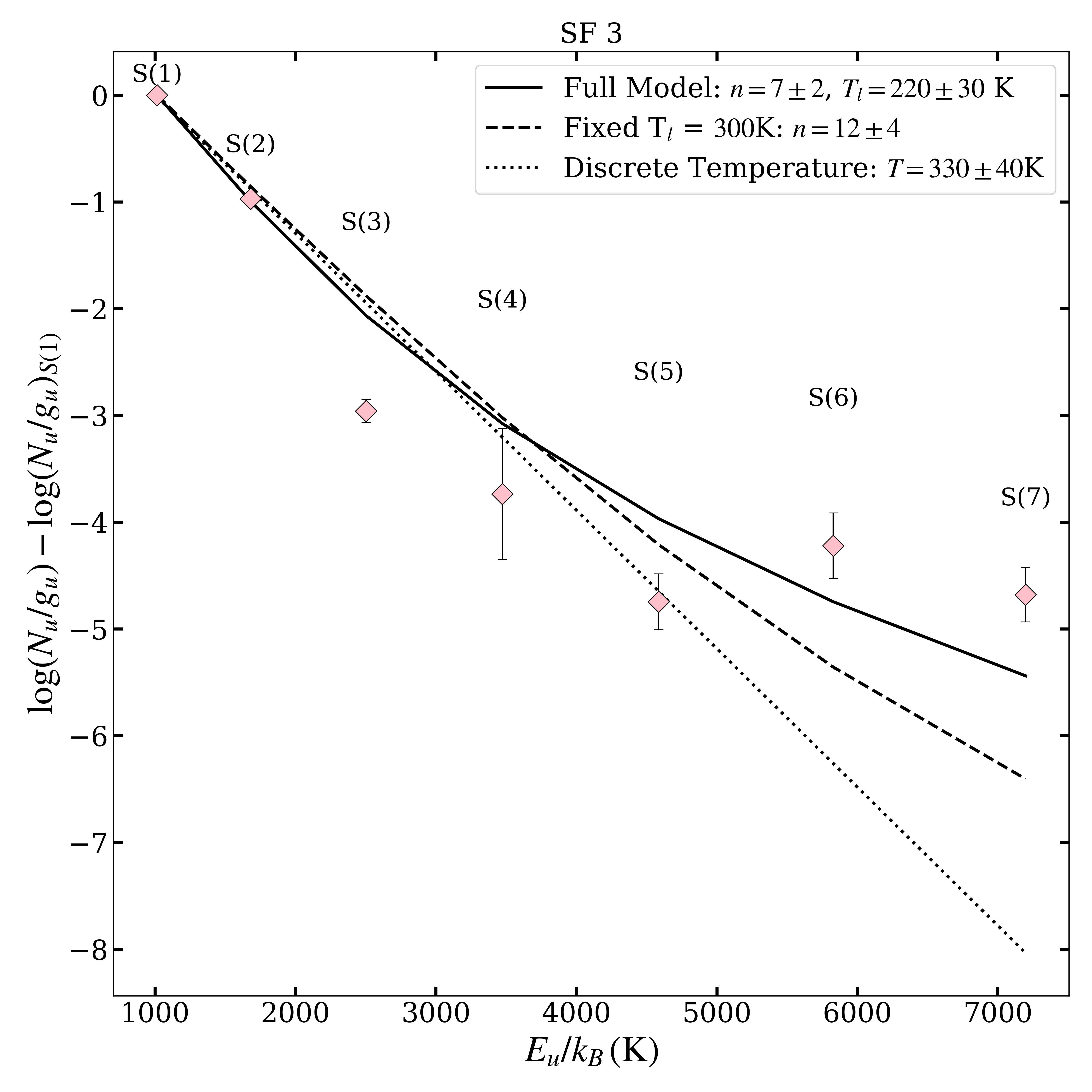}
    \includegraphics[width=0.42\linewidth]{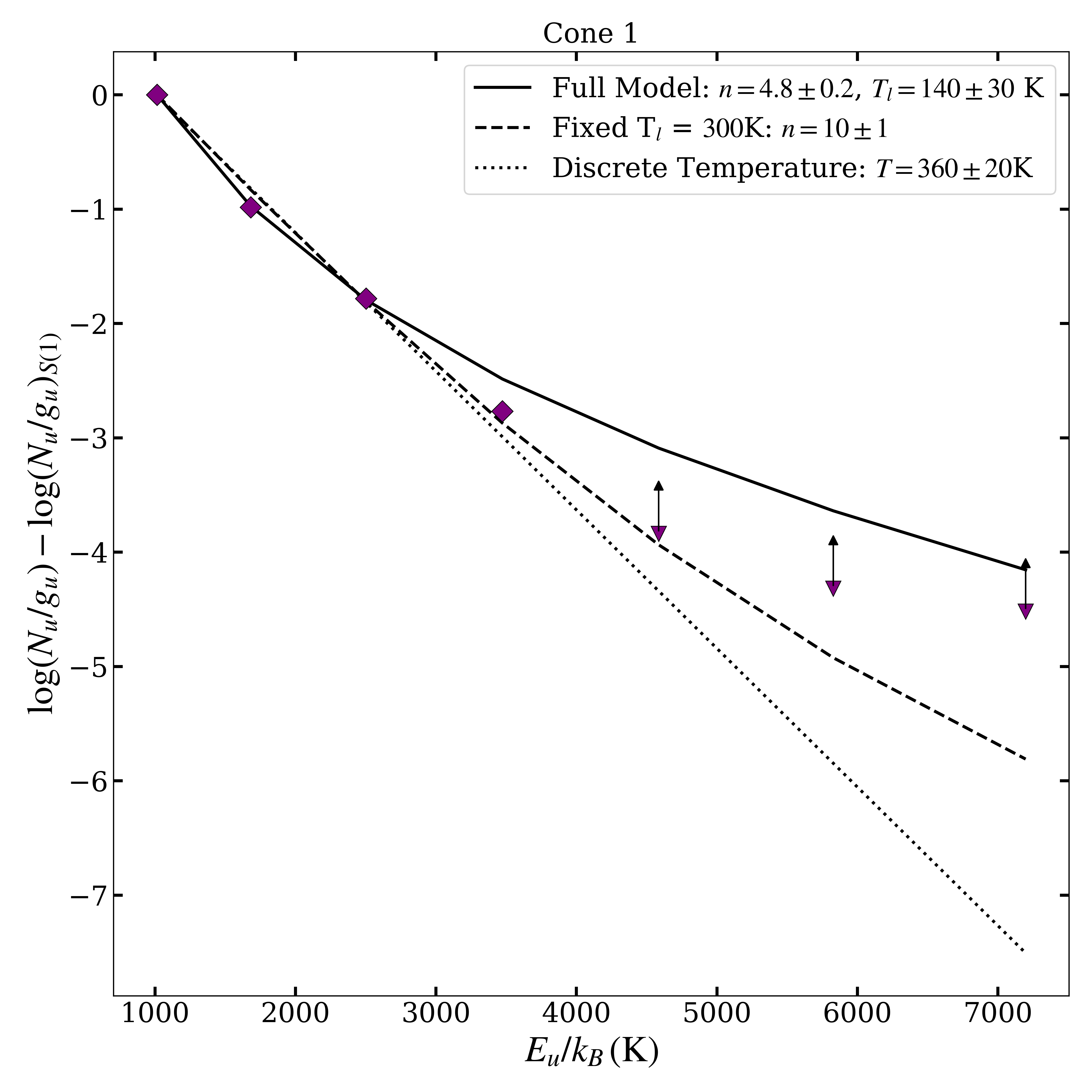}
    \includegraphics[width=0.42\linewidth]{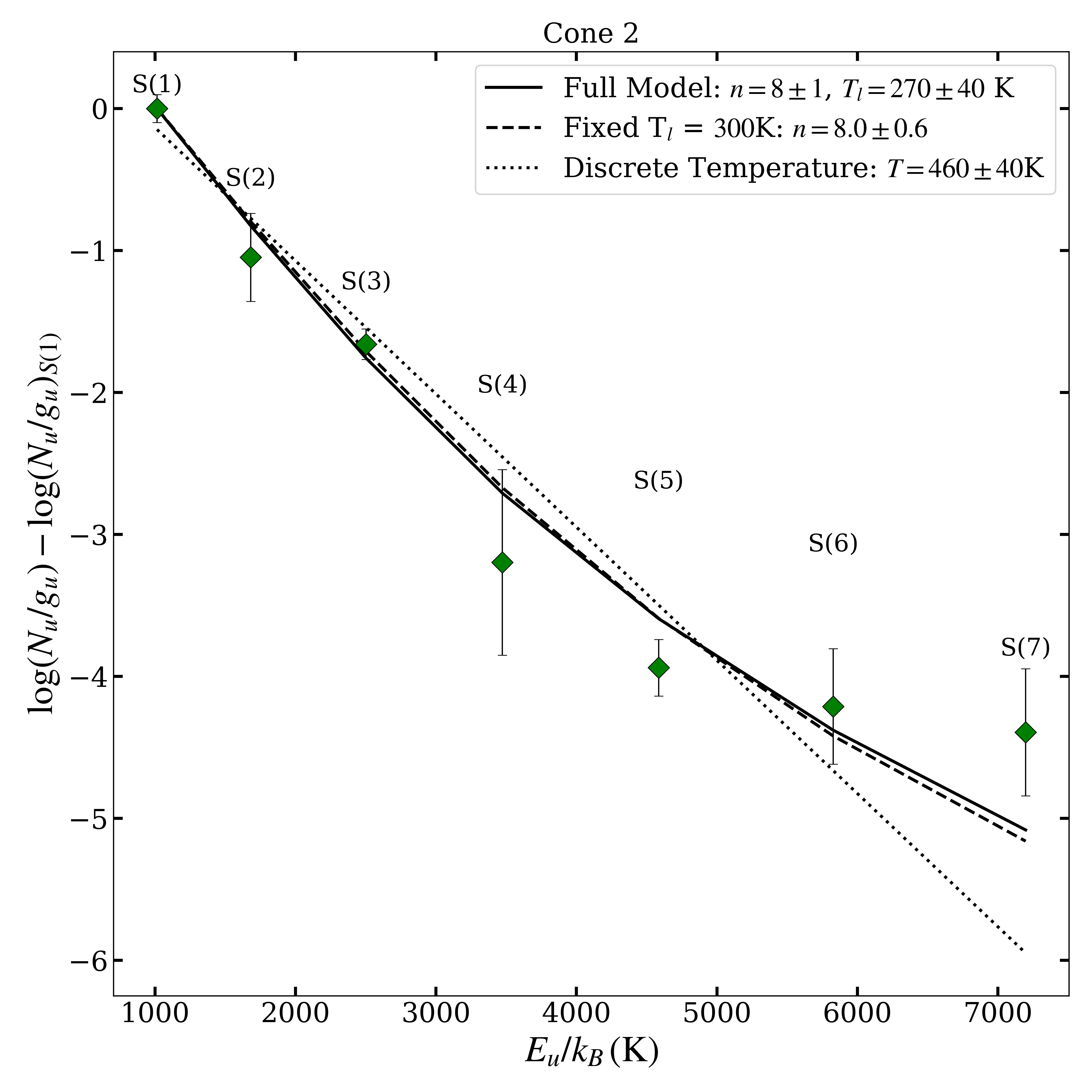}
    \caption{excitation diagram fits shown for each aperture separately.}
    \label{fig:separate_rotation_fits}
\end{figure*}

We also provide the excitation diagrams for every region plotted separately in Fig.~\ref{fig:separate_rotation_fits}, as opposed to Fig.~\ref{fig:TS16_fits} in which every region was plotted together for each model. Plotting based on region instead (as in Fig.~\ref{fig:separate_rotation_fits}) allows for easier comparison between models for each region.

\bsp	% typesetting comment
\label{lastpage}
\end{document}